# Synthesis, structure and electric conductivity of higher hydrides of ytterbium at high pressure

Tomasz Jaroń,[a,b,c*] Jianjun Ying,[b,d] Marek Tkacz,[e] Adam Grzelak,[a] Vitali B. Prakapenka,[f] Viktor. V. Struzhkin[b,g*], Wojciech Grochala[a*]

[a]Centre of New Technologies, University of Warsaw, Banacha 2c, 02-097 Warsaw, Poland, [b]Geophysical Laboratory, Carnegie Institution of Washington, 5251 Broad Branch Road NW, Washington, DC 20015, USA, [c]Faculty of Chemistry, University of Warsaw, Pasteura 1, 02-089 Warsaw, Poland, [d]HPCAT, Geophysical Laboratory, Carnegie Institution of Washington, Argonne, IL 60439, USA, [e]Institute for Physical Chemistry, Polish Academy of Science, Warsaw, Poland, [f]Consortium for Advanced Radiation Sources, The University of Chicago, Chicago, IL 60637, USA, [g]Center for High Pressure Science and Technology Advanced Research, Shanghai 201203, China.

*e-mail: tjaron@uw.edu.pl, viktor.struzhkin@hpstar.ac.cn, w.grochala@cent.uw.edu.pl

ABSTRACT

While most of the rare earth metals readily form trihydrides, due to increased stability of the filled 4$f$ electronic shell for Yb(II), only YbH$_{2.67}$, formally corresponding to Yb$^{II}$(Yb$^{III}$H$_4$)$_2$ (or Yb$_3$H$_8$), remains the highest hydride of ytterbium. Utilizing diamond anvil cell methodology and synchrotron powder x-ray diffraction we have attempted to push this limit further $via$ hydrogenation of metallic Yb (at room temperature and heated $in\ situ$) and of Yb$_3$H$_8$. Compression of the latter has also been investigated in a neutral pressure transmitting medium (PTM). While the $in\ situ$ heating of Yb facilitates the formation of YbH$_{2+x}$ hydride, we have not observed the clear qualitative differences between the systems compressed in H$_2$ and He or Ne PTM. In all these cases a sequence of phase transitions from the unit cells of $P$-31m symmetry to the $I$4/m and $I$4/mmm systems occurred within $ca.$ 13–18 GPa and around 27 GPa, respectively. At the same time, the molecular volume of the systems compressed in H$_2$ PTM is $ca.$ 1.5% larger than of those compressed in inert gases, suggesting a small hydrogen uptake. Nevertheless, hydrogenation towards YbH$_3$ is incomplete, and polyhydrides do not form up to the highest pressure studied here ($ca.$ 75 GPa). As pointed out by our electronic transport measurements under compression in NaCl PTM, the mixed-valence Yb$_3$H$_8$ retains its semiconducting character up to at least 50 GPa, although the very low remnant activation energy of conduction (<5 meV) suggests that the metallization under further compression should be achievable. Finally, we provide a theoretical description of a hypothetical stoichiometric YbH$_3$.

1. Introduction

The enormous interest in diverse hydrogen-rich systems manifested in the last decades can be linked mainly to the two topics – energy storage in chemical compounds and phonon-driven superconductivity (SC), depending on the nature of systems studied. Due to the high gravimetric efficiency required for the energy storage in most potential applications, the first area has been dominated by the materials composed predominantly of light elements such as Li, B, Al, etc.[1–4] On the other hand, the hydrides which have been studied as potential superconductors are not limited in this respect and their components cover virtually the entire periodic table.[5–8]

The recent upsurge in the search for high-temperature SC in hydrides under high pressure has been stimulated by Neil Ashcroft's idea of "chemical precompression" of hydrogen contained in hydrogen-rich chemical compounds.[9] In such hydrogen-dominated systems (in terms of an atomic fraction) the hydrogen sub-lattice is expected to contribute to the metallic state at drastically lower pressures compared to elemental hydrogen. This would finally allow for taking an advantage of the large phonon frequencies related to the lightweight hydrogen atoms to raise the critical temperature of SC. SC might occur provided that the system features a high density of electronic states at the Fermi level and strong electron-phonon coupling. As a consequence of this concept, many binary and even more complex hydride-based systems have been screened theoretically[10] leading ultimately to experimental discovery of superconductivity in some of them. Several hydrides, especially those containing the $p$-block and rare-earth ($RE$) elements,[7] reveal strikingly-high critical temperatures of superconductivity ($T_C$): compressed $H_2S$ forming superconducting $H_3S$ phase of $T_C$ = 203 K at 155 GPa,[11] an incompletely identified C-S-H material synthesized in a diamond anvil cell (DAC) of $T_C$ = 288 K at 267 GPa,[12] $LaH_{10}$ of $T_C$ = 250–260 K upon compression to $ca.$ 200 GPa,[13–15] or $YH_6$ of $T_C \approx$ 220 K at 166–183 GPa[16,17] and $YH_9$ of $T_C$ = 243–262 K at 182–201 GPa.[17,18]

Considering the $RE$-based systems, in these superhydrides the hydrogen sub-lattice is usually composed of clathrate-like and other complex geometrical motifs or contain mixed molecular and atomic units.[5,10,19] While the polyhydrides of exotic compositions may form in a megabar pressure range, they are preceded by more usual stoichiometries appearing at much lower pressure. Some of the latter also superconduct, as it has been verified experimentally or as expected based on computational screening.[7] Obviously, lowering the pressure of synthesis of a superconducting hydride or metastability of such phases under as close to ambient conditions as possible would be beneficial for more thorough macroscopic studies and possible applications. Recently, such high-temperature superconducting phases have been found for the hydrides based on lanthanides below the megabar pressure region.[20,21]

With the exception of Eu and Yb, all lanthanides are easily hydrogenated at room temperature, absorbing up to 300 mol% of hydrogen to form trihydrides, $LnH_3$, already under the pressure of a few atmospheres,[22] see also *Fig. S1* in the Supplementary Information (SI). However, due to the stabilization of half-filled or filled *4f* electronic shells for Eu(II) and Yb(II), respectively, oxidation of these elements to the trivalent state by hydrogen is much more difficult. While the corresponding dihydrides can be readily prepared,[22] further absorption of hydrogen requires significantly higher pressure. The uptake of hydrogen by $EuH_2$ occurs only above 8.7 GPa, as has been reported by Matsuoka *et al.* based on x-ray diffraction and Eu-Mössbauer spectroscopy measurements.[23] At this pressure range a tetragonal $EuH_{2+x}$ containing trivalent $Eu^{3+}$ starts to form, and the latter remains the only detected oxidation state of Eu above 12.5 GPa.[24] Very recently, Semenok *et al.*[25] reported that $EuH_9$ is present in the samples already at 74 GPa, while the other europium superhydrides were detected within the range 74–130 GPa, some of them showing antiferromagnetic ordering. Unfortunately, the lower-pressure regime has not been studied in this work, leaving the gap between *ca.* 50 GPa for which $EuH_{2+x}$ phases were reported and 74 GPa with the superhydride phases already formed.

In the case of ytterbium, some early reports indicated preparation of cubic $YbH_{2+x}$, x <0.7, after reaction of this metal with hydrogen pressurized to *ca.* 120 bar.[26–28] These findings have been clarified by Auffermann, who prepared the metastable higher hydride of ytterbium *via* a high-pressure, high-temperature reaction (200 – 3200 bar, >600 K).[29] While the major features of the powder x-ray diffraction pattern of $YbH_{2+x}$ are compatible with an *fcc* unit cell reported earlier, application of powder neutron diffraction allowed for proper identification and refinement of its crystal structure. It appeared that the prepared compound can be described as $YbH_{2.67}$, *i.e.* $Yb_3H_8$ or $Yb^{II}(Yb^{III}H_4)_2$, and it crystallizes in a trigonal unit cell with close packing of the nine ytterbium atoms (stacking sequence: …ABC…), while hydrogen occupies all the tetrahedral and 2/3 of the octahedral holes, and no disorder is observed in the structure. The crystal structures of the higher hydrides of Yb and Eu relevant to the current study were summarized in *Figure 1*.

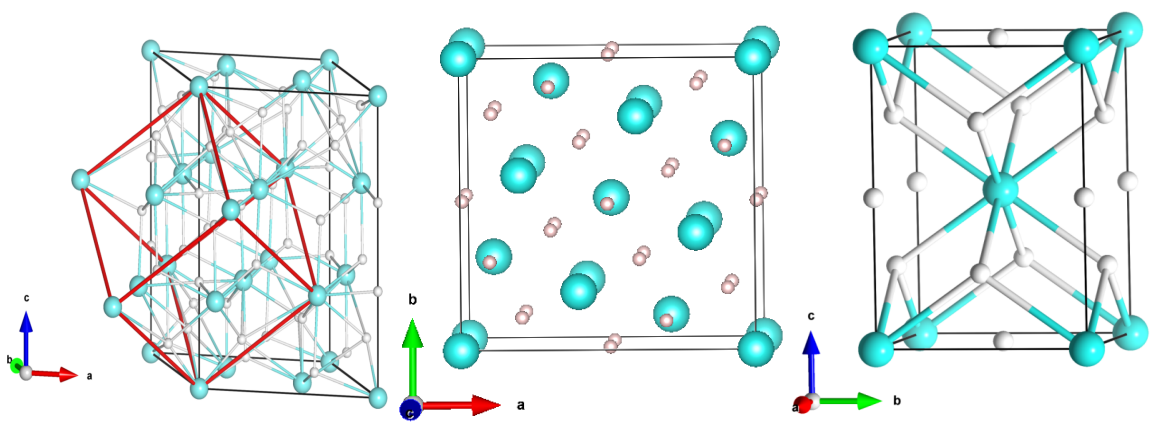

*Figure 1. Overview of the crystal structures of $LnH_{2+x}$, Ln = Yb, Eu: (a) $Yb_3H_8$ ($YbH_{2.67}$) P-31m;[29] (b) $EuH_{2+x}$ I4/m; (c) $EuH_{2+x}$ I4/mmm.[23]*

The case of Yb(III) is special.[2] The vicinity of 4f shell closure for $4f^{13}$ electron configuration of $Yb^{III}$ resembles the $3d^9$ configuration of parent oxocuprate superconductors, and the early DFT calculations indicated the possibility of significant contribution of H 1s states to the electron density at the Fermi level even for the stoichiometric Yb(III) hydrides.[30] Markedly, the only hole in the f electron set usually resides on the orbital which forms sigma* states with H(1s), which is analogous to a hole in the d set of Cu(II) residing in sigma* states with O(2p) in cuprates. Moreover, Yb(III) features a large magnetic moment of *ca.* 4.3–4.9 $m_B$ corresponding formally to spin-½ which again renders it similar to Cu(II) in parent compounds of oxocuprate superconductors; thus, one might expect that electron doping could affect magnetism and lead to SC, just like for Cu(II) oxides. Last but not the least, more recent theoretical considerations suggest that besides yttrium and the lightweight lanthanides (Ln)[31–33] the late lanthanides (Yb and Lu) should also form the superhydrides of $T_C$ >100 K.[34]

The above findings have stimulated the present work in which we expand the chemistry of Yb-H system far beyond the previously studied range of *ca.* 0.3 GPa. Using DAC methodology we investigate the reaction of metallic Yb with $H_2$ and attempt further hydrogenation of $Yb_3H_8$ ($YbH_{2.67}$). We discuss the observed phase transitions as well as the possibility of formation of the higher hydrides of ytterbium, *i.e.* $YbH_{2+x}$, especially with x>0.67. We have also studied the electronic transport of the mixed-valence $Yb_3H_8$ in the function of external pressure.

2. **Methods**

Experimental

Metallic Yb (99.9%, Sigma-Aldrich) and Au pressure standard (99.999%, Alfa Aesar) have been loaded into the DAC. Diamonds of 200–300 μm culets and rhenium gaskets have been used. $Yb_3H_8$ has been prepared according to the literature procedure,[29] and analyzed at ambient conditions using powder X-ray diffraction, *cf. Fig. S2,* and *Tab. S1* in the SI. Ne, He, and $H_2$, loaded at *ca.* 170 MPa in the custom-designed gas loading systems of GL CIW and GSE CARS APS,[35] were applied as the pressure-transmitting medium (PTM).

High-pressure angle-dispersive x-ray diffraction (XRD) measurements have been performed using Advanced Photon Source synchrotron facility of Argonne National Laboratory. The measurements were carried out on sectors 13ID-D and 16ID-B operating at wavelengths from 0.2952 Å to 0.4066 Å. The sample to detector distance and other geometrical parameters were calibrated using $LaB_6$ or CeO$_2$ standards. One of the samples of Yb in H$_2$ PTM was heated by the double-sided laser systems available at the beamlines, to the maximum temperatures of about 2100 K measured by fitting grey body thermal radiation.[36] Heating was carried out in several pulses lasting a few seconds, however, the overall amount of energy absorbed by the sample was not monitored. The heating was not uniform across the sample, which has manifested as variable amounts of crystalline phases present in various areas of the sample.

The electronic transport properties were carried out in a custom, miniaturized diamond anvil cell in a classical four-electrode geometry, using a Quantum Design Physical Property Measurement System. NaCl has been used as a PTM, and the pressure has been measured using ruby fluorescence.

The two-dimensional diffraction images were analyzed and integrated using the DIOPTAS software.[37] X-Cell has been used for pattern indexing.[38] The structures were refined in JANA2006.[39] The pseudo-Voigt function with Berar-Baldinozzi correction for asymmetry has been utilized for modeling of diffraction peak shape. The background was corrected by Legendre polynomials. The third-order Birch-Murnaghan equations of state (EoS) were fitted using EoSFit7-GUI program.[40,41]

Computational

Density Functional Theory (DFT) calculations were performed using CASTEP.[42] Generalized Gradient Approximation (GGA) was used with PBE functional adjusted for solids (PBEsol).[43] As hydrides usually require a large cutoff, here the value of 700 eV was applied to lead to very good energy convergence. The density of the k-point grid was set at 0.04 Å$^{-1}$; ultrasoft generated on the fly pseudopotentials were used as they provide more accurate lattice parameters. YbH$_3$ stoichiometry was assumed and the *I*4/mmm structure type exhibited by YbH$_{2+x}$ at elevated pressures was adopted. The sqrt2 x sqrt2 x 1 supercell (Z=4) was applied to account for various magnetic ordering schemes; the formal spin of Yb(III) was used as the initial one in the first SCF step. For the magnetic structures, the DFT+U formalism has been used, with U = 5 eV for the 4*f* electrons of Yb. Enthalpies of various magnetic models and that of a nonmagnetic one, were calculated at several pressure values. Electronic density of states (DOS) was computed for the lowest enthalpy solutions.

3. Results and discussion

As it could be expected, the reaction between metallic ytterbium and hydrogen proceeds already at room temperature immediately after hydrogen loading to the cell. However, the unreacted metal is present in the whole pressure range studied, contributing to the strongest diffraction peaks (*cf.* Table 1), even for the largest pressure achieved for the sample compressed at room temperature to *ca.* 40

GPa, *Fig. S3–S4*. The evolution of the crystalline phases detected in this sample well corresponds to that described earlier, and YbH$_2$ remains the main hydridic phase.[44–46] The crystalline phase which can be attributed to a higher hydride of ytterbium, YbH$_{2+x}$, is barely detected even at *ca.* 40 GPa, *Fig. S3*. However, the composition is variable from sample to sample, and in the case of the second DAC the clear, although relatively weak signals of the *P*-31m phase of YbH$_{2+x}$ are visible already at 3.6 GPa, before heating of the sample, *cf. Fig. 2a*. This may be attributed to a variable degree of surface oxidation which should hamper further hydrogenation of the sample (this problem has been recently addressed by using a thin Pd coating for the synthesis of yttrium superhydrides).[18]

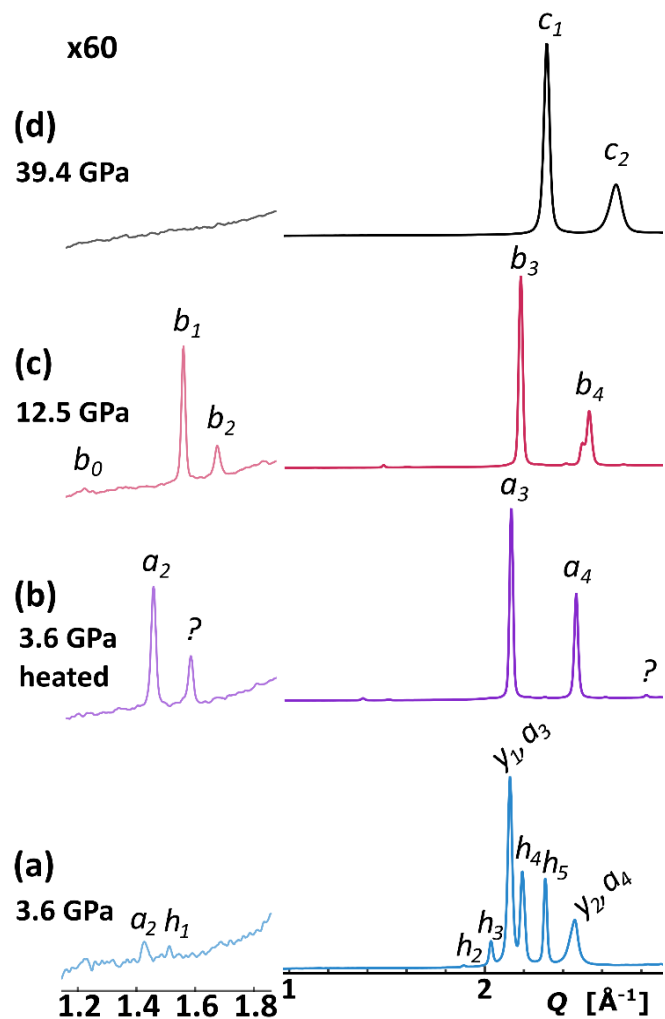

*Figure 2. The diffraction patterns for the sample of Yb compressed in H$_2$: (a) before and (b) after the laser heating at ca. 3.6 GPa; (c) after several rounds of laser heating (ca. 12.5 GPa); (d) at ca. 39.4 GPa. The low-Q part has been additionally presented in 60x magnification of intensity. λ = 0.3344 Å. a – YbH$_{2+x}$ P-31m: 2-(1 0 1), 3-(0 0 3) and (2 -1 1), 4-(1 1 2); b – YbH$_{2+x}$ I4/m: 0-(1 1 0), 1-(1 0 1), 2-(2 0 0), 3-(2 1 1), 4-(0 0 2) and (3 1 0); c – YbH$_{2+x}$ I4/mmm: 1-(1 0 1), 2-(1 1 0) and (0 0 2); h – YbH$_2$ Pnma: 1-(101), 2-(0 0 2), 3-(0 1 1), 4-(1 0 2) and (2 0 0), 5-(1 1 1); y – Yb Fm-3m: 1-(1 1 1), 2-(0 0 2).*

Table 1. Selected low-angle reflections originating from four distinct phases, as marked in Figures 2–4.

| phase | reflections |
|---|---|
| YbH$_{2+x}$ P-31m | a$_1$-(1 0 0), a$_2$-(1 0 1), a$_3$-(0 0 3) and (2 -1 1), a$_4$-(1 1 2), a$_5$-(2 -1 4) and (3 0 0) |
| YbH$_{2+x}$ I4/m | b$_1$-(1 0 1), b$_2$-(2 0 0), b$_3$-(2 1 1), b$_4$-(0 0 2) and (3 1 0), b$_5$-(3 1 2) and (4 2 0) |
| YbH$_{2+x}$ I4/mmm | c$_1$-(1 0 1), c$_2$-(0 0 2) and (1 1 0), c$_3$-(1 1 2) and (2 0 0), c$_4$-(1 0 3) and (2 1 1) |
| YbH$_2$ Pnma | h$_1$-(101), h$_2$-(0 0 2), h$_3$-(0 1 1), h$_4$-(1 0 2) and (2 0 0), h$_5$-(1 1 1) |
| Yb Fm-3m | y$_1$-(1 1 1), y$_2$-(0 0 2) |

The *in situ* laser heating of the partially hydrogenated ytterbium sample up to *ca.* 2100 K, with several pulses lasting a few seconds each, facilitates further hydrogenation and delivers the *P*-31m phase of YbH$_{2+x}$ (*Fig. 1a*) which predominates the powder diffraction pattern, *Fig. 2b* and *Fig. S5–S7*. The YbH$_2$ is detected in the sample in a minor amount, together with very weak signals from an unidentified crystalline phase(s), *Fig. S7*. Although the sample remains not fully homogenous, it can be concluded that both these phases disappear after a few additional heating cycles on larger compression, *Fig. S8*. Above *ca.* 12.5 GPa the *P*-31m phase of YbH$_{2+x}$ undergoes a transition to the *I*4/m phase which is isostructural to that reported for EuH$_{2+x}$ at 8.7–9.7 GPa,[23] as indicated by the peak splitting (*e.g.* the peak *b$_4$* of (0 0 2) and (3 1 0) reflections, while the (1 1 2) reflection of the trigonal phase contributes solely to the related *a$_4$* peak ) and the low-intensity reflections from the tetragonal superstructure (*e.g.* (1 1 0), (1 0 1), (2 0 0), marked as *b$_0$*, *b$_1$,* and *b$_2$*, respectively), *Fig. 2c*. During further compression, the latter phase symmetrizes to the *I*4/mmm structure, *Fig. 2d*, similar to EuH$_{2+x}$ analog reported for the hydrogen pressures exceeding 9.7 GPa.[23] That second phase transition manifests itself as the vanishing superstructure signals and as a less pronounced peak splitting for the main peaks (*e.g. c$_2$* corresponding to (0 0 2) and (1 1 0) reflections).

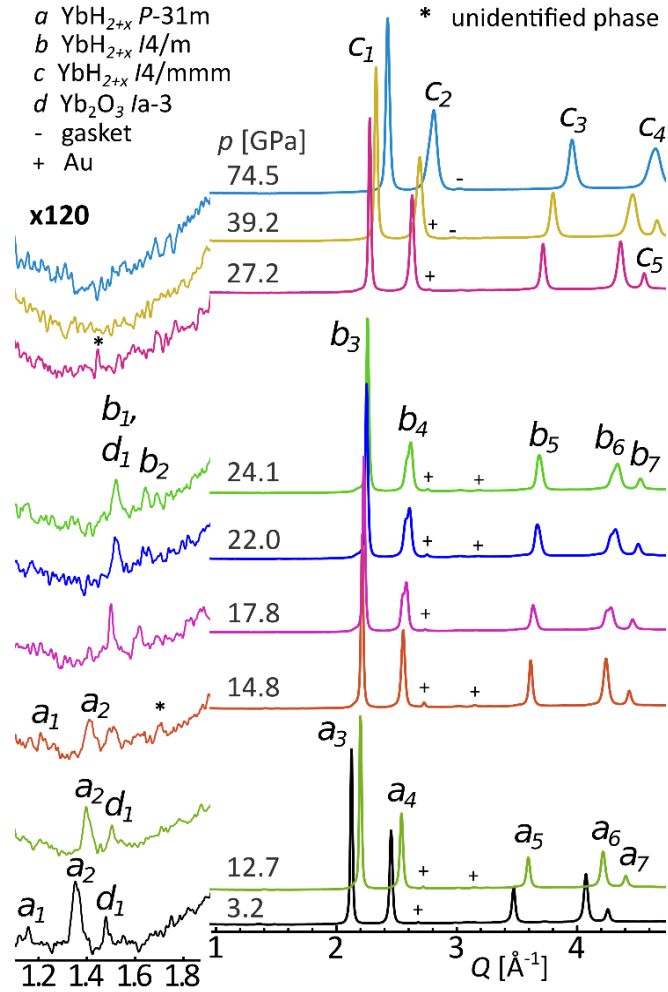

*Figure 3. The diffraction patterns for the sample of $Yb_3H_8$ compressed in $H_2$ at room temperature. The low-Q part has been additionally presented in 120x magnification of intensity. $\lambda = 0.4066$ Å. a – $YbH_{2+x}$ P-31m: 1-(1 0 0), 2-(1 0 1), 3-(0 0 3) and (2 -1 1), 4-(1 1 2); b – $YbH_{2+x}$ I4/m: 1-(1 0 1), 2-(2 0 0), 3-(2 1 1), 4-(0 0 2) and (3 1 0); c – $YbH_{2+x}$ I4/mmm: 1-(1 0 1), 2-(1 1 0) and (0 0 2).*

To study the more homogenous samples including those of fixed content of hydrogen we have also investigated $Yb_3H_8$ ($YbH_{2+x}$, x = 0.67) which was prepared *ex-situ* according to the Auffermann's procedure[29]; here, $Yb_3H_8$ was studied in DAC under compression in $H_2$ or in He or Ne pressure-transmitting media (PTM). The evolution of the crystalline phases for $Yb_3H_8$ compressed in $H_2$ or an inert PTM well corresponds to that observed for the sample of ytterbium heated in $H_2$, *Fig. 3*, and *Fig. S10–S23*. It appears that the *P*-31m and *I*4/m phases coexists within the range of *ca.* 13–18 GPa regardless of the PTM. The superstructure signals of the *I*4/m phase vanish above *ca.* 27 GPa indicating the transition to the *I*4/mmm phase. However, minor low-angle diffraction peaks have still been observed for some of the $Yb_3H_8$ samples compressed in the neutral PTM above 35 GPa (Fig. S20). The high-symmetry *I*4/mmm phase of $YbH_{2+x}$ remains stable up to the highest pressures reached in our investigations of *ca.* 75 GPa, *Fig. 3*.

The diffraction patterns from Yb$_3$H$_8$ samples in H$_2$ and Ar PTMs (Fig. 4) are quite similar to each other. All key reflections may be assigned to the same crystalline phase in both datasets. Such qualitative picture alone may suggest that an additional H$_2$ uptake does not occur for Yb$_3$H$_8$ even at 74.5 GPa, the highest pressure reached in our experiments. However, for the *I*4/*mmm* phase, the corresponding diffraction signals are shifted towards lower Q values for the sample of Yb$_3$H$_8$ compressed in H$_2$ PTM indicating larger unit cell volume. This phenomenon is noticeable around 30 GPa, becoming obvious for larger pressures, *Fig. S23*.

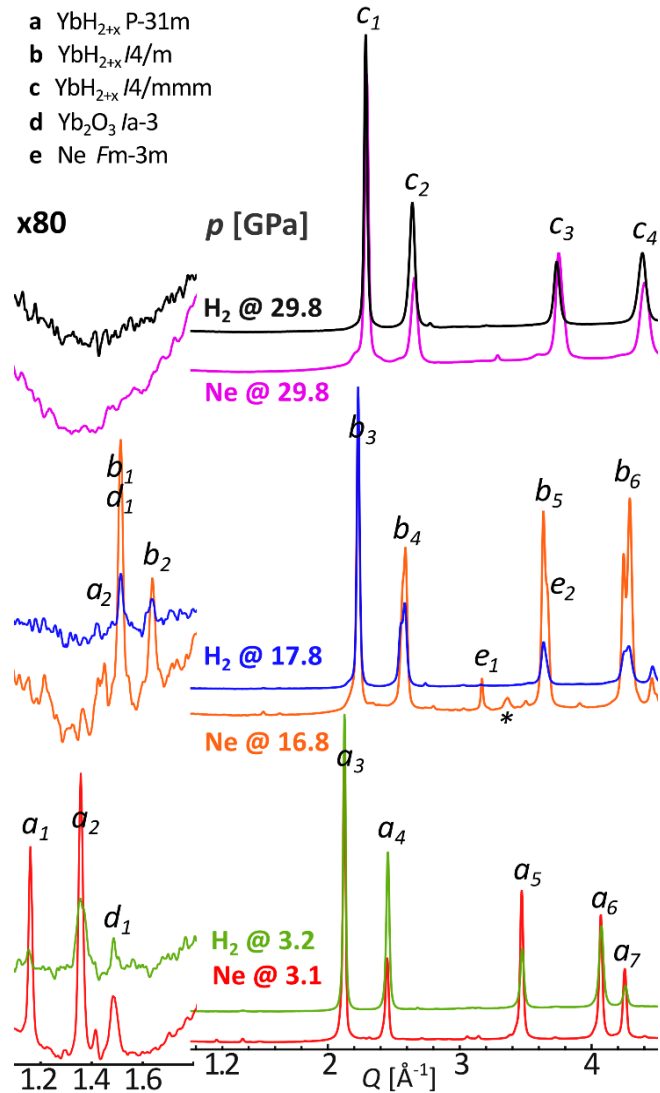

*Figure 4. A comparison of the integrated diffraction data for Yb$_3$H$_8$ compressed in Ne and in H$_2$. The low-Q region has been expanded (left). λ = 0.3344 Å (Ne at 3.1 and 16.8 GPa), 0.2952 Å (Ne at 29.8 GPa) and 0.4066 Å (H$_2$). The most visible reflections of the YbH$_{2+x}$ and Yb$_2$O$_3$ phases were marked: a – YbH$_{2+x}$ P-31m: 1-(1 0 0), 2-(1 0 1), 3-(0 0 3) and (2 -1 1), 4-(1 1 2); b – YbH$_{2+x}$ I4/m: 1-(1 0 1), 2-(2 0 0), 3-(2 1 1), 4-(0 0 2) and (3 1 0); c – YbH$_{2+x}$ I4/mmm: 1-(1 0 1), 2-(0 0 2) and (1 1 0), 3-(1 1 2) and (2 0 0), 4-(1 0 3) and (2 1 1). \* - the strongest signal from the unidentified phase.*

EoS for Yb$_3$H$_8$ has been derived independently for the YbH$_{2+x}$ crystalline phases present in the samples compressed in H$_2$ and Ne PTMs for the experimental runs covering a sufficient number of *V(p)* data points, *Fig. S24–S29*. The obtained fit parameters are presented in *Table 2*, while some of the available EoS parameters for the related lanthanide hydrides were summarized in *Table S2*. In general, the $V_0$ of the low-pressure trigonal form is slightly lower than the high-pressure *I*4/mmm phase, and the inverse trend is observed for their bulk moduli. The *P*–31m form of Yb$_3$H$_8$ reveals $V_0$ comparable to the *P*nma YbH$_2$ of $V_0$ = 35.63(7) Å$^3$, but the former remains significantly less compressible, *cf. Fig. 5*, (for *P*nma YbH$_2$ $B_0$ = 40.2(22) GPa).[45] The *I*4/mmm phase of YbH$_{2+x}$ present above *ca.* 27 GPa has a slightly smaller molecular volume with increased bulk modulus. These valuesfulfill a similar trend for the europium hydrides: the *I*4/mmm phase of EuH$_{2+x}$ of $V_0$ = 38.8(1) Å$^3$ (slightly bulkier than the Yb analog, according to the relation of Eu and Yb ionic radii) shows significantly higher bulk modulus in comparison to the dihydride ($B_0$ = 69(2) GPa vs. $B_0$ = 40 – 45 GPa for the two polymorphs of EuH$_2$).[47] Contrastingly, the trihydrides of the neighboring lanthanides (especially the Sm for Eu and Er or Tm for Yb) reveal rather larger values of the bulk moduli, while the $V_0$ of the *I*4/mmm forms of EuH$_{2+x}$ and YbH$_{2+x}$ falls close to the values reported for hexagonal and cubic polymorphs of the respective LnH$_3$, Ln = Sm (for Eu), Er and Tm (for Yb), see *Table S2*.

*Table 2. Summary of the parameters of the third-order Birch-Murnaghan EoS obtained from the compression data. The estimated standard deviations (E.S.D.) are given in parentheses. $B_0$′ has been fixed for the final refinement. N – number of data points.*

| phase of YbH$_{2+x}$ | PTM, sample | $V_0$ per Yb atom [Å$^3$] | $B_0$ [GPa] | $B_0$′ | max Δp [GPa] | $\chi^2$, N |
|---|---|---|---|---|---|---|
| *P*–31m | **Ne**, (g) | 35.033(32) | 56.13(63) | 5.5 | 0.31 | 1.82, 19 |
|  | **Ne**, (h) | 35.31(12) | 53.4(19) | 5.5 | -0.43 | 6.82, 11 |
|  | **H$_2$**, (d) | 35.292(88) | 55.0(14) | 5.5 | 0.22 | 3.26, 8 |
| *I*4/m | **Ne**, (g) | 35.82(23) | 63.1(22) | 3.0 | -0.31 | 1.15, 8 |
| *I*4/mmm | **Ne**, (i) | 34.05(34) | 64.0(33) | 4.5 | 1.03 | 7.95, 10 |
|  | **H$_2$**, (d) | 34.98(27) | 59.7(23) | 4.5 | 1.67 | 7.21, 19 |

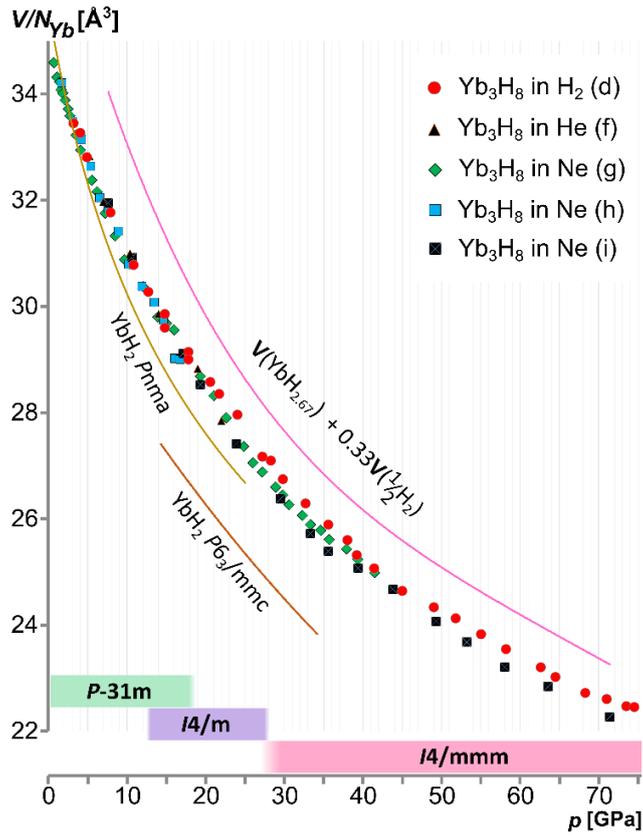

*Figure 5. The volume per one ytterbium atom in the function of pressure for the compression of $Yb_3H_8$ in $H_2$, He, and Ne at room temperature. The molecular volumes of $YbH_2$ (both polymorphs)[45] and the volume estimated for $YbH_3$ have also been plotted. The crystallographic phases of $YbH_{2+x}$, detected under a specific pressure range, are indicated at the bottom.*

One key issue is that of the composition of the $YbH_{2+x}$ specimen studied and that of a possible $H_2$ uptake by the sample in the $H_2$ atmosphere. Inspection of *Fig. 5* reveals that the differences in the molar volume of the $Yb_3H_8$ specimen compressed in various PTM are only marginal, which is also reflected by the respective EoS curves, *Fig. S30*. However, on larger compression, the *V(p)* curve seems to be slightly less steep for the sample pressurized in $H_2$. Although the discrepancy is rather small and it may be caused by the errors in the determination of pressure in both types of samples, it is fairly consistent across the pressures for which the tetragonal phases are favored, especially 25–75 GPa. The EoS curves fitted for the *I4/mmm* $YbH_{2+x}$ in $H_2$ and Ne PTM fall apart by *ca.* 1.5% in terms of *V(p)*, *Fig. S31*. At the same time, a simple addition of 1/3 atomic volume of hydrogen would cause a raise in the molecular volume of *I4/mmm* $YbH_{2+x}$ by *ca.* 4.8% at this pressure range. Altogether, this might suggest that the composition of the sample compressed in excess $H_2$ is similar to $YbH_{2.77}$ rather than $YbH_{2.67}$.

Indeed, one might speculate that amount of 1/3 of vacancies is not a natural value for the tetragonal *I4/mmm* cell and that ¼ would be a more natural value (hence, $YbH_{2.75}$ would be a more appropriate composition for such specimen, very close to YbH$_{2.77}$ suggested above from the EOS analyses). However, at lack of direct support for such a scenario one must refrain from making any firm conclusions here. In any case, the stoichiometry of the samples studied must be still far from the one expected for Yb(III)H$_3$ stoichiometry (*Fig. 5*). Resistance of Yb to be fully oxidized to Yb(III) in the hydride environment is remarkable; it turns out that only if a negative charge density on hydride is reduced, *e.g.* by binding it to B(III) in the form of BH$_4^-$ anion, stoichiometric Yb(III) species may form and yet they are thermodynamically metastable at ambient (p,T) conditions.[48–51]

The electric conductivity of formally mixed-valence species, be it YbH$_{2.67}$, YbH$_{2.75}$ or YbH$_{2.77}$, is of great interest in the context of potential generation of superconductivity in this material. *Figure 6* (top) shows the electric resistivity versus temperature plot for a sample of Yb$_3$H$_8$ embedded in NaCl PTM. It is clear that the low-pressure phases of Yb$_3$H$_8$ show a semiconducting behavior as their resistivity is strongly decreasing with the temperature raising. Large drop of resistivity may be seen between 20.2 and 25.2 GPa, roughly corresponding to the second structural phase transition from the *I*4/m to the *I*4/mmm polymorphs.

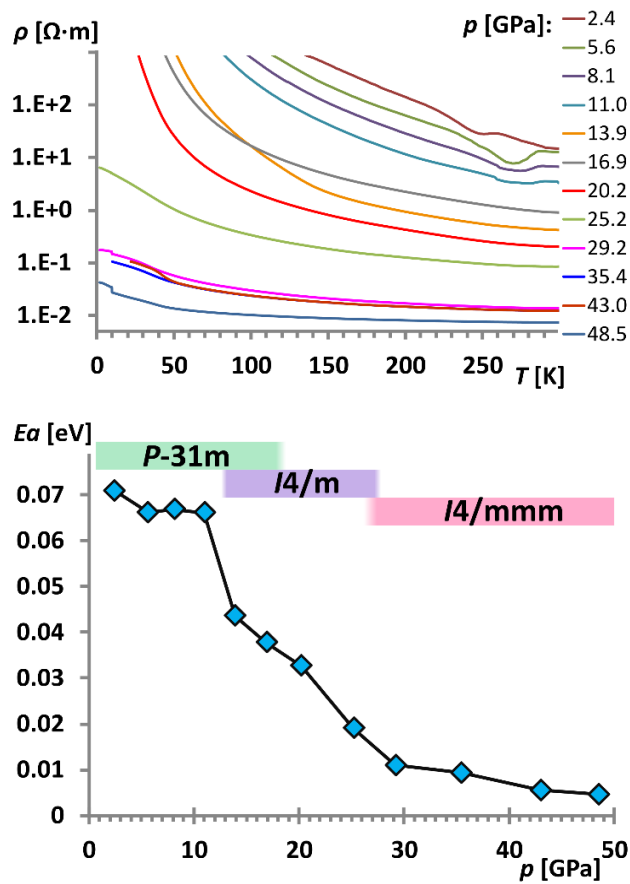

*Figure 6. The resistivity vs. T [K] for Yb$_3$H$_8$ compressed in NaCl presented in a logarithmic scale indicating semiconductor-like behavior (top). Evolution of the activation energy of the electronic transport in the function of pressure (bottom). Despite the significant decrease of the E$_a$, the system avoids full metallization till 50 GPa (the highest pressure in our experiment).*

Analysis of the activation energy for conductivity, $E_a$, (*Fig. 6* bottom) suggests that $E_a$ is already quite small (*ca.* 0.07 eV or 812 K) for $Yb_3H_8$ at low pressure (2.4 GPa). This feature comes certainly from the mixed-valence nature of the compound. One sharp drop of the $E_a$ value is seen between 11.0 and 13.9 GPa, roughly corresponding to the first crystallographic phase transition. The second drop of the $E_a$ value between 20 and 29 GPa is more subtle which suggests that the second phase transition affects $E_a$ less considerably, and it is followed by the flattening of the curve above *ca.* 29 GPa. The $E_a$ value determined at 48.5 GPa is as small as 0.005 eV (58 K) yet not null. Thus, $Yb_3H_8$ retains its semiconducting character even at *ca.* 50 GPa; this makes it similar to $Fe_3O_4$.[52] Metallicity, which would correspond to the formulation of $Yb(III)_3H_8(e^-)$ might be within the reach of the 1 Mbar experiments.

To learn what properties might be expected from fully stoichiometric $YbH_3$, we have performed quantum mechanical calculations for this compound; we have assumed that it would adopt the *I*4/*mmm* structure observed here, similar to the europium analog for which solely the $Eu^{3+}$ oxidation state has been detected above 12.5 GPa.[24] Both the magnetic and nonmagnetic models were considered. Optimization of geometry resulted in the molar volumes approaching the experimental results, especially for the larger pressures, while the EOS for $YbH_4$ polyhydride of the same symmetry (as proposed for $YH_4$ and other $RE$H$_4$)[53] is obviously different, *Fig. S33*, and it does not match experimental data.

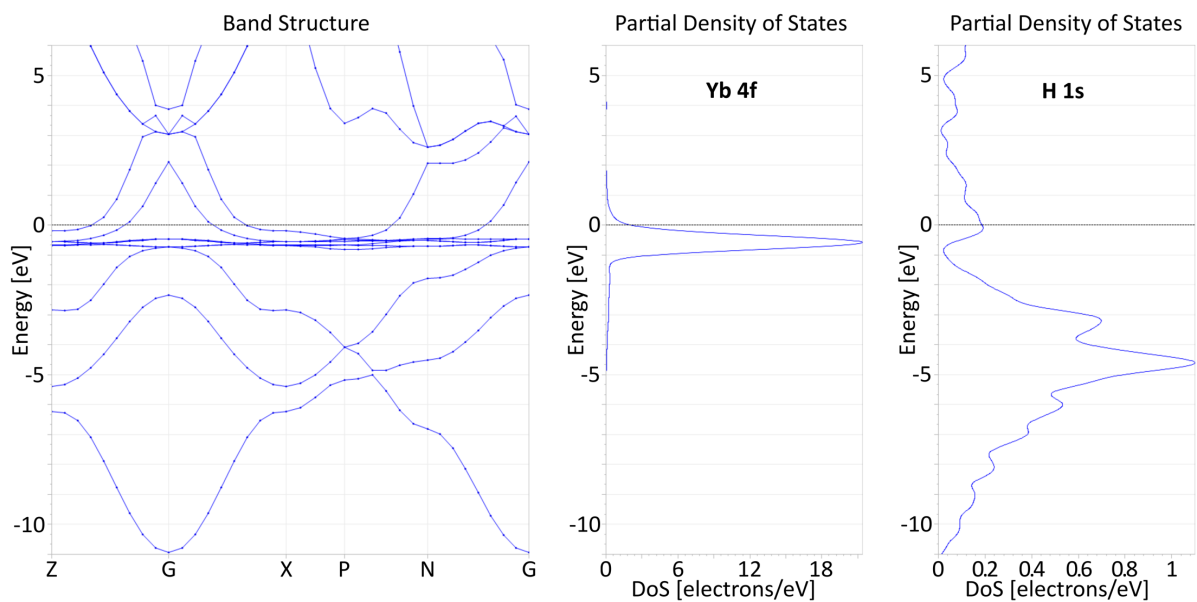

*Figure 7. Electronic band structure (left) and DOS (Yb 4f center, H 1s right) for the spin-unpolarized solution, for YbH$_3$ in the I4/mmm structure type at 50 GPa. Note the partial DOS scale difference between those for Yb and H electrons.*

The DFT calculations for all magnetic states converge to solutions characterized by extremely small remnant magnetic moments. This implies that YbH$_3$ if prepared, could be a nonmagnetic solid. The corresponding spin-nonpolarized band structure and DOS (Figure 7) suggest that the compound would have a metallic nature, with several broad bands crossing the Fermi level. Both Yb(f) and H(s) states contribute to the states at the Fermi level; such strong hybridization was noticed previously.[30] The fact that the spin of Yb(III) would not localize but instead give rise to itinerant electrons may seem surprising but this could be due to strong hybridization with broad H(s) states (and this was the reason why we were targeting YbH$_3$ composition in experiments).

**CONCLUSIONS**

We have attempted synthesis of yet elusive ytterbium trihydride utilizing both metallic Yb and its currently most hydrogen-rich hydride, YbH$_{2.67}$, as the precursors in high-pressure hydrogenation (up to *ca.* 75 GPa). Our results point at the lack of a marked qualitative difference between these systems compressed in H$_2$ and the samples of YbH$_{2.67}$ pressurized in He or Ne PTM. In all these cases a sequence of phase transitions from the unit cells of *P*-31m symmetry to the *I*4/m and *I*4/mmm systems occurred within *ca.* 13–18 GPa and around 27 GPa, respectively. The same tetragonal phases have been observed upon compression of EuH$_2$ in H$_2$ PTM: the *I*4/m system formed from the *P*6$_3$/mmc dihydride above 8.7 GPa and the high-symmetry *I*4/mmm phase containing Eu$^{3+}$ has been detected above 9.7 GPa.[23] The fact that Eu$^{3+}$ in a hydride environment forms at pressure as large as 9.7 GPa, while analogous species of Yb3+ may be prepared at several kPa, is following the standard redox potentials of the Eu$^{3+}$/Eu$^{2+}$ and Yb$^{3+}$/Yb$^{2+}$ pairs, which are –0.35 V and –1.05 V, respectively. In other words, Yb$^{2+}$ can be easier oxidized by H$_2$ to Yb$^{3+}$ than its Eu$^{2+}$ analog.

For the highest pressure investigated here (25–75 GPa) corresponding to the *I*4/mmm phase, the molecular volume of the systems compressed in H$_2$ PTM consistently remains *ca.* 1.5% larger than that for the systems compressed in inert gas media. This indicates incremental hydrogenation towards YbH$_3$ and could correspond to the formation of YbH$_{2.75-2.77}$. However, additional research is necessary to fully identify the stoichiometry achieved upon hydrogenation, and even rather larger pressure is needed to push the system towards stoichiometric YbH$_3$ or ytterbium polyhydrides. Such compounds should be interesting due to their metallic nature, as expected on the basis of DFT calculations for YbH$_3$. At the same time, while the mixed-valence Yb$_3$H$_8$ retains its semiconducting character up to at least 50 GPa, the very low remnant activation energy of conduction (<5 meV) suggests that the metallization under further compression of this mixed-valence compound should also be achievable.

**Acknowledgements**

T. J. thanks Polish Ministry of Science and Higher education for funding (project "Mobility Plus" No. 1064/MOB/13/2014/0). W.G. acknowledges Polish National Science Center (NCN) for the Harmonia project 2012/06/M/ST5/00344. T. J. and V. V. S. acknowledge Dr. Ross Hrubiak for the help in the XRD measurements, and Dr. Sergey Tkachev for Ne loading using GSECARS gas loading system at Advanced Photon Source (APS). Use of the COMPRES-GSECARS gas loading system was supported by COMPRES under National Science Foundation (NSF) Cooperative Agreement EAR 11-57758 and by GSECARS through NSF grant EAR-1128799 and DOE grant DE-FG02-94ER14466. Portions of this work were performed at GeoSoilEnviroCARS (The University of Chicago, Sector 13), APS, Argonne National Laboratory. GeoSoilEnviroCARS is supported by NSF – Earth Sciences (EAR-1128799) and DOE-GeoSciences (DE-FG02-94ER14466). Portions of this work were performed at HPCAT (Sector 16), APS Argonne National Laboratory. HPCAT operations are supported by DOE-NNSA under Award No. DE-NA0001974 and DOE-BES under Award No. DE-FG02-99ER45775, with partial instrumentation funding by NSF. This research used resources of the APS, a U.S. DOE Office of Science User Facility operated by Argonne National Laboratory under Contract No. DE-AC02-06CH11357.

**Synthesis, structure and electric conductivity of higher hydrides of ytterbium at high pressure**

Tomasz Jaroń,[a,b,c*] Jianjun Ying,[b,d] Marek Tkacz,[e] Adam Grzelak,[a] Vitali B. Prakapenka,[f] Viktor. V. Struzhkin[b,g*], Wojciech Grochala[a*]

[a]Centre of New Technologies, University of Warsaw, Banacha 2c, 02-097 Warsaw, Poland, [b]Geophysical Laboratory, Carnegie Institution of Washington, 5251 Broad Branch Road NW, Washington, DC 20015, USA, [c]Faculty of Chemistry, University of Warsaw, Pasteura 1, 02-089 Warsaw, Poland, [d]HPCAT, Geophysical Laboratory, Carnegie Institution of Washington, Argonne, IL 60439, USA, [e]Institute for Physical Chemistry, Polish Academy of Science, Warsaw, Poland, [f]Consortium for Advanced Radiation Sources, The University of Chicago, Chicago, IL 60637, USA, [g]Center for High Pressure Science and Technology Advanced Research, Shanghai 201203, China.

*e-mail: tjaron@uw.edu.pl, viktor.struzhkin@hpstar.ac.cn, w.grochala@cent.uw.edu.pl

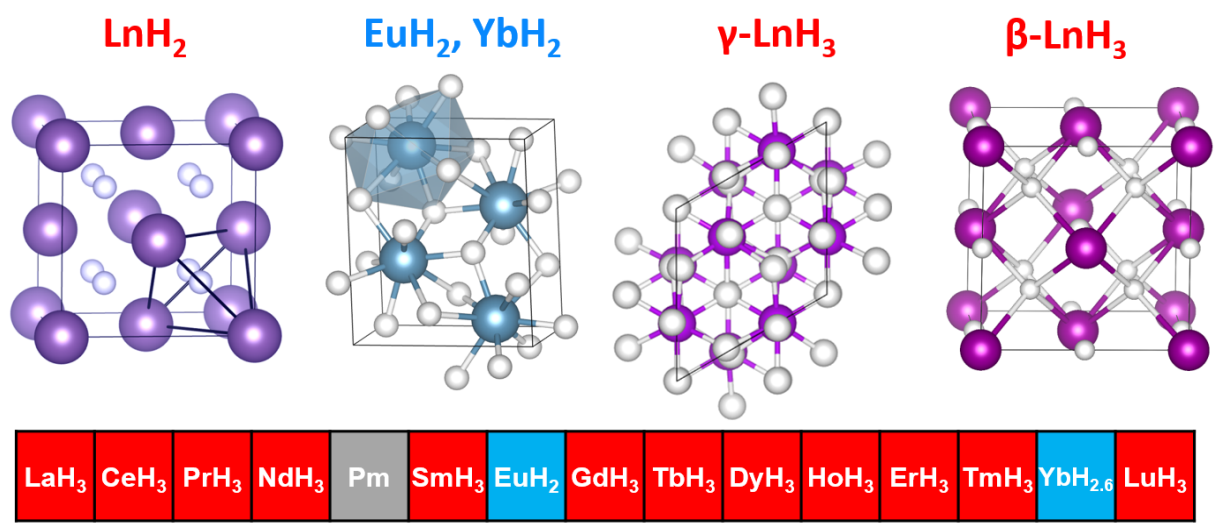

*Figure S1. The crystal structures adopted by lanthanide hydrides near ambient pressure. EuH$_2$ and YbH$_2$ of ionic character crystallize in PbCl$_2$ structure (Pnma), while the other lanthanide dihydrides (metallic) – in fcc CaF$_2$ structure. The early LnH$_3$ crystallize in an fcc structure of LnH$_2$, but with filled octahedral sites, while the later LnH$_3$ – in an hcp structure of HoH$_3$ type. The stoichiometry of LnH$_x$ achievable at the pressures up to ca. 0.3 GPa has been listed at the bottom.*

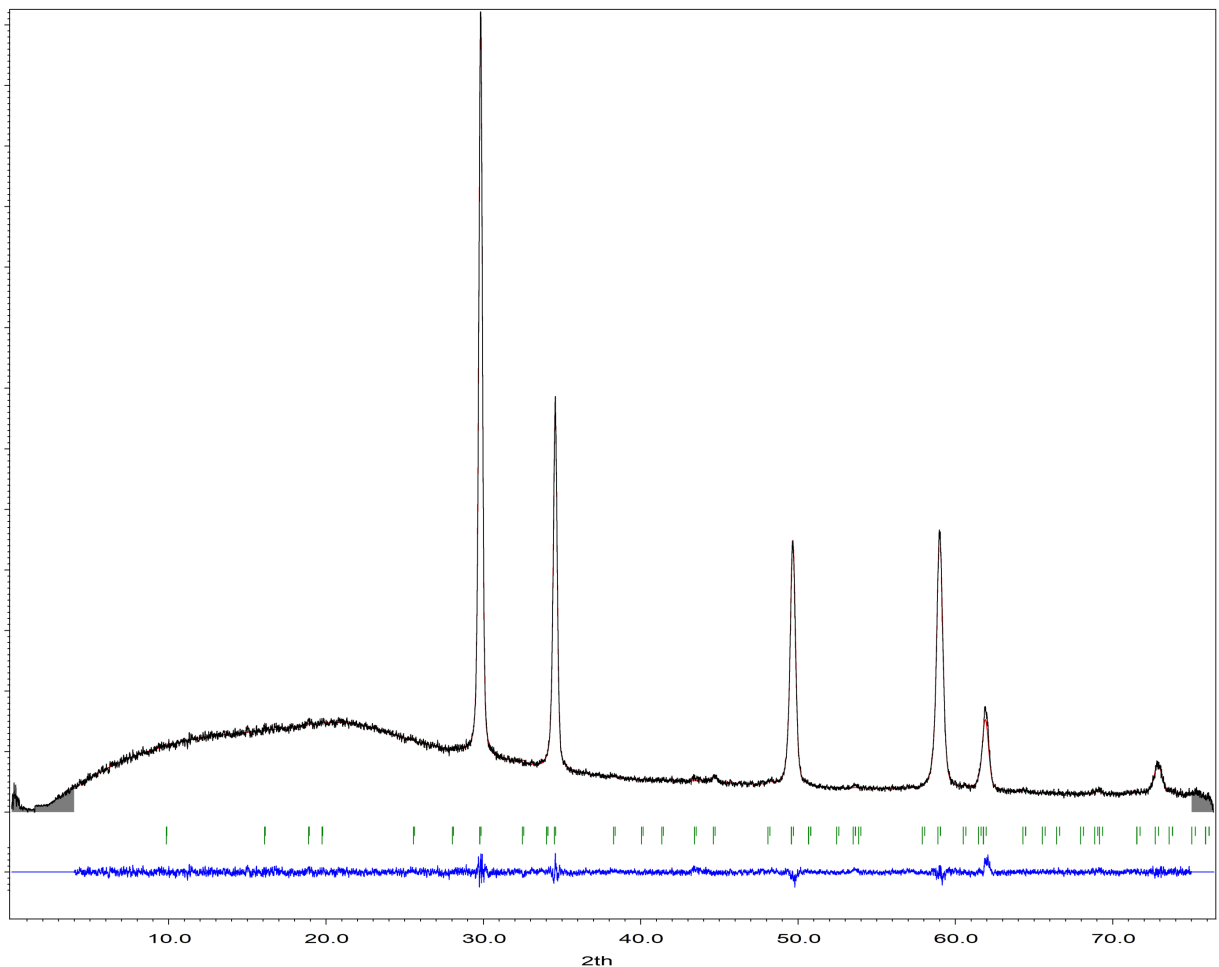

*Figure S2. The PXD pattern of Yb$_3$H$_8$ measured under ambient conditions. Black curve – measured data, red – calculated, bottom – difference plot (Rietveld refinement). Cu Kα radiation has been used (λ≈1.5406 Å).*

*Table S1. The results of Rietveld refinement of **Yb$_3$H$_8$** measured under ambient conditions.*

| p [atm] | 1 | | | |
|---|---|---|---|---|
| T [°C] | RT, *ca.* 25 | | | |
| space group | *P*-31m (162) | | | |
| a [Å] | 6.3699(18) | | | |
| c [Å] | 9.007(5) | | | |
| V [Å$^3$] | 316.50(10) | | | |
| 3Z | 9 | | | |
| V/3Z [Å$^3$] | 35.167(11) | | | |
| wRp; cwRp [%] | 4.00; 10.51 | | | |
| d$_{calc}$ [g cm$^{-3}$] | 8.298(3) | | | |
| Yb1 | 0 | 0 | 0 | Biso 2.39(6)* |
| Yb2 | 1/3 | 2/3 | 0 | Biso 2.39(6)* |
| Yb3 | 0.3484(5) | 0 | 0.3298(10) | Biso 2.39(6)* |
| H1 | 0 | 0 | 0.644 | Biso 3** |
| H2 | 1/3 | 2/3 | 0.219 | Biso 3** |
| H3 | 0.322 | 0 | 0.5812 | Biso 3** |
| H4 | 0.356 | 0 | 0.0757 | Biso 3** |
| H5 | 0.2364 | 0 | 0.8319 | Biso 3** |

\* Uiso of all Yb atoms has been set as equal, ** the parameters of H atoms were not refined

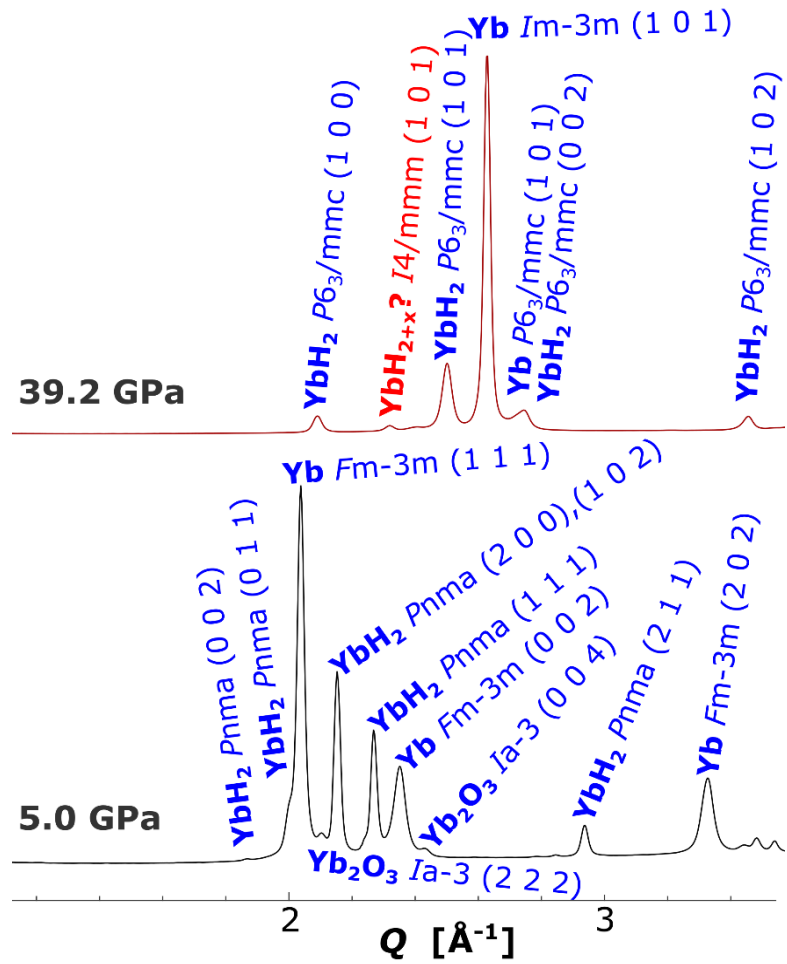

Figure S3. The diffraction data for the products of the reaction between Yb and $H_2$ performed at room temperature: p = 5.0 GPa and 39.2 GPa. The low-Q area has only been shown for better visibility. λ = 0.3344 Å. The most visible signal possibly originating from the higher ytterbium hydride has been marked with the red font. $YbH_{2+x}$: I4/mmm, a ≈ 3.307 Å, c ≈ 4.709 Å, V ≈ 51.49 Å$^3$.

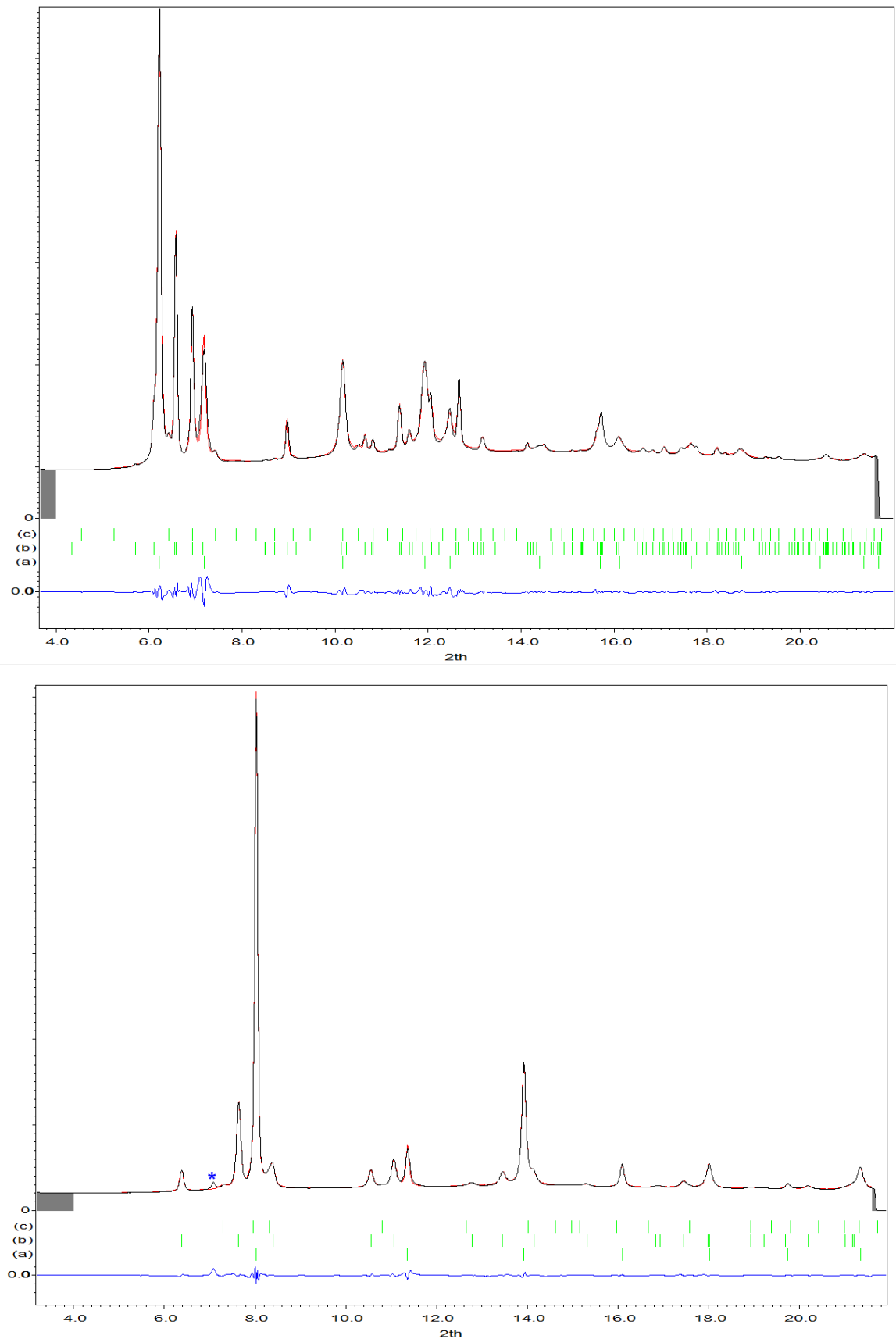

*Figure S4. The typical diffraction data for the products of the reaction between Yb and $H_2$ performed at room temperature. **Top**: p = 5.0 GPa, (a) – Yb Fm-3m, (b) – $YbH_2$ Pnma, (c) – $Yb_2O_3$ Ia-3. **Bottom**: p = 39.2 GPa, (a) – Yb Im-3m, (b) – $YbH_2$ $P6_3/mmc$, (c) – Yb $P6_3/mmc$, \* – (1 0 1) reflection of $YbH_{2+x}$ I4/mmm. LeBail fit has been marked with a red line, while the positions of Bragg reflections and the difference curve have been plotted at the bottom. λ = 0.3344 Å.*

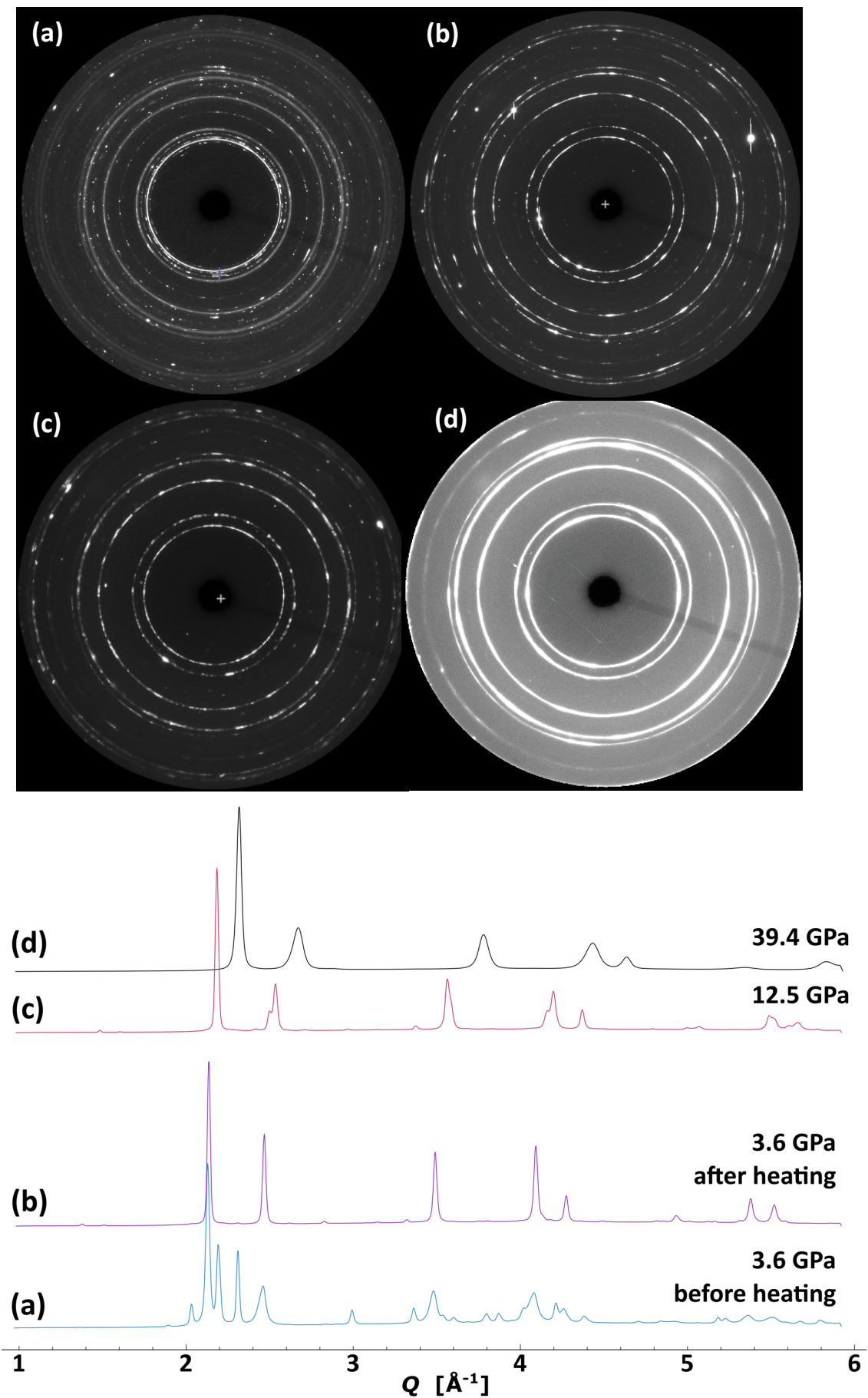

*Figure S5. The raw diffraction images for the sample of Yb compressed in $H_2$ (top), and the corresponding integrated diffraction patterns: (a) before and (b) after the laser heating at ca. 3.6 GPa; (c) after several rounds of laser heating (ca. 12.5 GPa); (d) at ca. 39.4 GPa. Notice the simplification of the pattern (due to disappearance of $YbH_2$ and Yb phases) and better averaging of the diffraction signals after heating. λ = 0.3344 Å.*

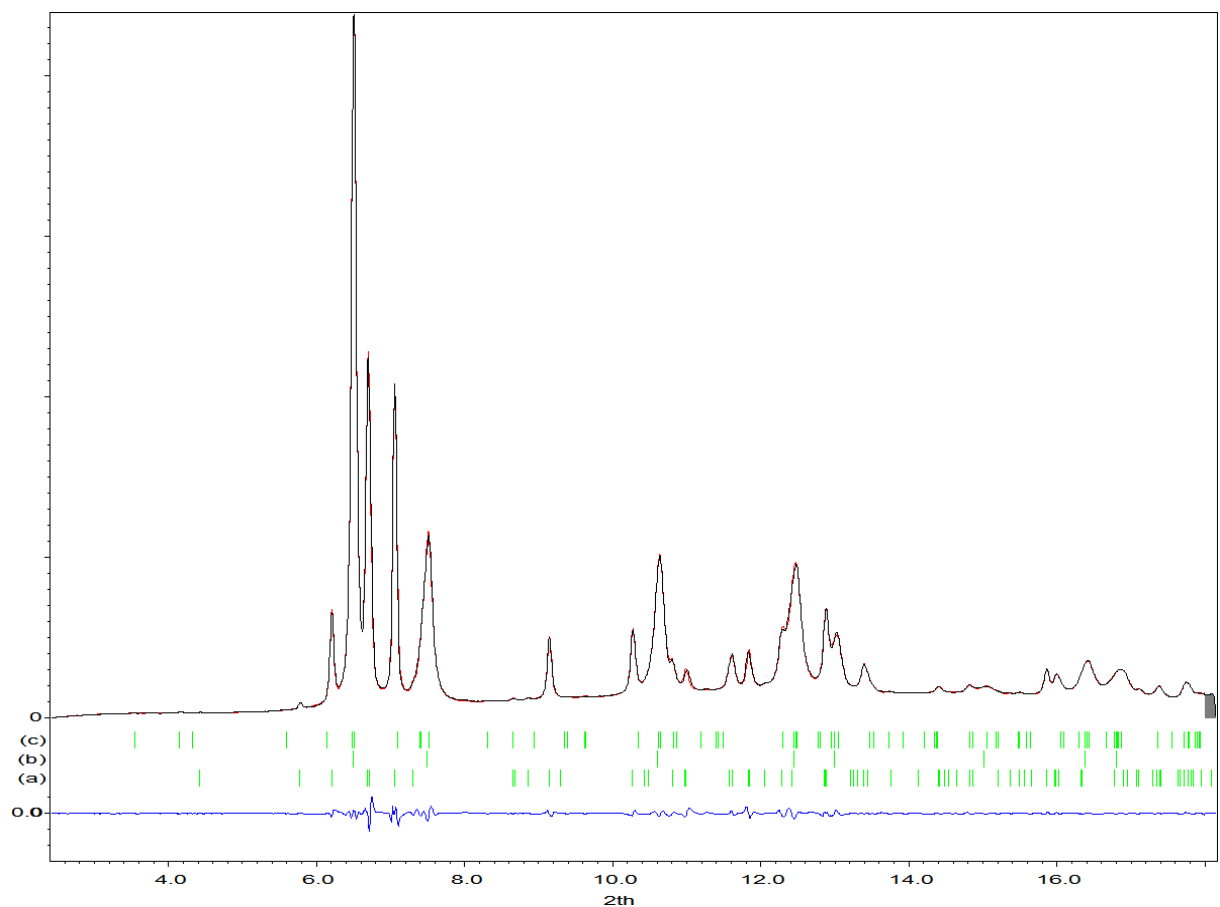
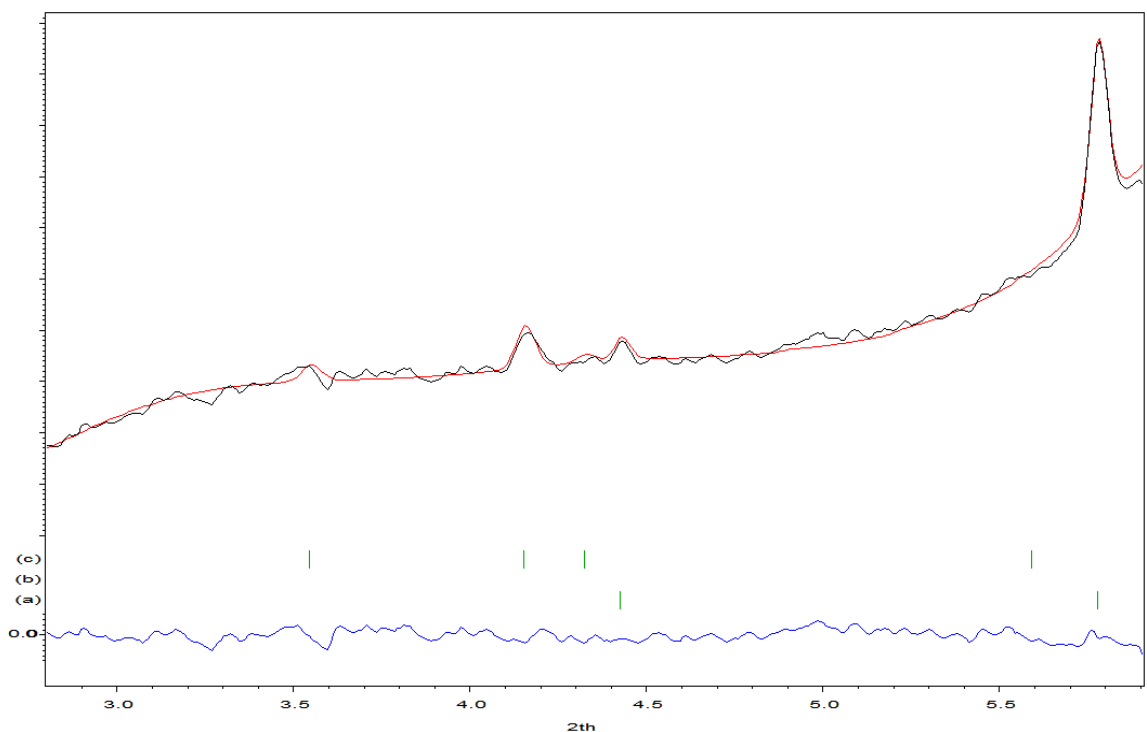

*Figure S6. The diffraction data for the products of the reaction between Yb and $H_2$ before the system was heated. **p** = 3.6 GPa, (a) – $YbH_2$ Pnma, (b) – Yb Fm-3m, (c) – $YbH_{2+x}$ P-31m. LeBail fit has been marked with a red line, while the positions of Bragg reflections and the difference curve have been plotted at the bottom. The low-angle region has been shown at the bottom, revealing i.a. a weak (1 0 1) reflection of the $YbH_{2+x}$ P-31m phase (ca. 4.15°). λ = 0.3344 Å.*

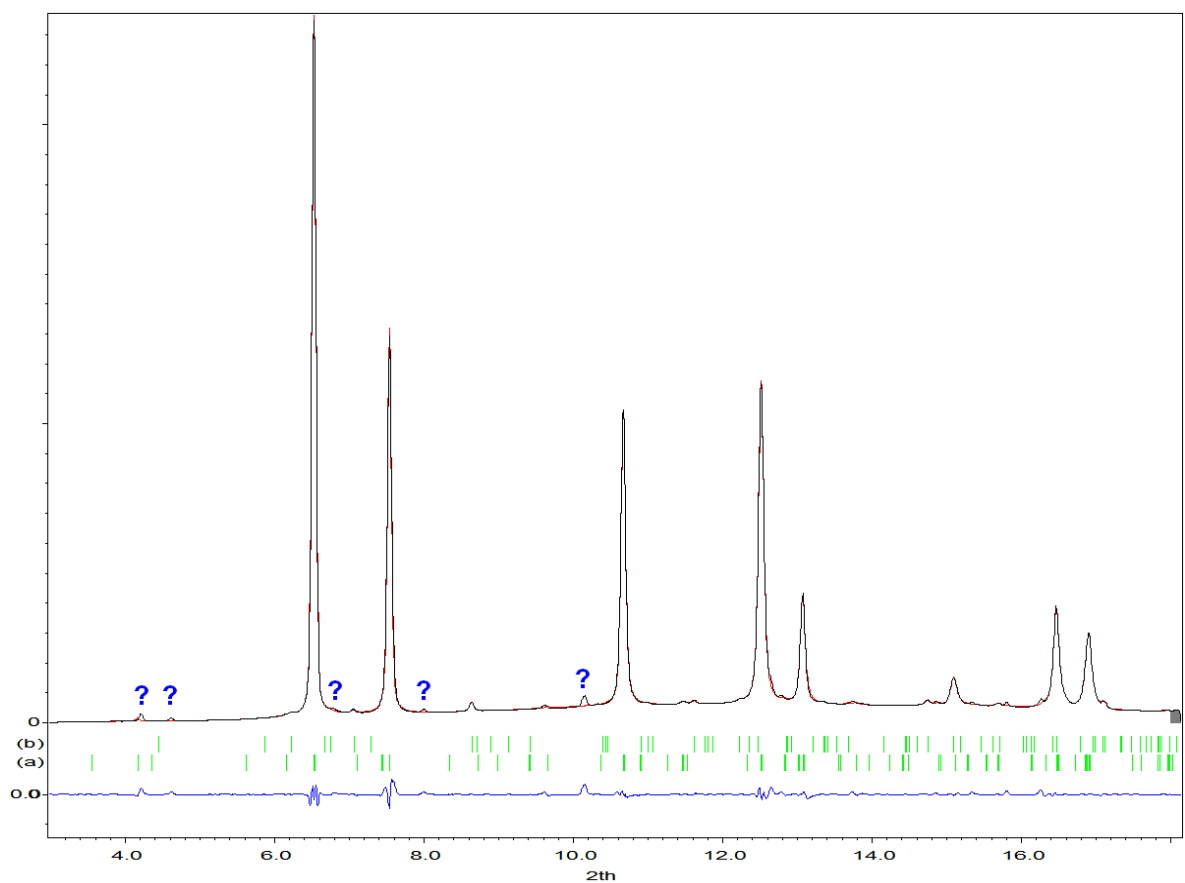

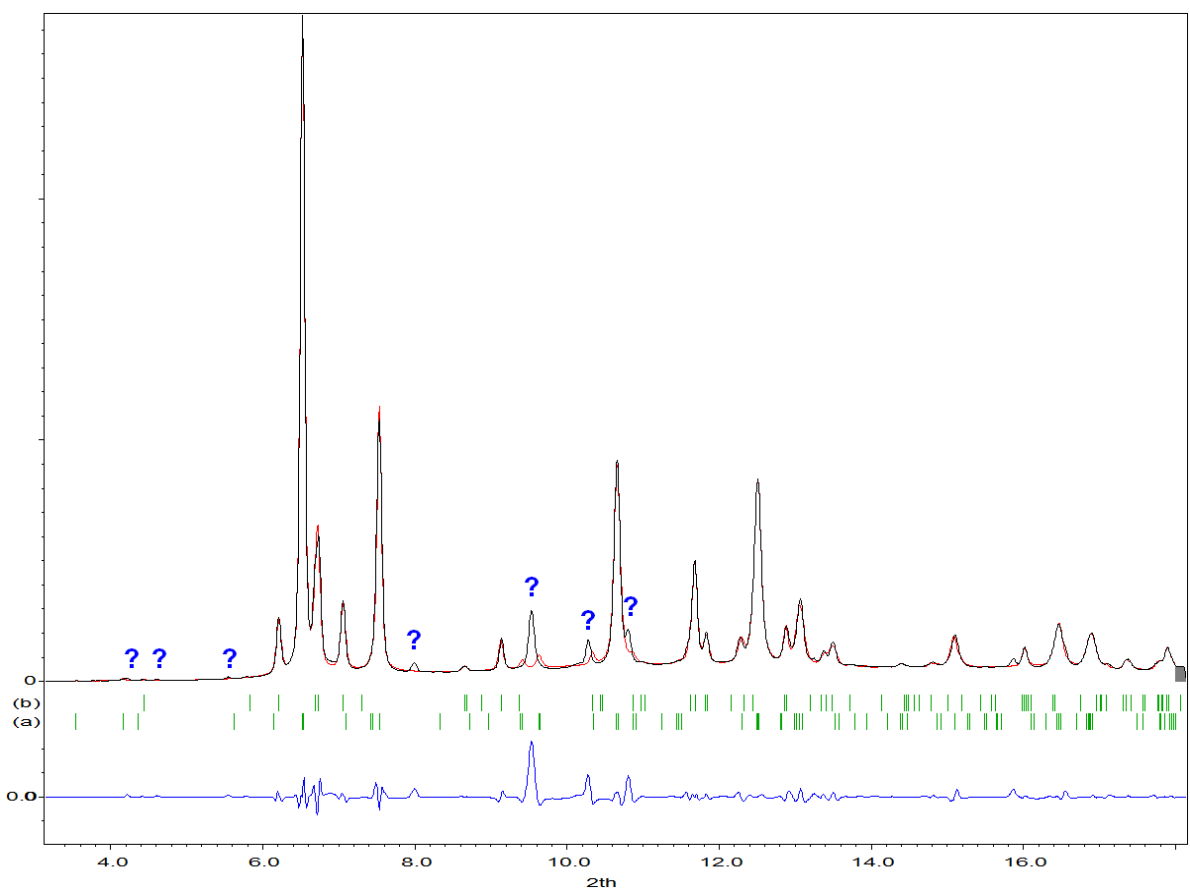

*Figure S7. The diffraction data for the products of the reaction between Yb and $H_2$ after laser heating at ca. 3.6 GPa, as measured in the two areas of the sample (the top and the bottom plots): (a) – $YbH_{2+x}$ P-31m, (b) – $YbH_2$ Pnma. LeBail fit has been marked with a red line, while the positions of Bragg reflections and the difference curve have been plotted at the bottom. The signals from the unidentified phase(s) have been marked with a "?". λ = 0.3344 Å. Note the variable contribution of the crystalline phases across the sample.*

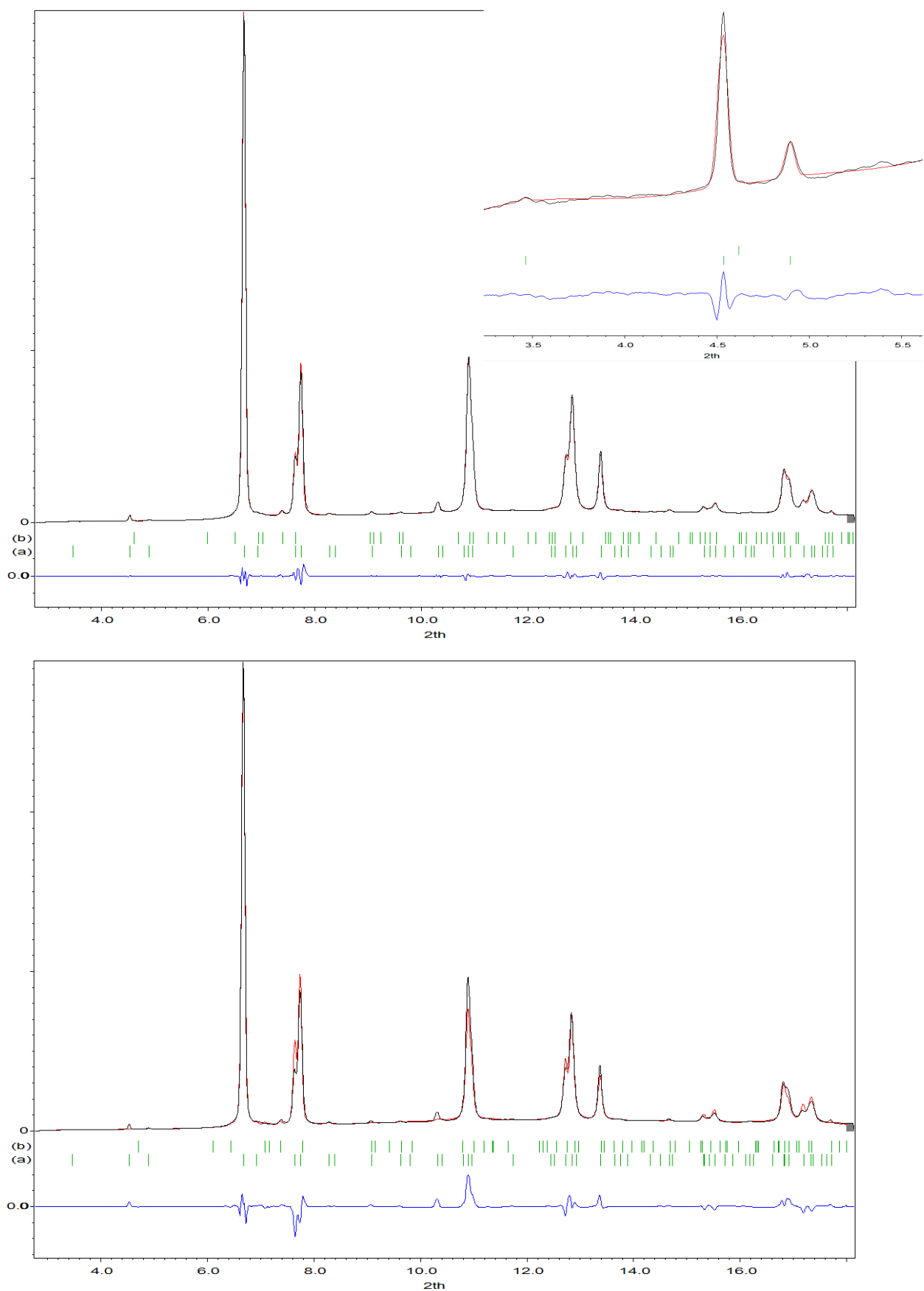

*Figure S8. The diffraction data for the products of the reaction between Yb and $H_2$ after several cycles of laser heating, compressed to ca. 12.5 GPa. Top – LeBail fit, bottom – **Rietveld** fit with no correction for the preferred orientation: (a) – $YbH_{2+x}$ P4/m, (b) – $YbH_2$ Pnma. The fit has been marked with a red line, while the positions of Bragg reflections and the difference curve have been plotted at the bottom. λ = 0.3344 Å.*

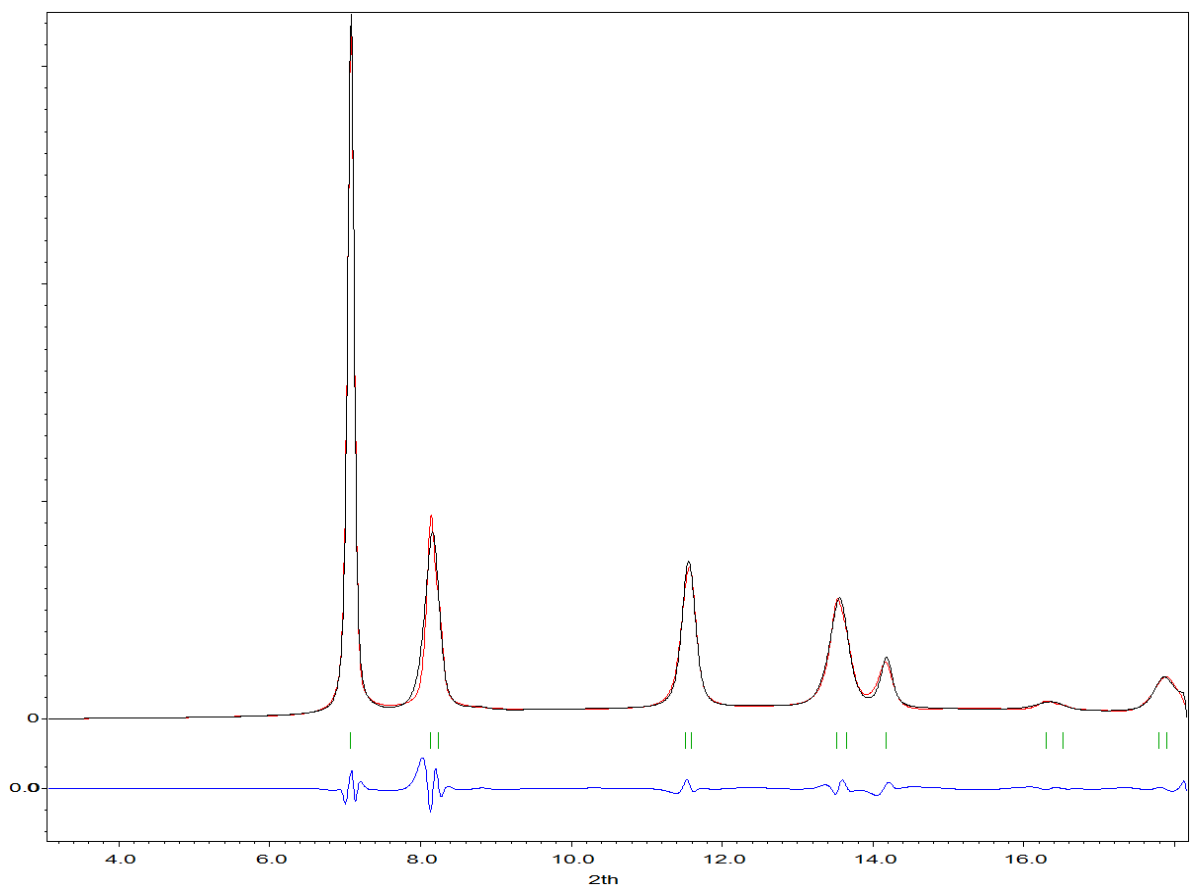

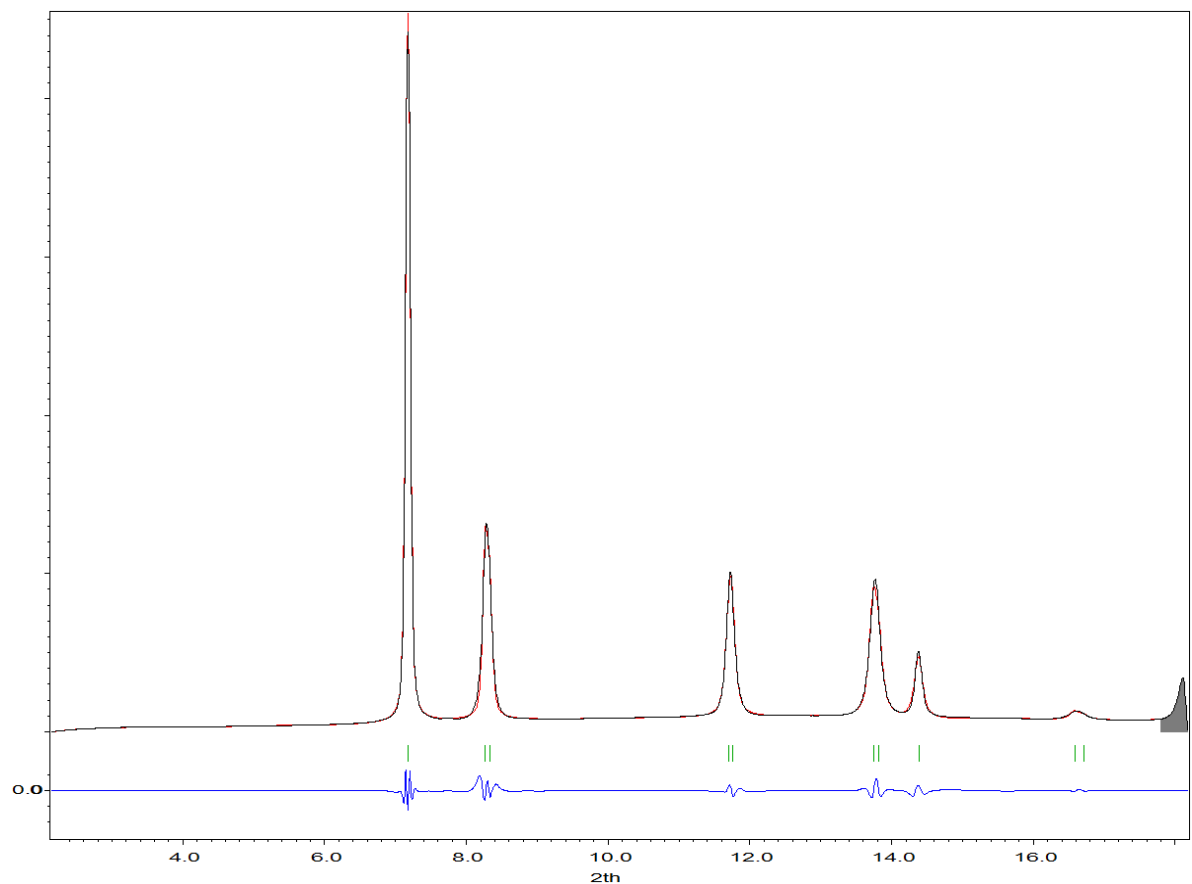

Figure S9. The diffraction data for the products of the reaction between Yb and $H_2$ after several cycles of laser heating, compressed to ca. 39.4 GPa, as measured in the two areas of the sample: $YbH_{2+x}$ P4/mmm. The LeBail fit has been marked with a red line, while the positions of Bragg reflections and the difference curve have been plotted at the bottom. $\lambda = 0.3344$

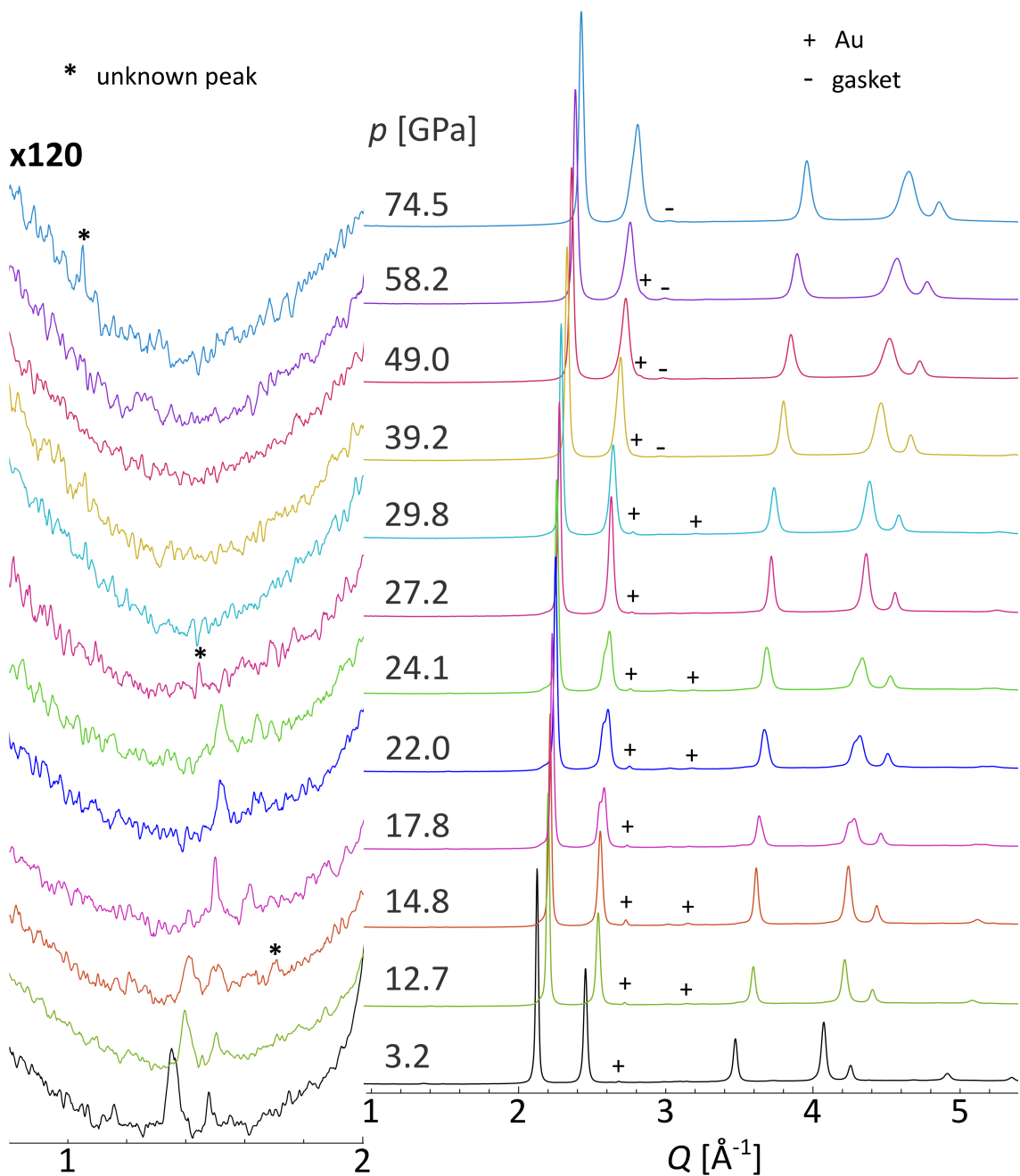

Figure S10. The integrated diffraction data for Yb$_3$H$_8$ compressed in H$_2$ (right) with the expanded low-Q region (left). λ = 0.4066 Å.

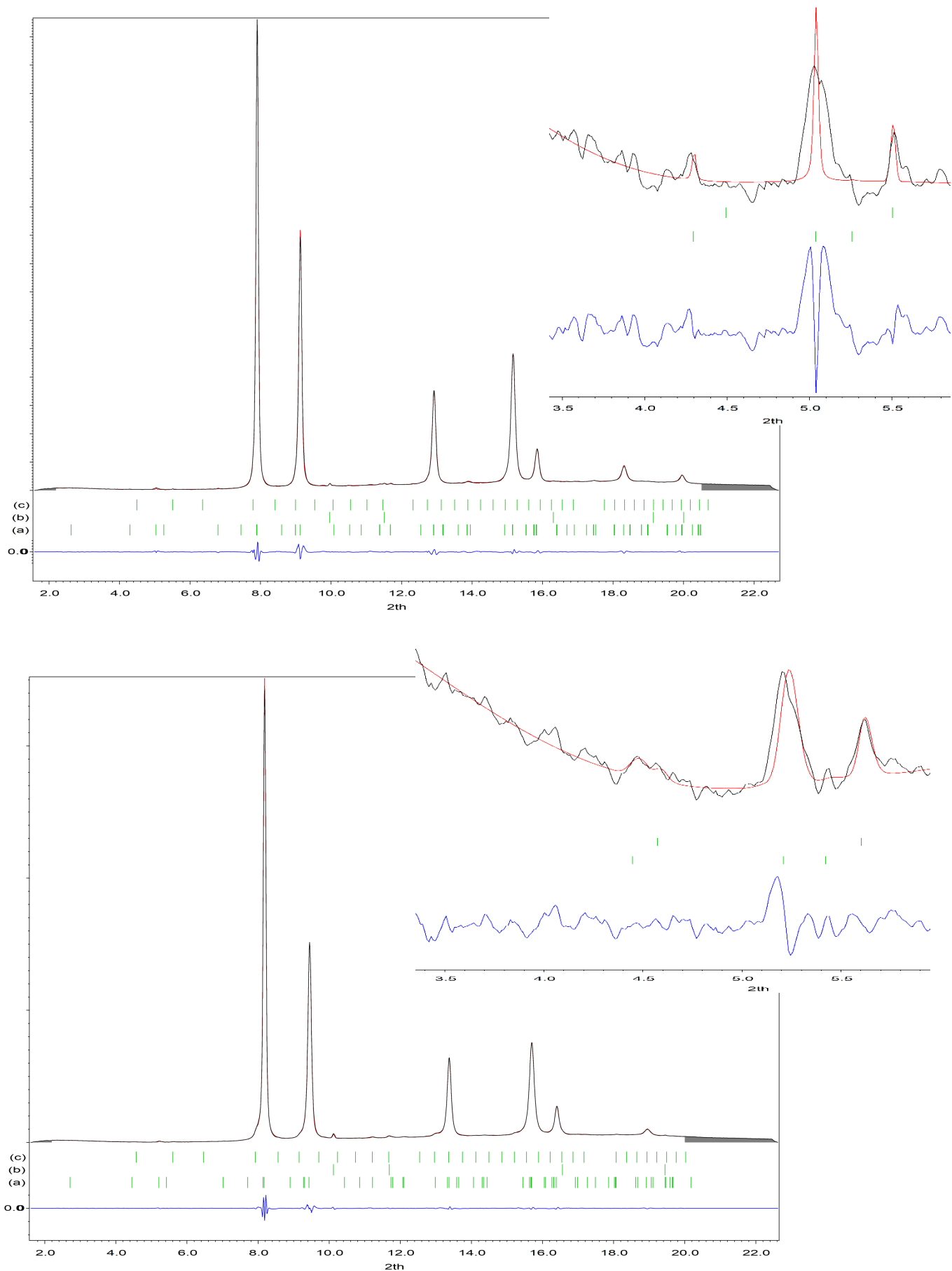

*Figure S11. The diffraction data for $Yb_3H_8$ compressed in $H_2$. **p** = 3.6 GPa (top) and 12.7 GPa (bottom), (a) – $YbH_{2+x}$ P-31m, (b) – Au Fm-3m, (c) – $Yb_2O_3$ Ia-3. LeBail fit has been marked with a red line, while the positions of Bragg reflections and the difference curve have been plotted at the bottom. The low-angle area is shown as an inset. λ = 0.4066 Å.*

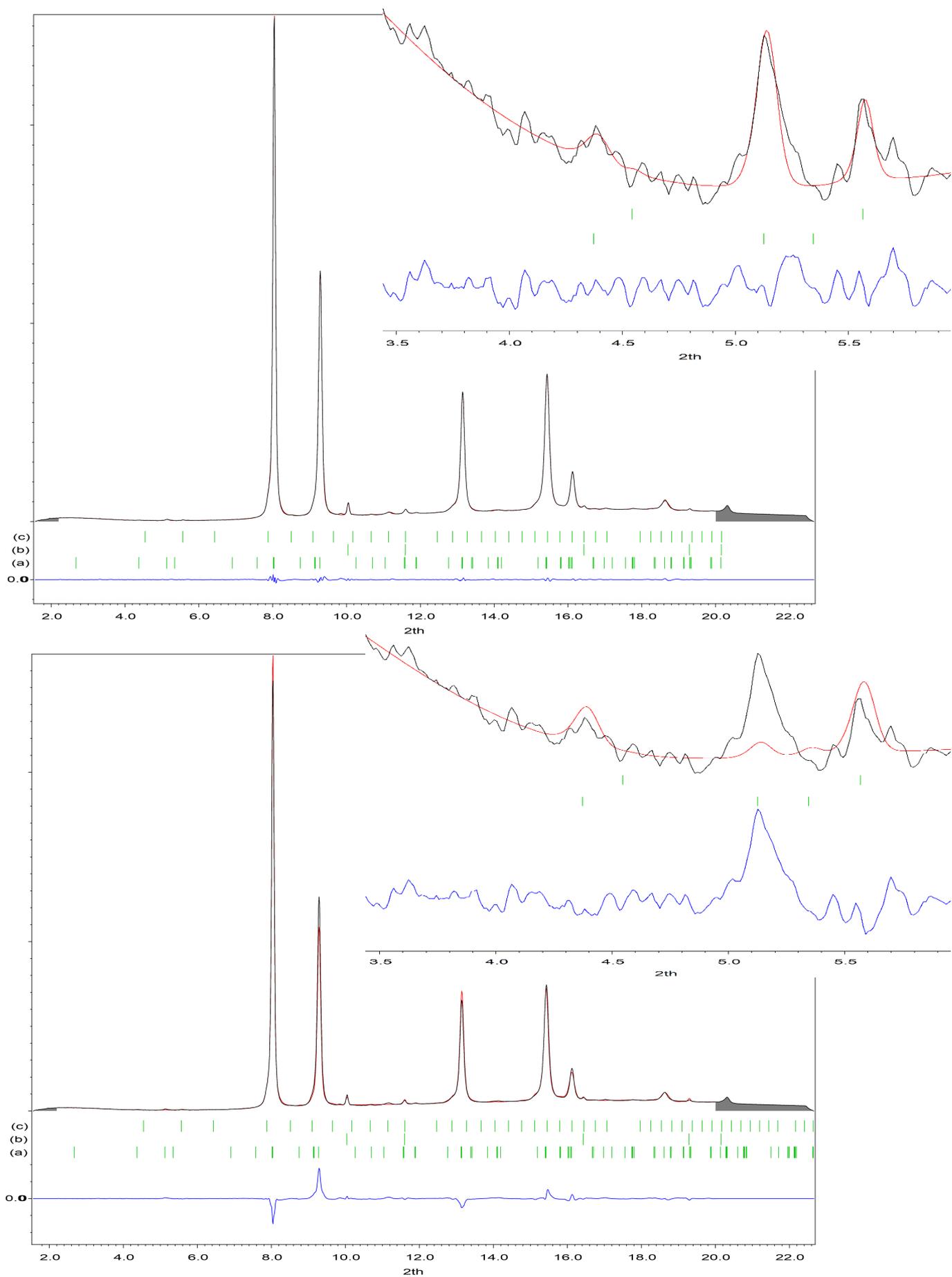

*Figure S12. A comparison of the LeBail (top) and **Rietveld** fits (bottom) for $Yb_3H_8$ compressed in $H_2$. **p** = 7.9 GPa, (a) – $YbH_{2+x}$ P-31m, (b) – Au Fm-3m, (c) – $Yb_2O_3$ Ia-3. LeBail (top plot) and Rietveld (bottom plot) fit have been marked with a red line, while the positions of Bragg reflections and the difference curve have been plotted at the bottom. The low-angle area is shown as an inset. $\lambda$ = 0.4066 Å. The relative phase amounts in wt.%: $YbH_{2+x}$ (as $Yb_3H_8$): 96.6(9); Au: 1.02(16); $Yb_2O_3$: 2.9(9). LeBail fit: wRp = 2.28%. Rietveld fit: wRp = 7.30%.*

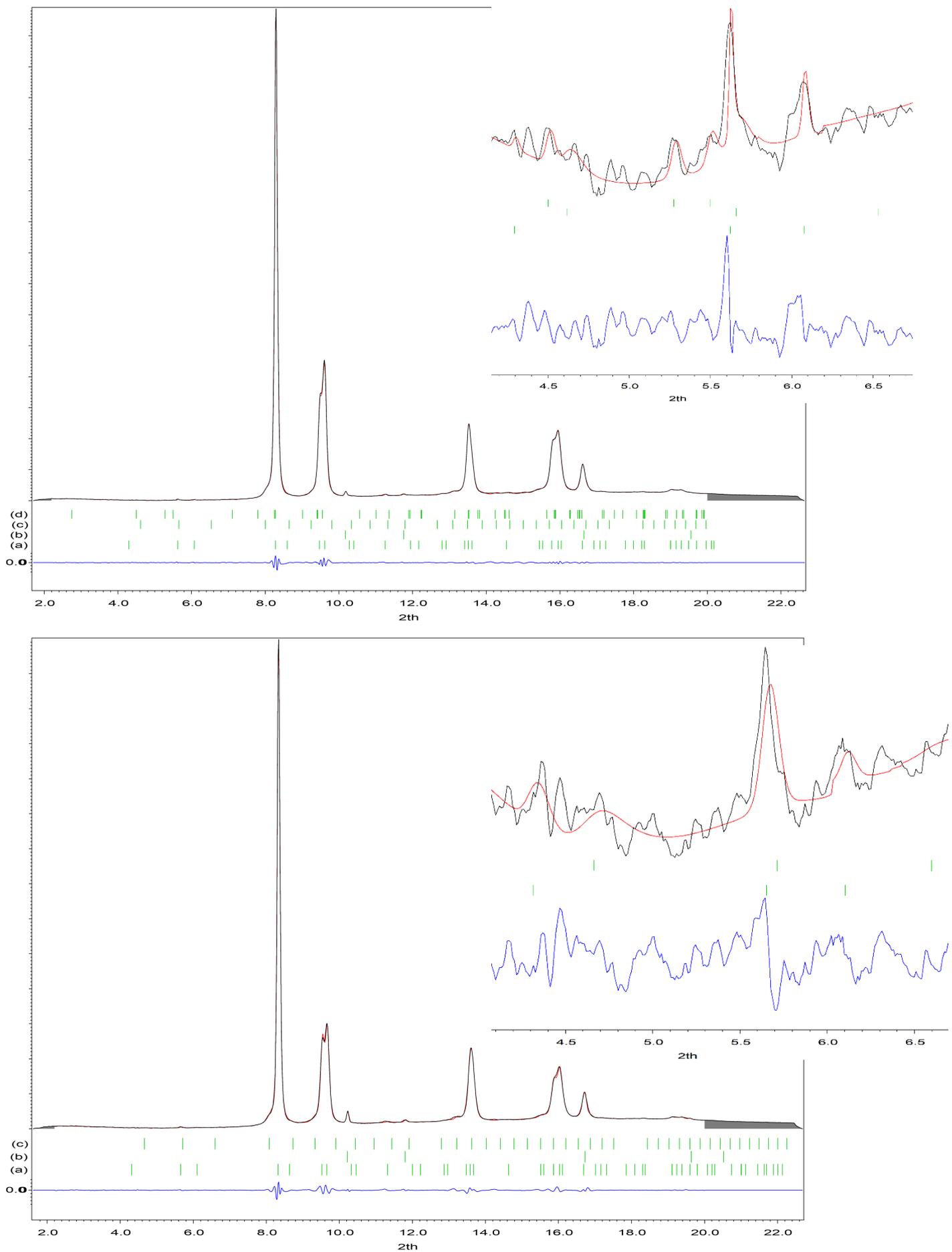

*Figure S13. The diffraction data for $Yb_3H_8$ compressed in $H_2$. **p** = 17.8 GPa (top) and 20.8 GPa (bottom), (a) – $YbH_{2+x}$ I4/m, (b) – Au Fm-3m, (c) – $Yb_2O_3$ Ia-3, (d) – $YbH_{2+x}$ P-31m. LeBail fit has been marked with a red line, while the positions of Bragg reflections and the difference curve have been plotted at the bottom. The low-angle area is shown as an inset. λ = 0.4066 Å.*

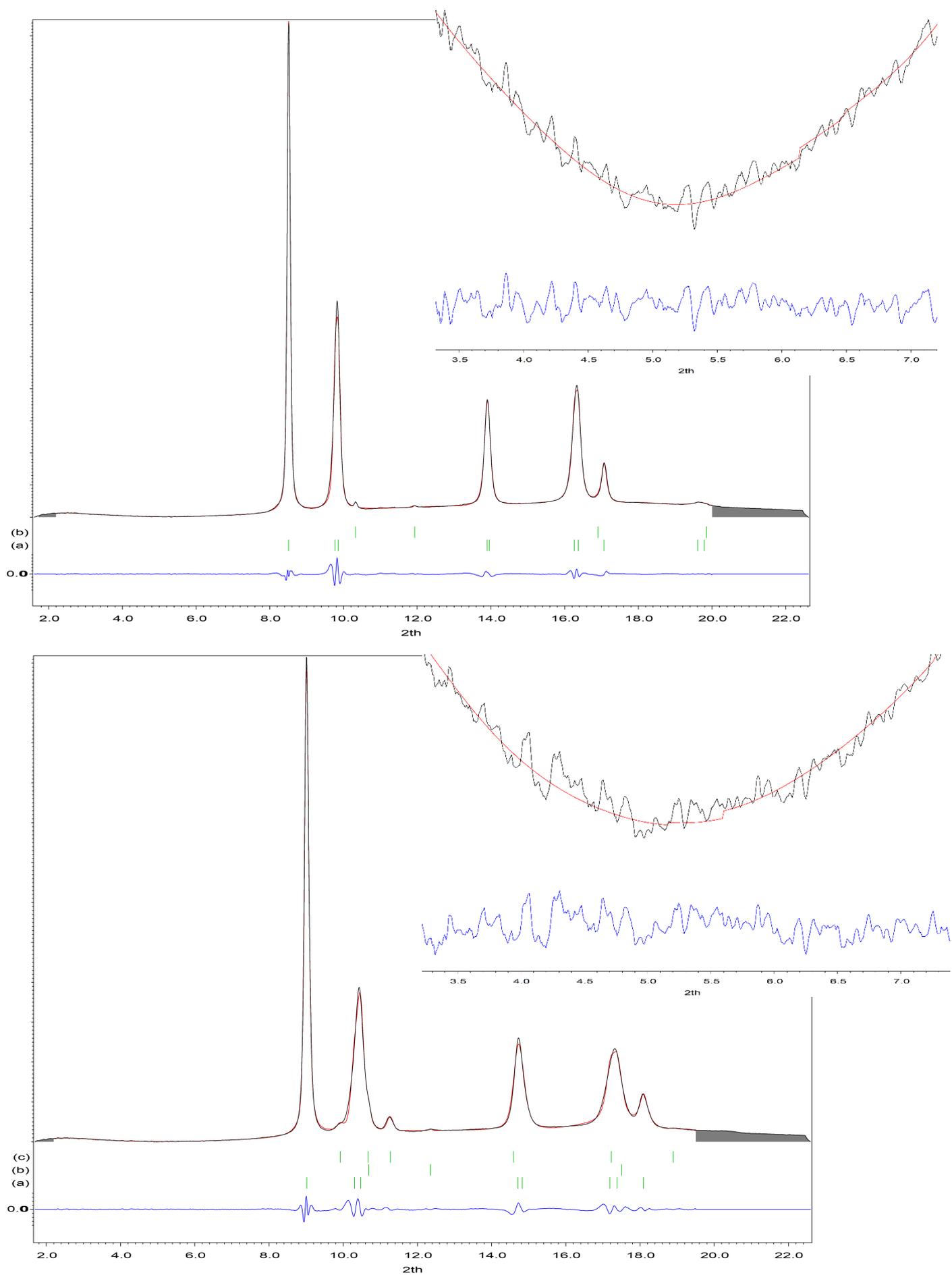

*Figure S14. The diffraction data for $Yb_3H_8$ compressed in $H_2$. **p** = 29.8 GPa (top) and 73.5 GPa (bottom), (a) – $YbH_{2+x}$ I4/mmm, (b) – Au Fm-3m, (c) – Re gasket $P6_3$/mmc. LeBail fit has been marked with a red line, while the positions of Bragg reflections and the difference curve have been plotted at the bottom. The low-angle area is shown as an inset. λ = 0.4066 Å.*

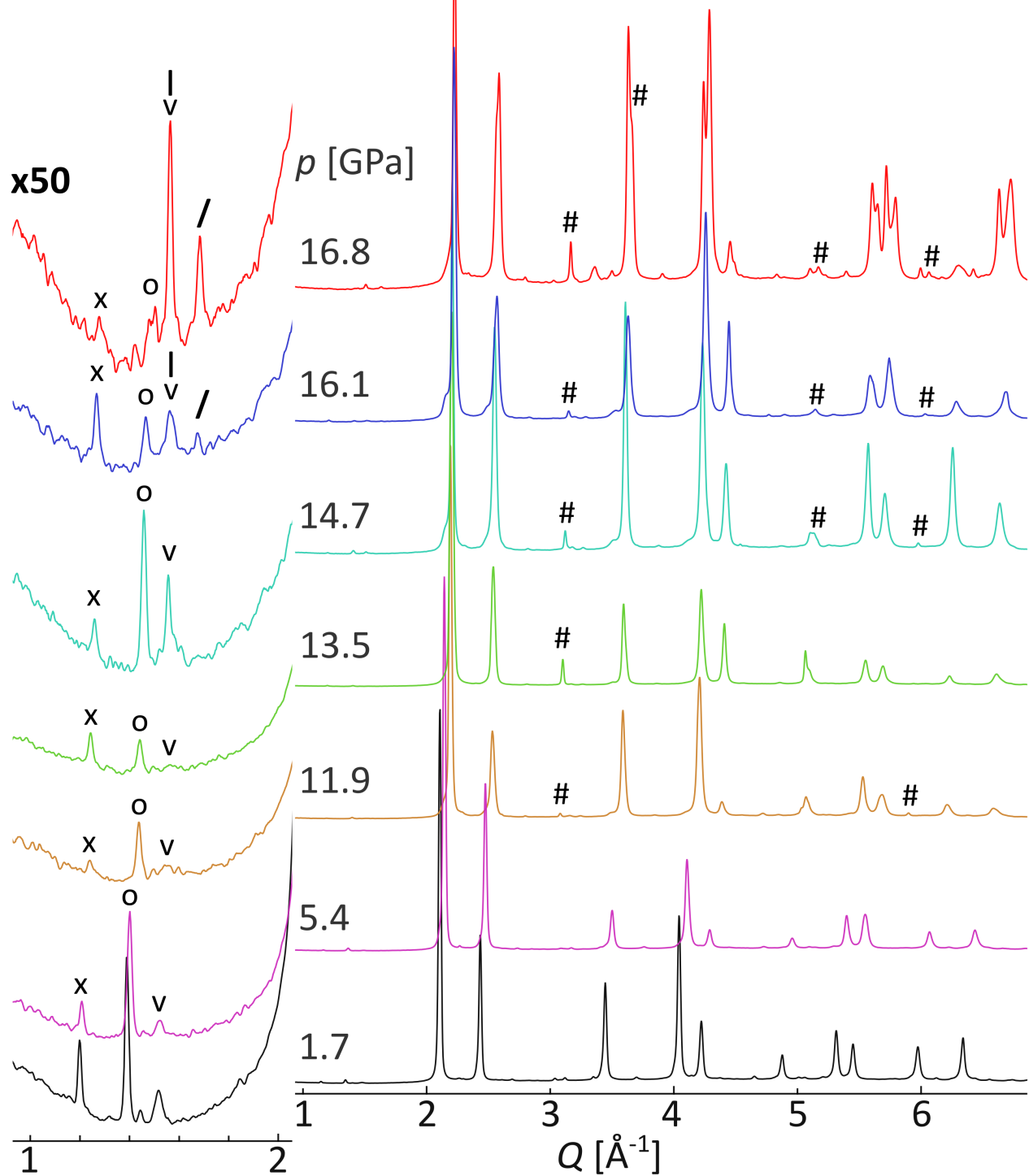

Figure S15. The integrated diffraction data for $Yb_3H_8$ compressed in Ne (right) with the expanded low-Q region (left). $\lambda$ = 0.3344 Å. The most visible reflections of the $YbH_{2+x}$ and $Yb_2O_3$ phases were marked on the low-Q region.

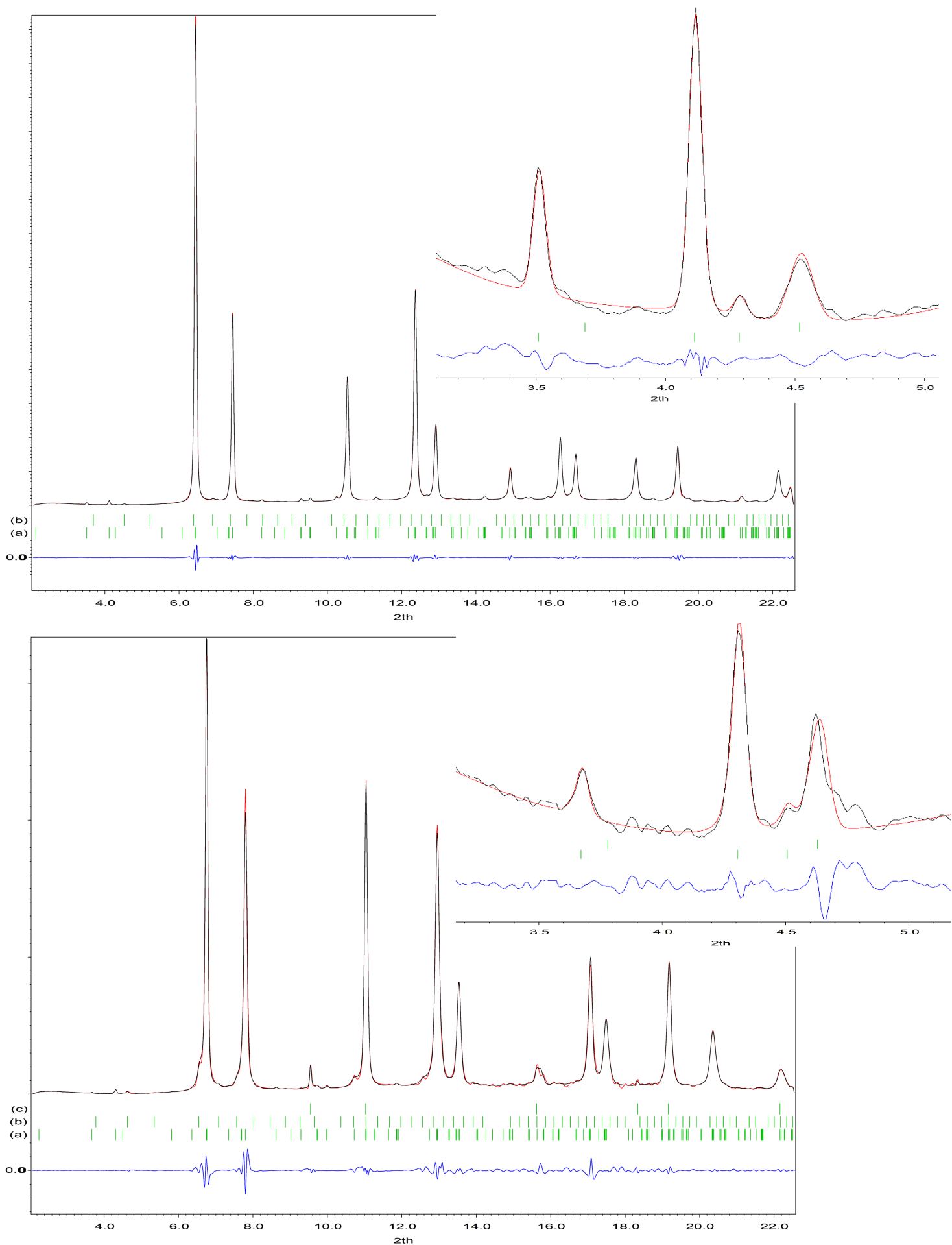

*Figure S16. The diffraction data for $Yb_3H_8$ compressed in Ne. **p** = 1.7 GPa (top) and 14.7 GPa (bottom), (a) – $YbH_{2+x}$ P-31m, (b) – $Yb_2O_3$ Ia-3, (c) – Ne Fm-3m. LeBail fit has been marked with a red line, while the positions of Bragg reflections and the difference curve have been plotted at the bottom. The low-angle area is shown as an inset. λ = 0.3344 Å.*

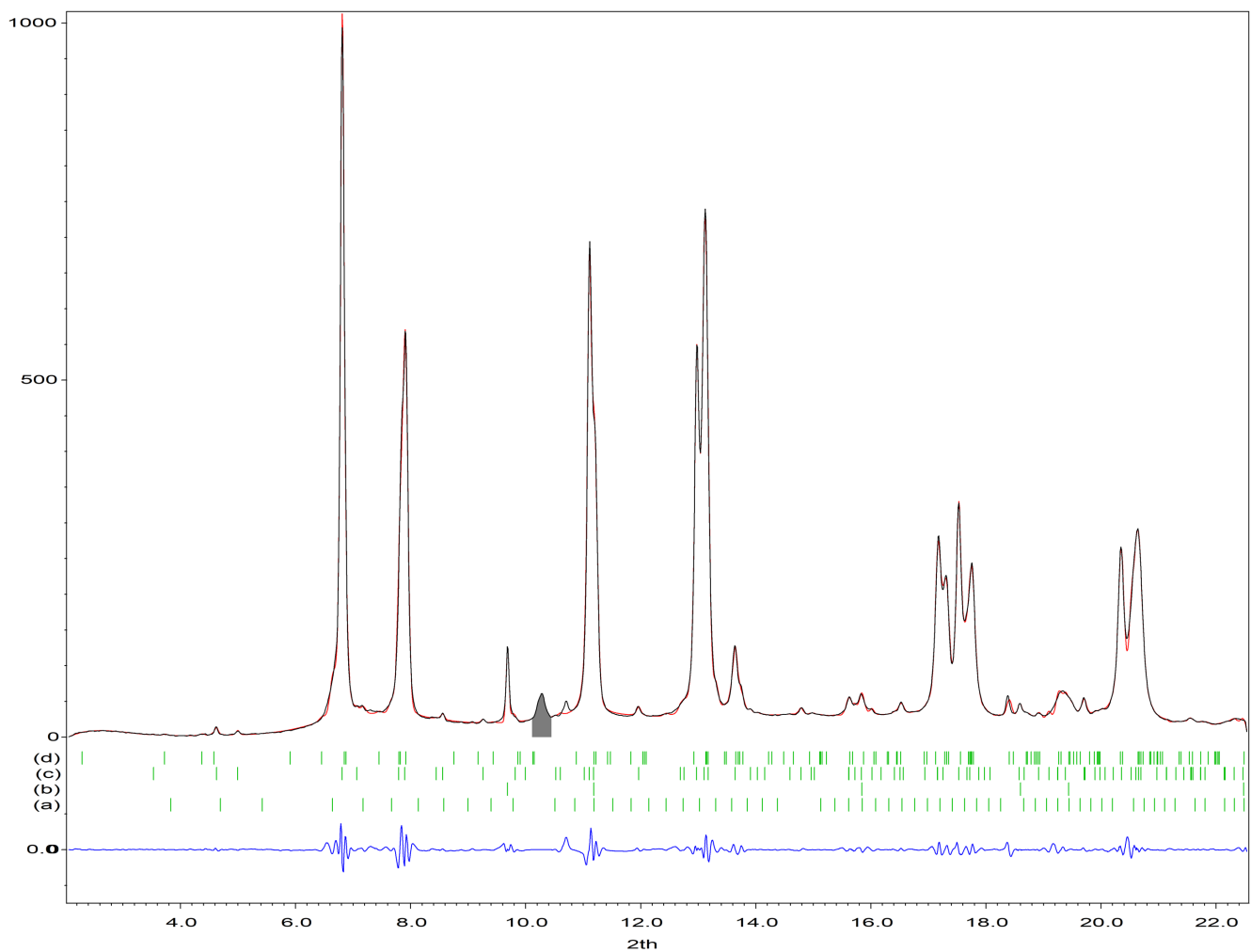
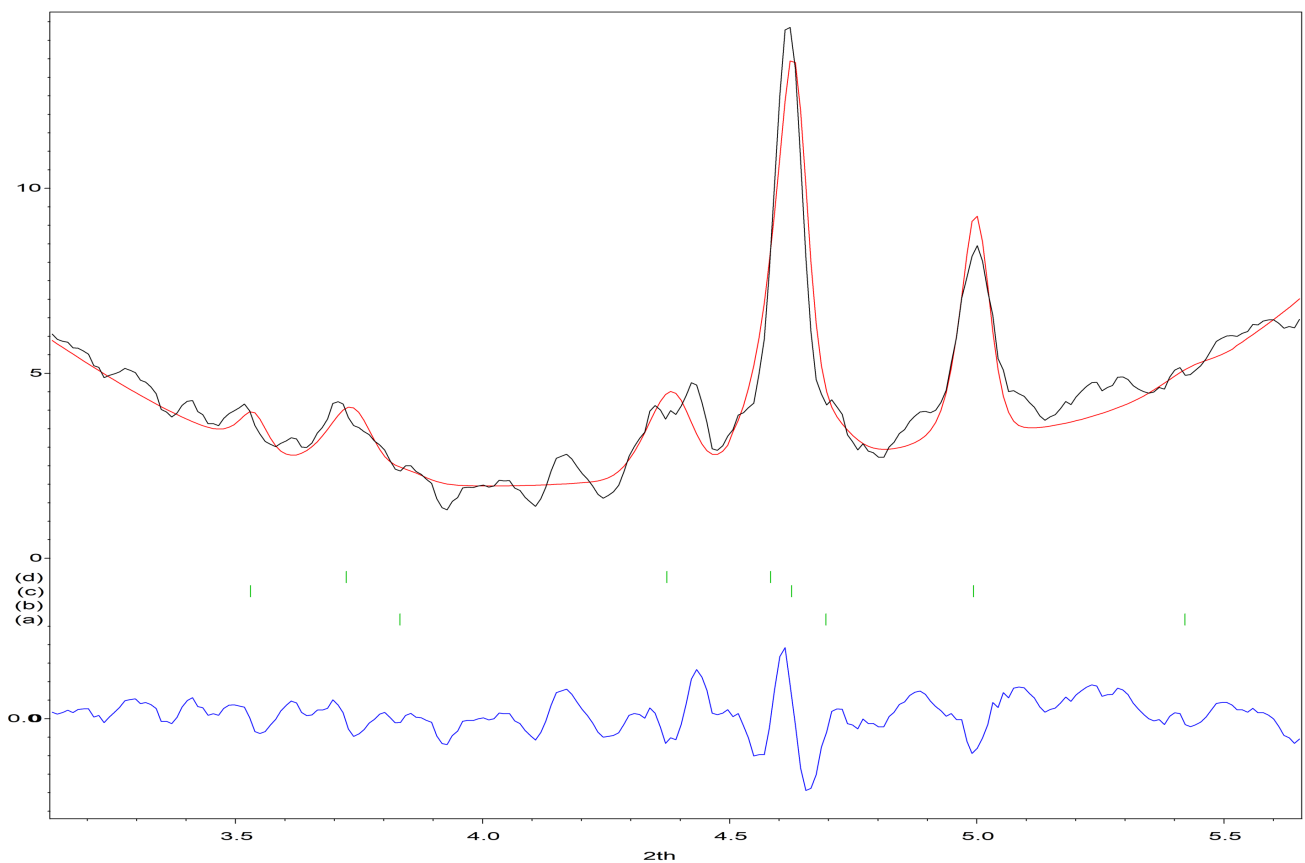

Figure S17. The diffraction data for $Yb_3H_8$ compressed in Ne. **p** = 16.8 GPa, (a) – $Yb_2O_3$ Ia-3, (b) – Ne Fm-3m, (c) – $YbH_{2+x}$ I4/m, (d) – $YbH_{2+x}$ P-31m. LeBail fit has been marked with a red line, while the positions of Bragg reflections and the difference curve have been plotted at the bottom. The low-angle area is shown at the bottom. $\lambda = 0.3344$ Å.

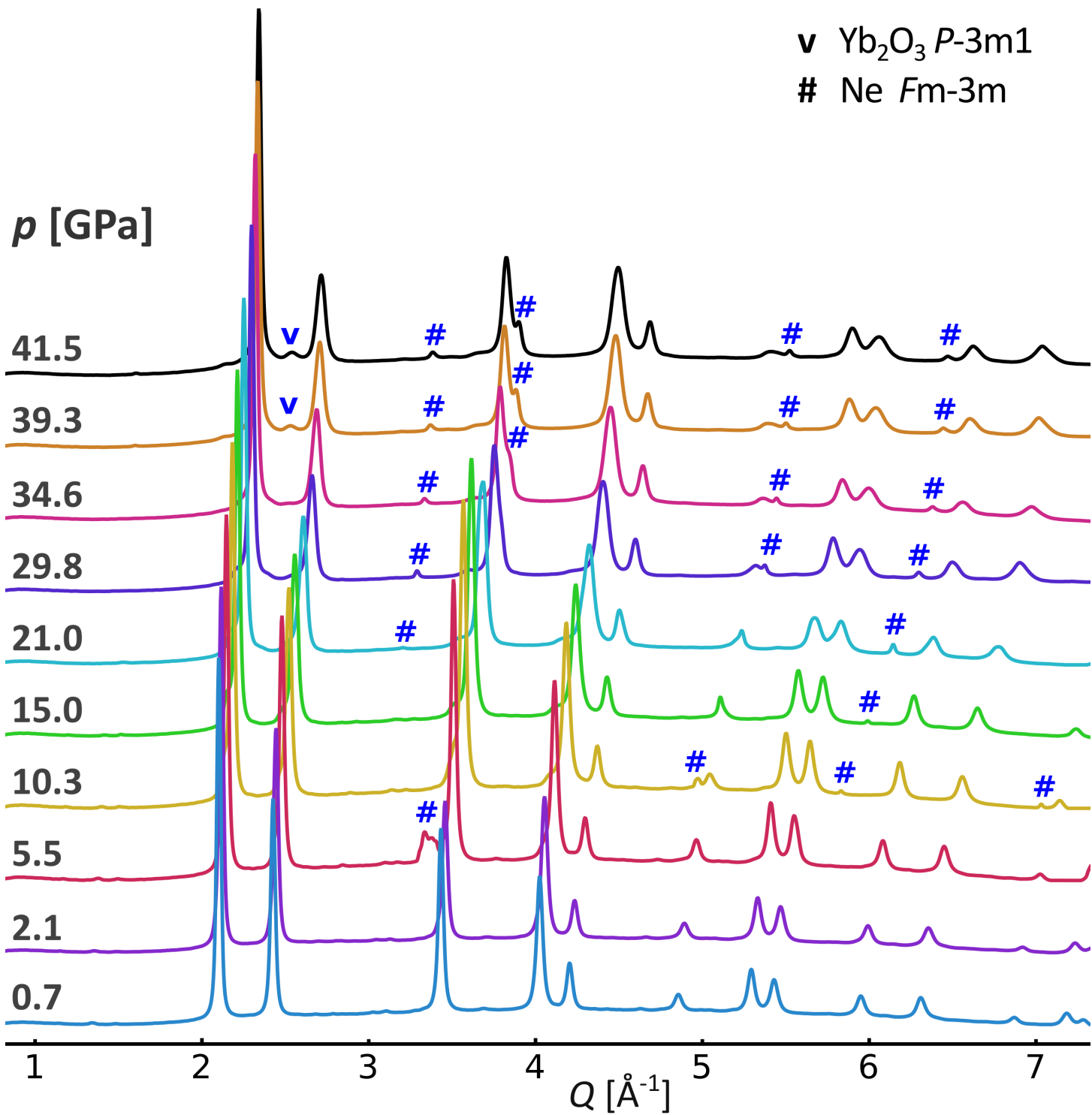

Figure S18. The integrated diffraction data for Yb$_3$H$_8$ compressed in Ne. λ = 0.2952 Å. The most visible reflections of the Yb$_2$O$_3$ and Ne phases were marked.

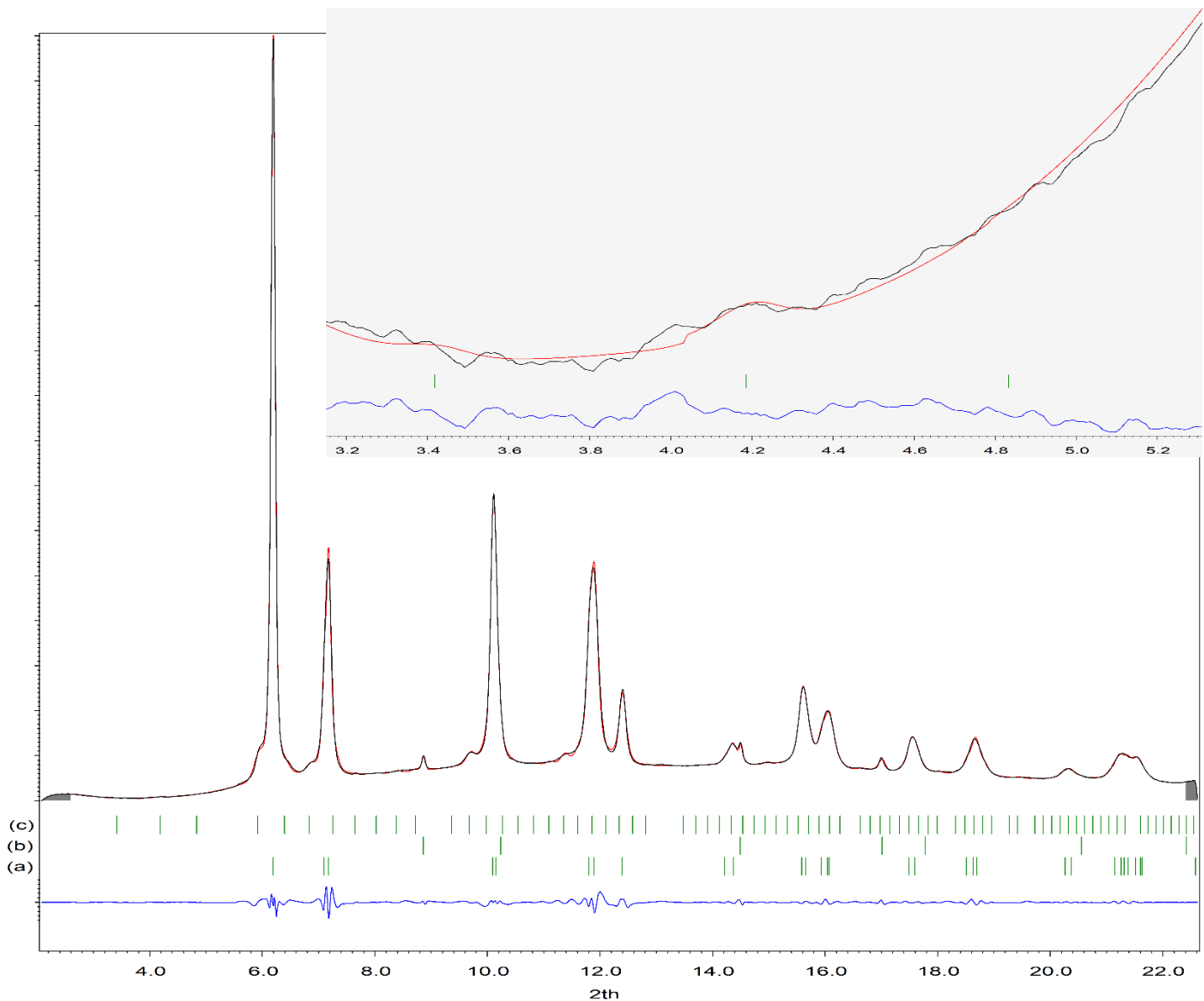

*Figure S19. The diffraction data for $Yb_3H_8$ compressed in Ne. **p** = 28.9 GPa, (a) – $YbH_{2+x}$ I4/mmm, (b) – Ne Fm-3m, (c) – $Yb_2O_3$ Ia-3. LeBail fit has been marked with a red line, while the positions of Bragg reflections and the difference curve have been plotted at the bottom. The low-angle area is shown as an inset (ca. 80x magnification of the ordinate). λ = 0.2952 Å.*

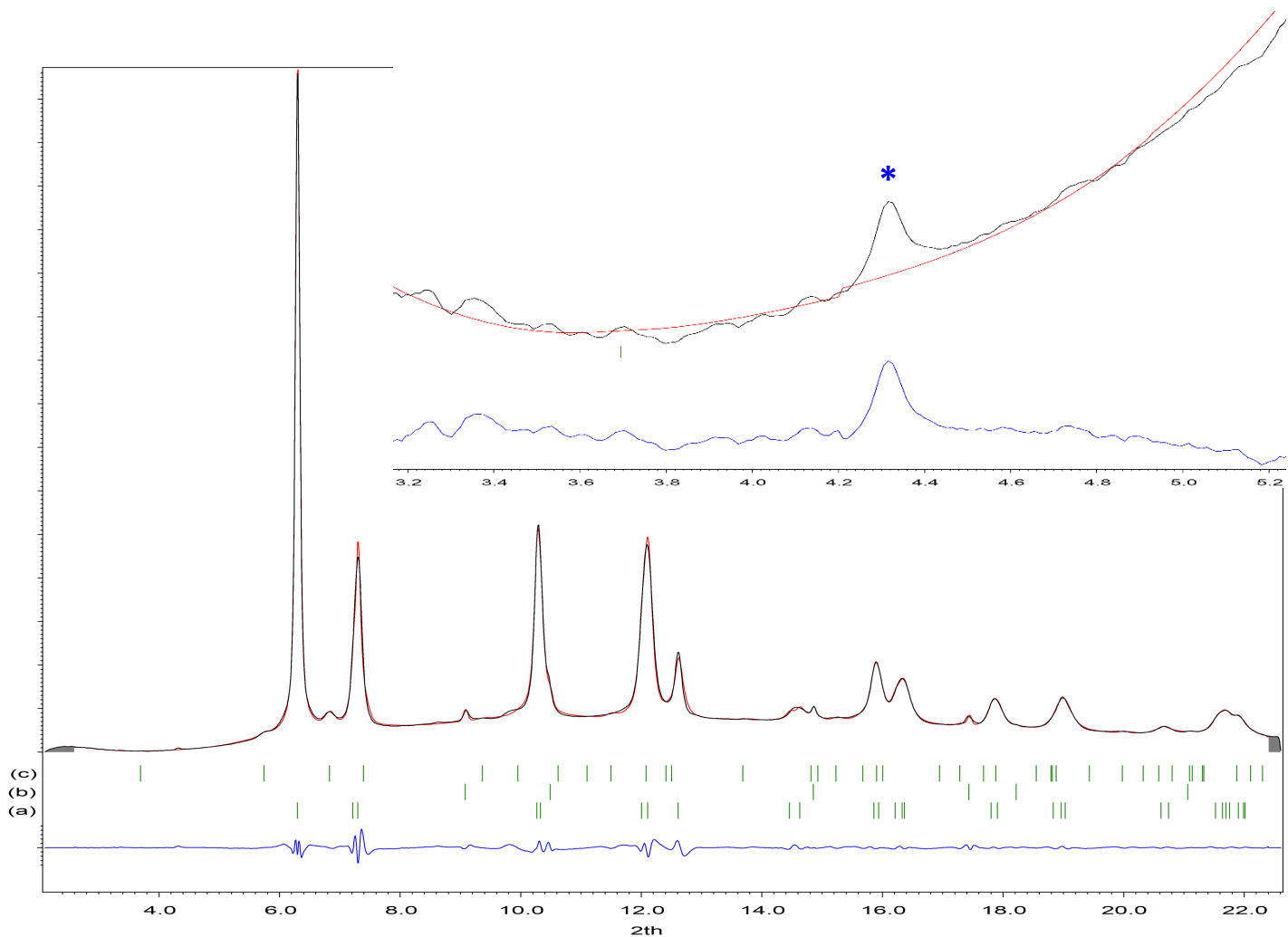

*Figure S20. The diffraction data for Yb$_3$H$_8$ compressed in Ne. **p** = 39.3 GPa, (a) – YbH$_{2+x}$ I4/mmm, (b) – Ne Fm-3m, (c) – Yb$_2$O$_3$ P-3m1; \* – unidentified signal. LeBail fit has been marked with a red line, while the positions of Bragg reflections and the difference curve have been plotted at the bottom. The low-angle area is shown as an inset (ca. 80x magnification of the ordinate). λ = 0.2952 Å.*

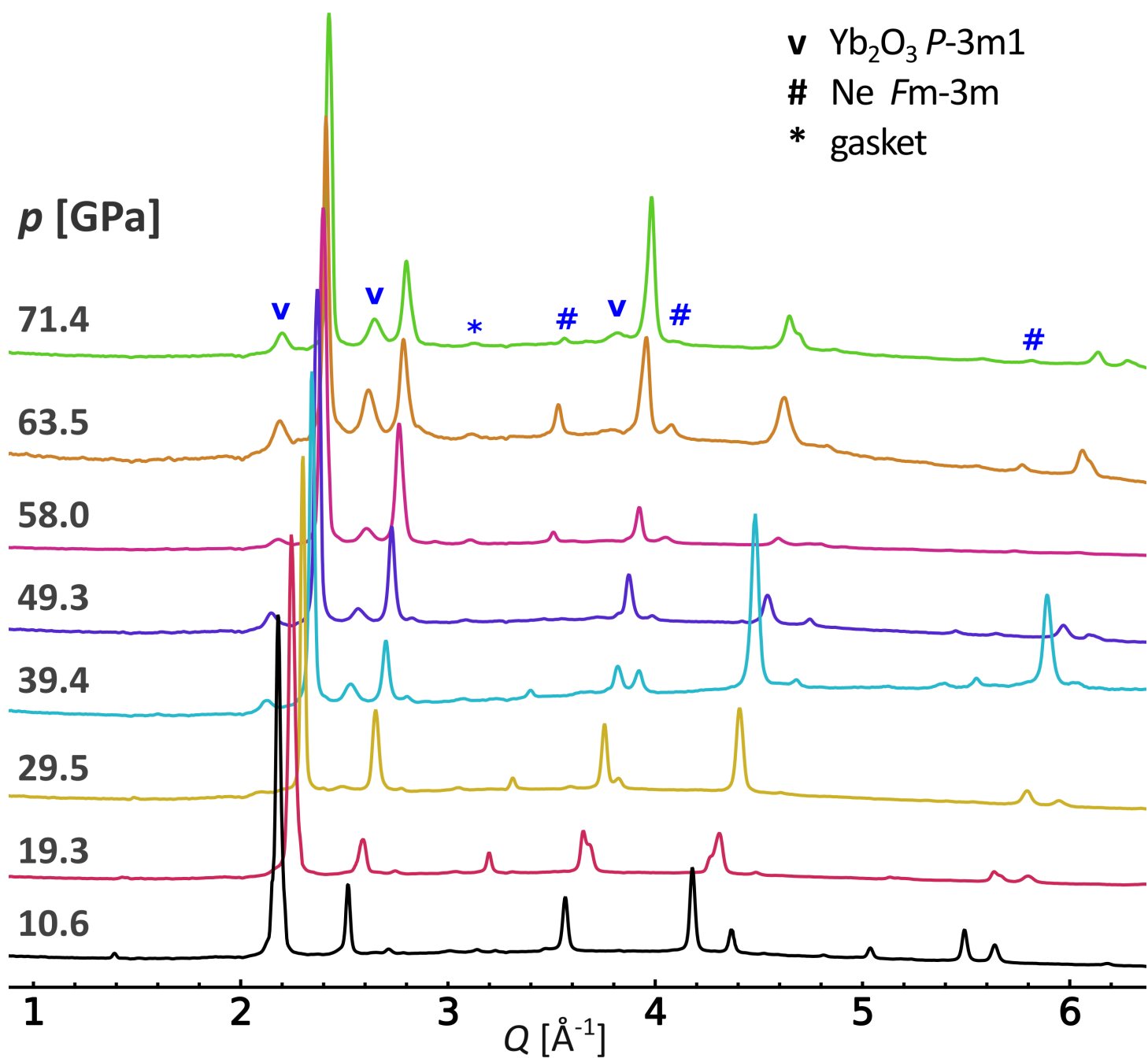

Figure S21. The integrated diffraction data for Yb$_3$H$_8$ compressed in Ne. λ = 0.3445 Å. The most visible reflections of the Yb$_2$O$_3$ and Ne phases were marked.

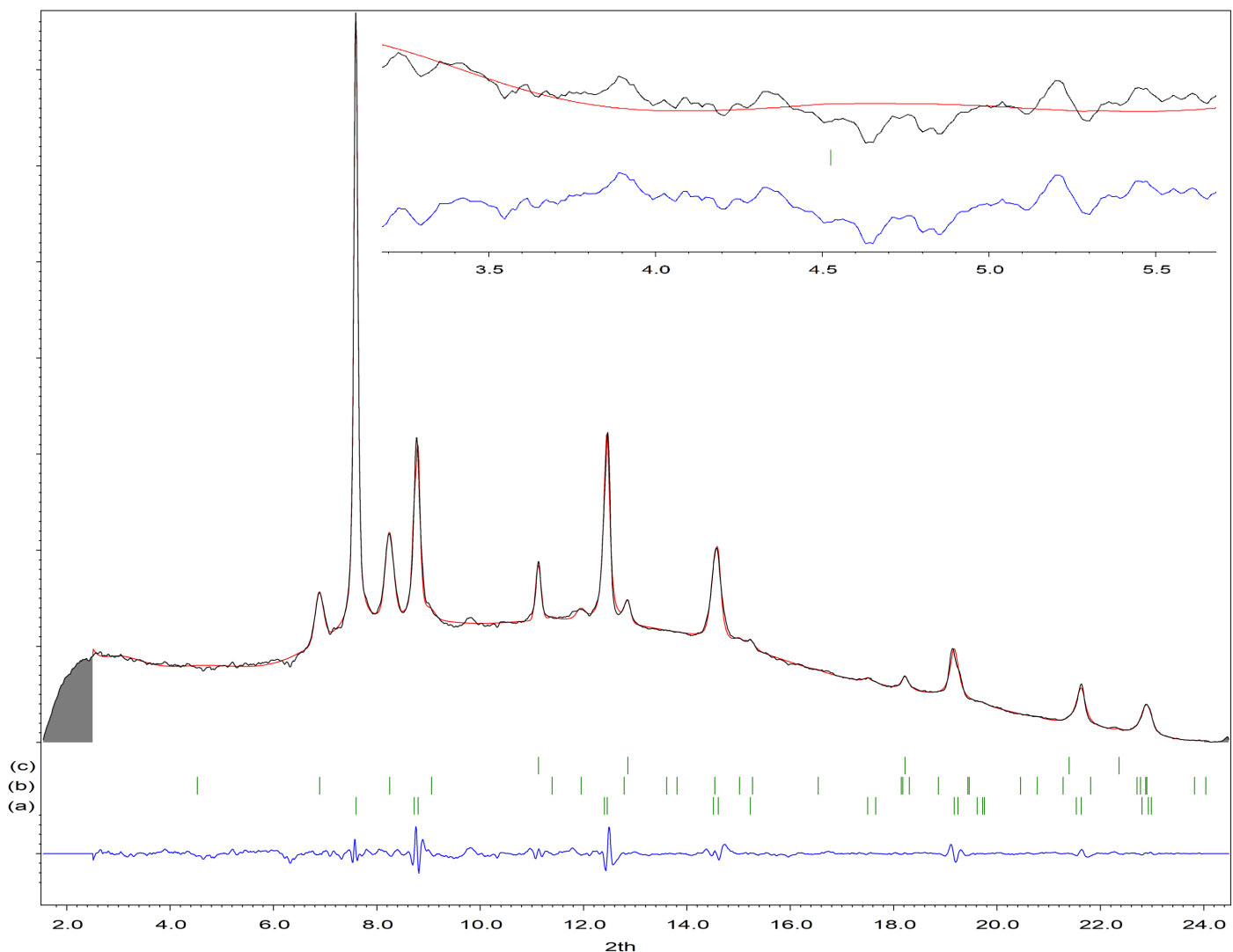

Figure S22. The diffraction data for $Yb_3H_8$ compressed in Ne. **p** = 63.5 GPa, (a) – $YbH_{2+x}$ I4/mmm, (b) – $Yb_2O_3$ P-3m1, (c)– Ne Fm-3m. LeBail fit has been marked with a red line, while the positions of Bragg reflections and the difference curve have been plotted at the bottom. The low-angle area is shown as an inset (ca. 80x magnification of the ordinate). λ = 0.3445 Å.

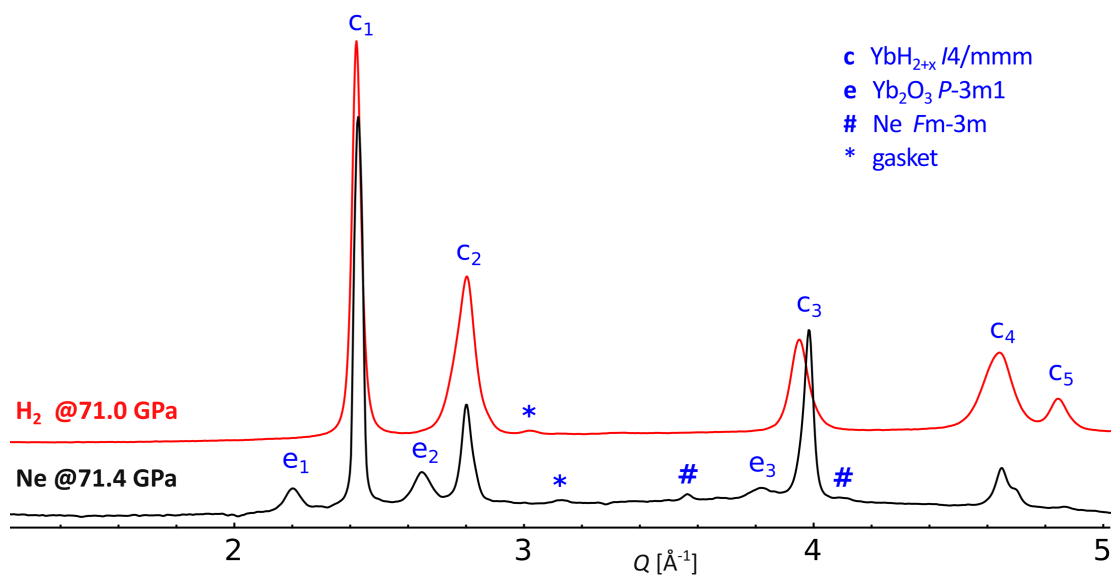

Figure S23. A comparison of the integrated diffraction data for $Yb_3H_8$ compressed to ca. 71 GPa in Ne and in $H_2$. λ = 0.3445 Å (Ne) and 0.4066 Å ($H_2$). The most visible reflections of the $YbH_{2+x}$ and $Yb_2O_3$ phases were marked: c – $YbH_{2+x}$ I4/mmm: 1-(1 0 1), 2-(0 0 2) and (1 1 0), 3-(1 1 2) and (2 0 0), 4-(1 0 3) and (2 1 1), 5-(2 0 2); e – $Yb_2O_3$ P-3m1: 1-(1 0 0), 2-(1 0 1), 3-(2 -1 0).

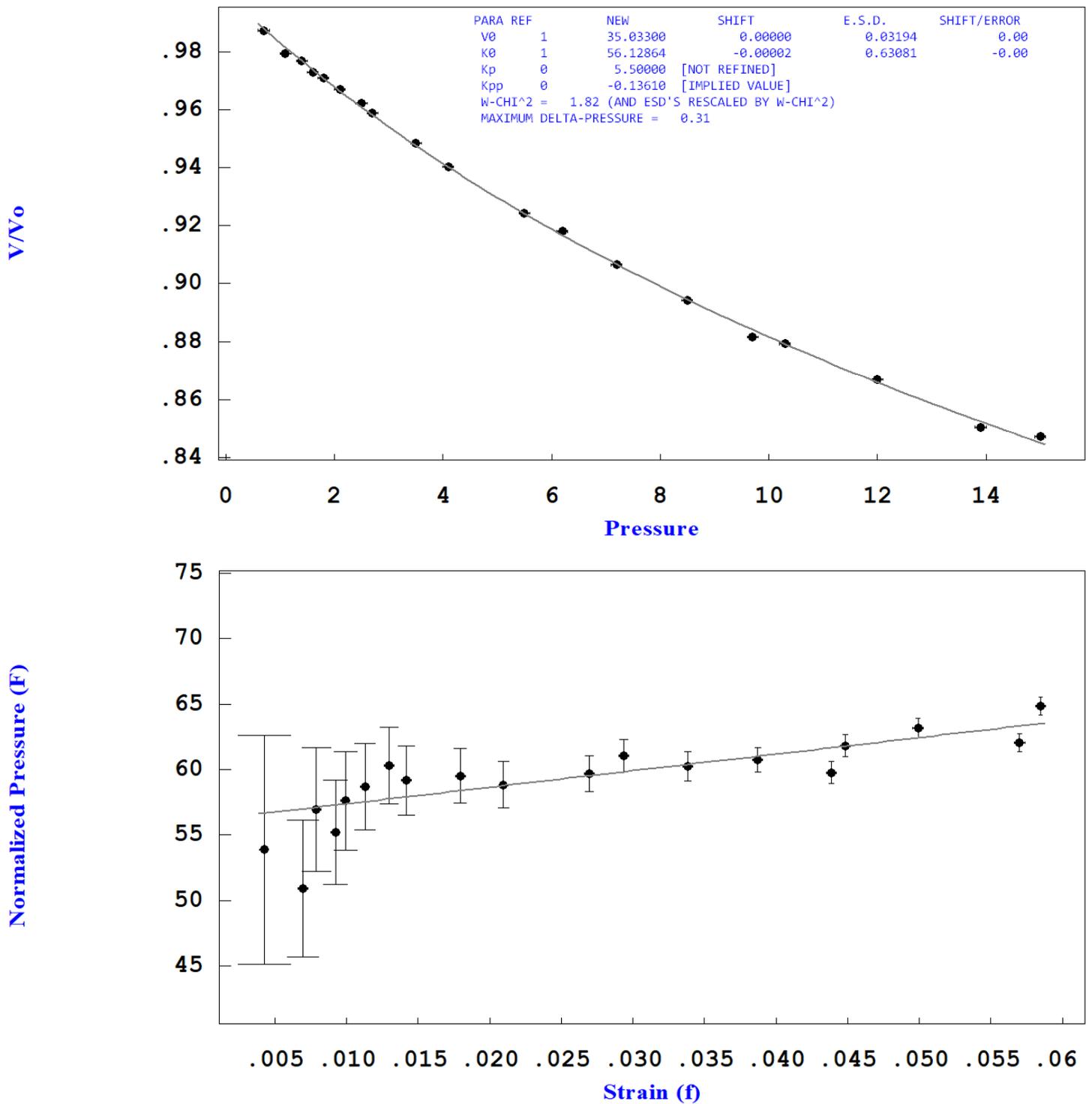

Figure S24. The fit of the volume per Yb atom to the third-order Birch-Murnaghan equation of state for P–31m phase of $Yb_3H_8$ compressed in Ne PTM (top) and the F–f plot (bottom). F – normalized pressure [GPa], f – Eulerian strain; $F = P/3f(1+2f)^{5/2}$, $f = [(V_0/V)^{2/3} – 1]/2$, $F = K_0 + [3K_0(K' – 4)/2]f$. The experimental data reveal the positive slope of the F–f plot, indicating $K' > 4$.[1] Run (g).

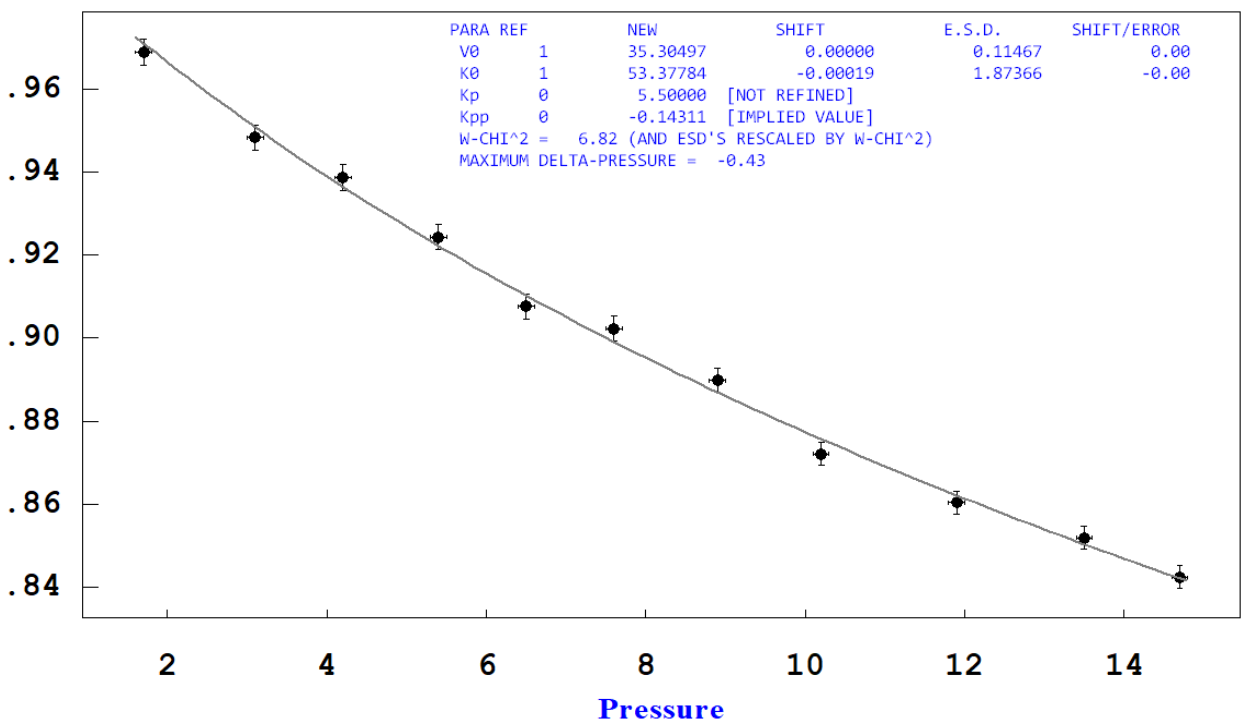

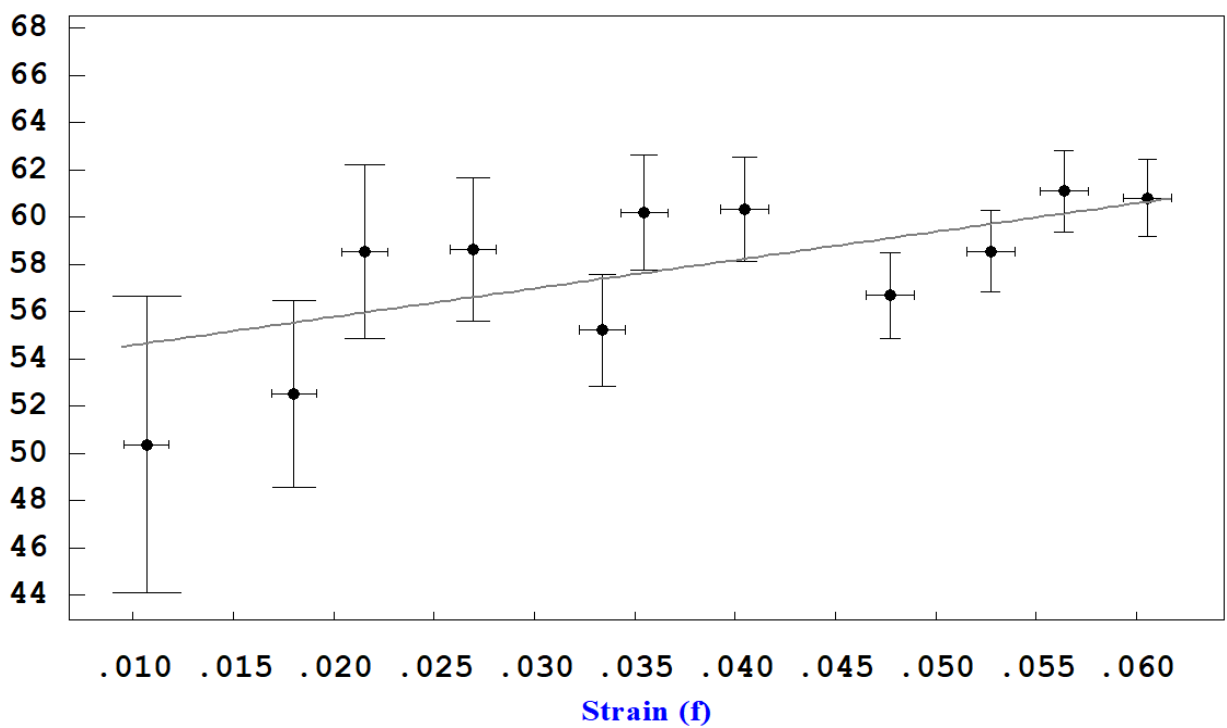

Figure S25. The fit of the volume per Yb atom to the third-order Birch-Murnaghan equation of state for P–31m phase of $Yb_3H_8$ compressed in Ne PTM (top) and the F–f plot (bottom). F – normalized pressure [GPa], f – Eulerian strain. Alternative run (h).

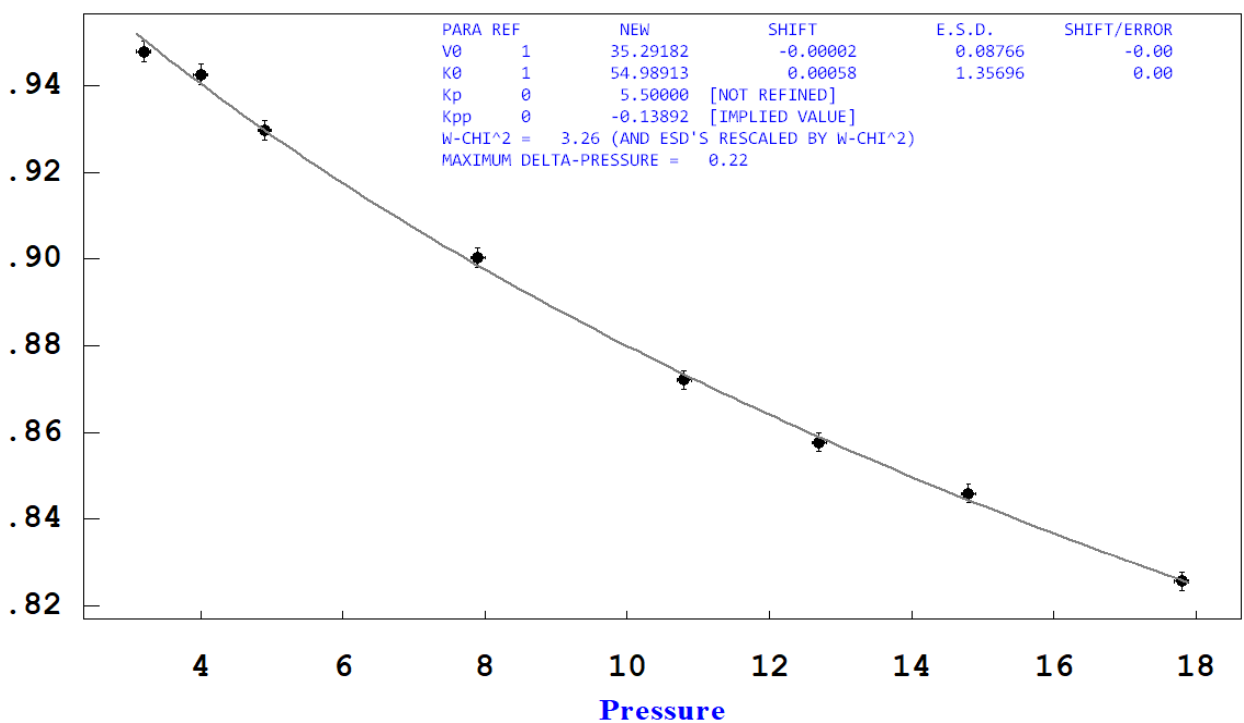

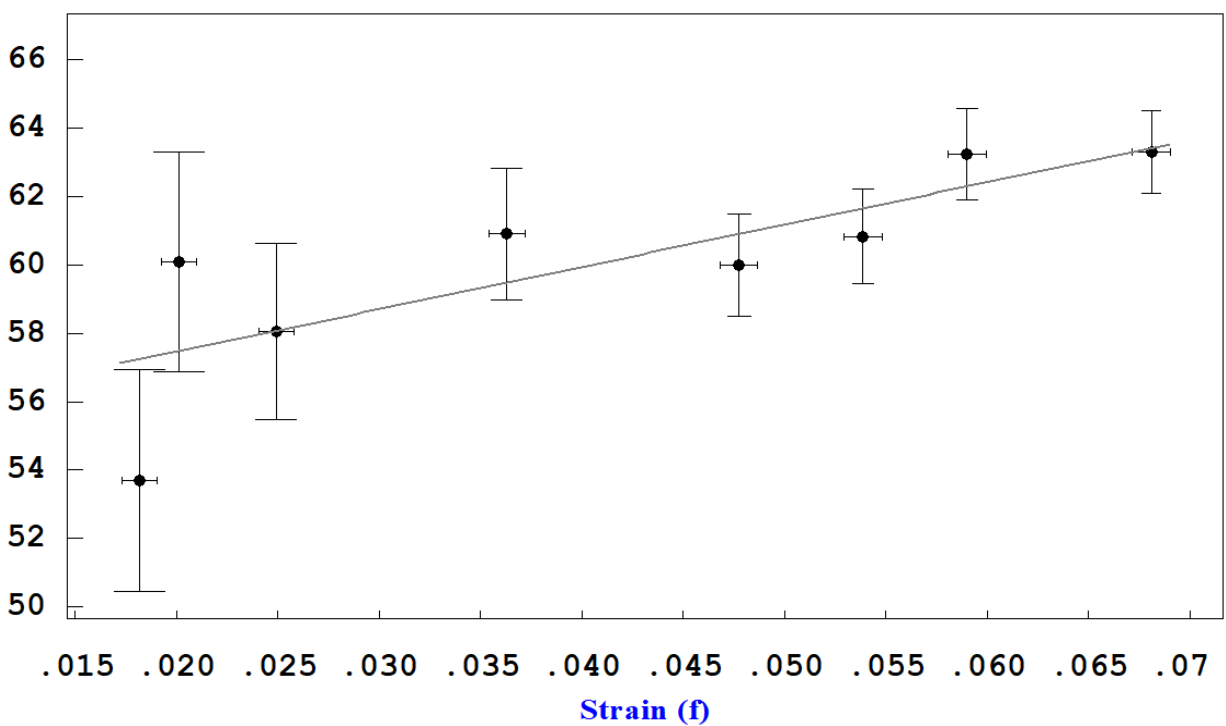

Figure S26. The fit of the volume per Yb atom to the third-order Birch-Murnaghan equation of state for P–31m phase of Yb$_3$H$_8$ compressed in H$_2$ PTM (top) and the F–f plot (bottom). F – normalized pressure [GPa], f – Eulerian strain. Run (d).

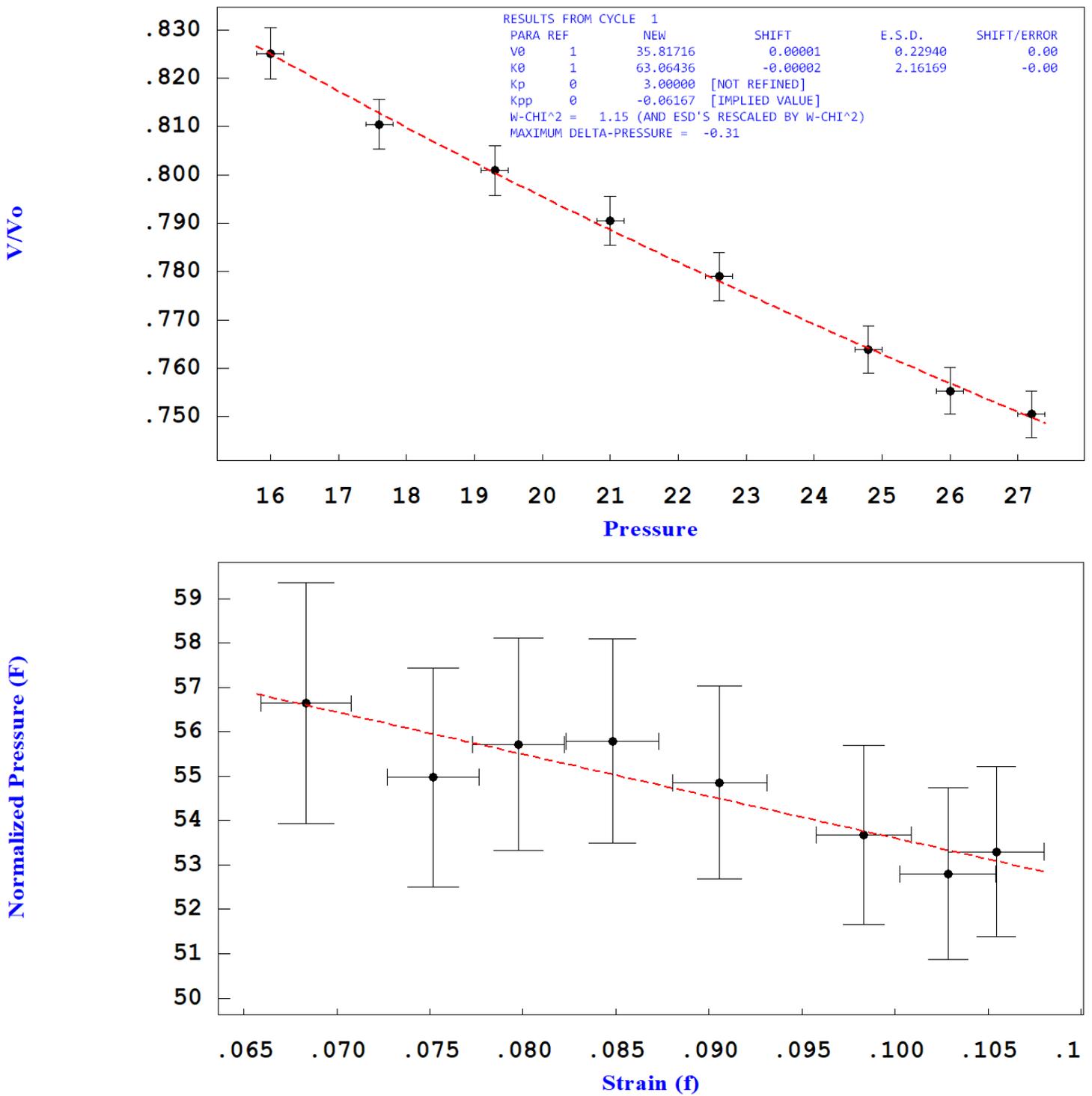

Figure S27. The fit of the volume per Yb atom to the third-order Birch-Murnaghan equation of state for I4/m phase of $Yb_3H_8$ compressed in Ne PTM (top) and the F–f plot (bottom). F – normalized pressure [GPa], f – Eulerian strain. Run (g).

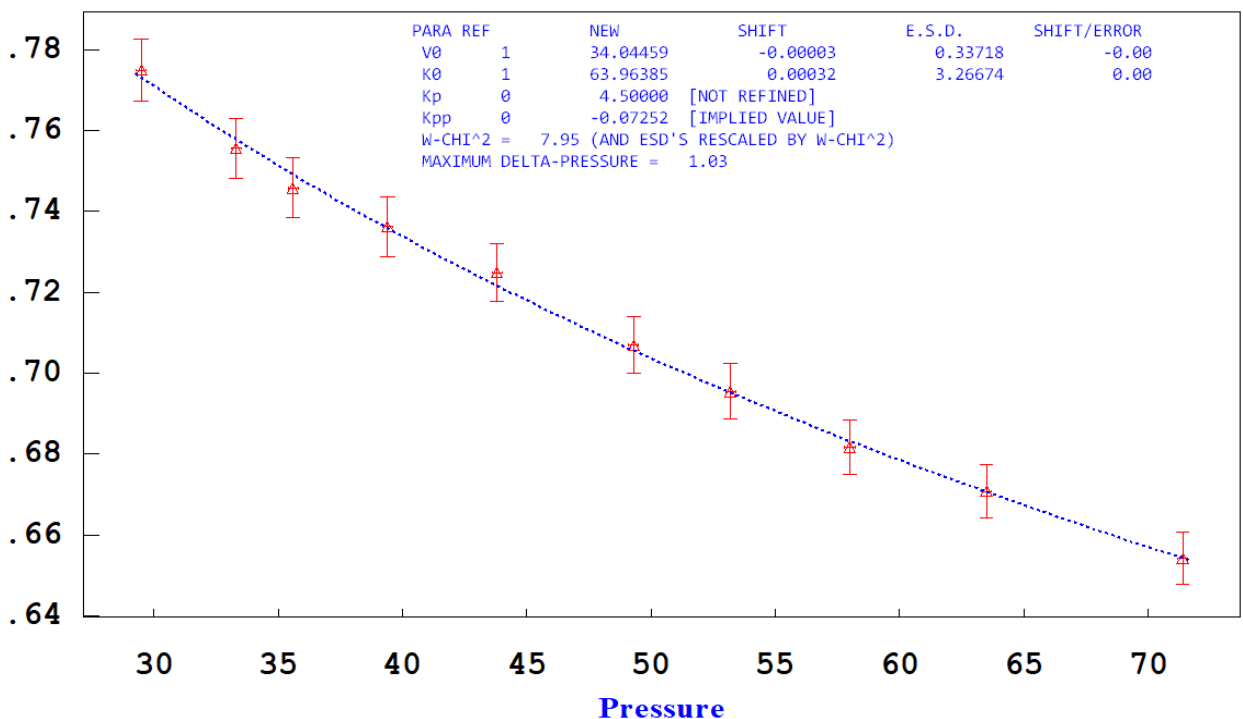

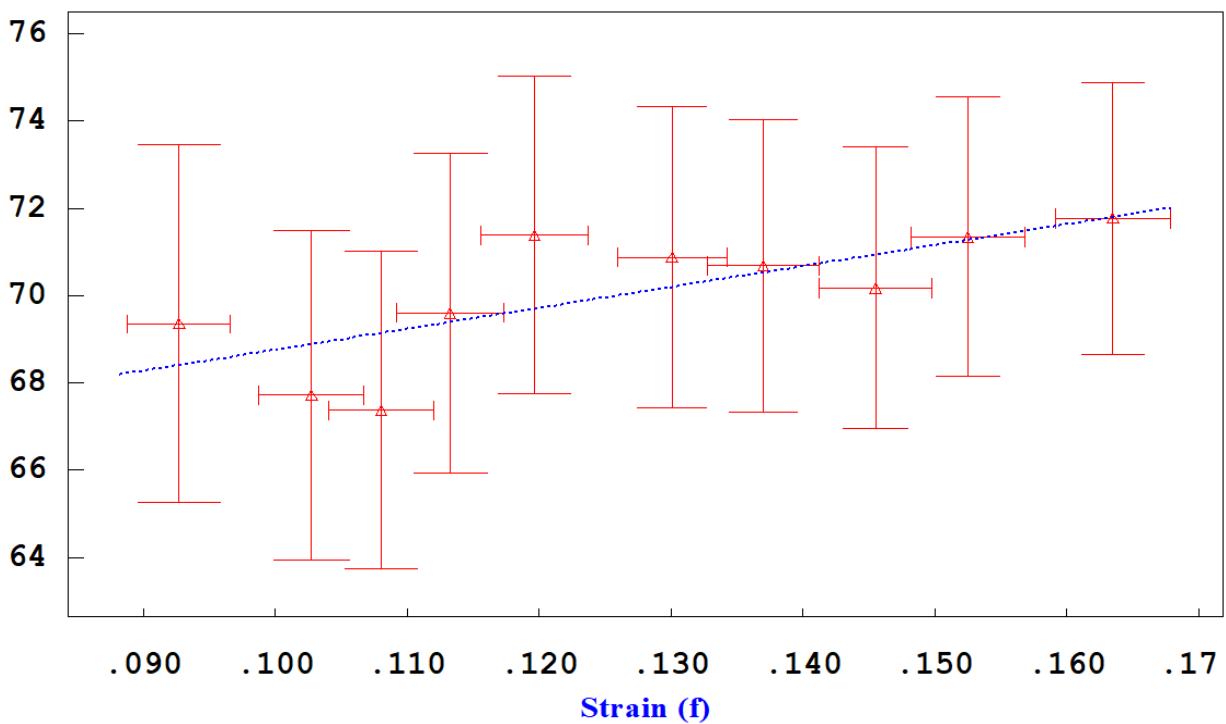

Figure S28. The fit of the volume per Yb atom to the third-order Birch-Murnaghan equation of state for I4/mmm phase of $Yb_3H_8$ compressed in Ne PTM (top) and the F–f plot (bottom). F – normalized pressure [GPa], f – Eulerian strain. run (i).

S27

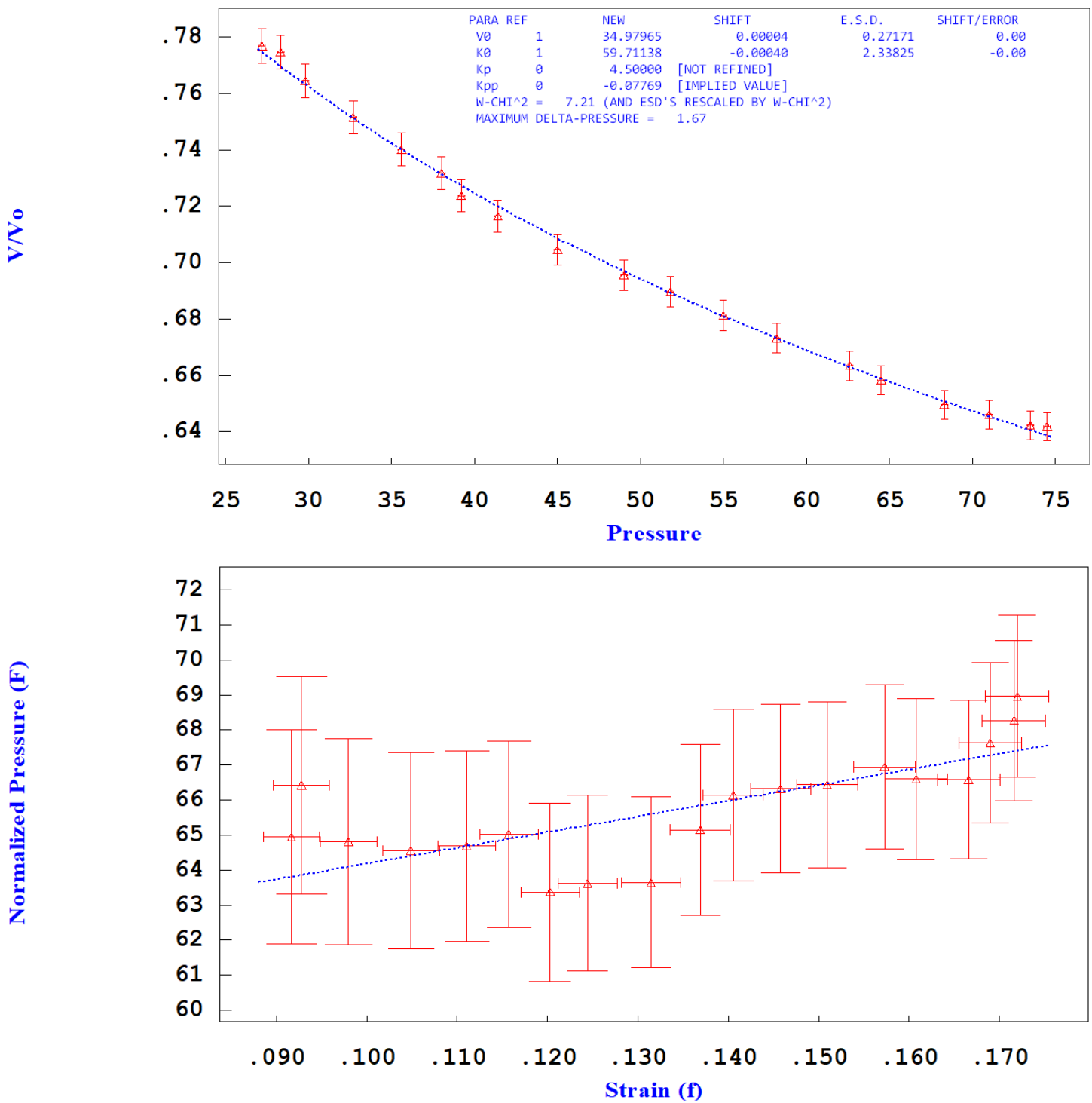

Figure S29. The fit of the volume per Yb atom to the third-order Birch-Murnaghan equation of state for I4/mmm phase of $Yb_3H_8$ compressed in $H_2$ PTM (top) and the F–f plot (bottom). F – normalized pressure [GPa], f – Eulerian strain. run (d).

Table S2. Summary of the EoS parameters reported for some of the lanthanide hydrides relevant for this study. The estimated standard deviations (E.S.D.) are given in parentheses. EoS type: B-M – Birch-Murnaghan, M – Murnaghan.

| phase | $V_0$ per metal atom [Å$^3$] | $B_0$ [GPa] | $B_0'$ | EoS type | notes | ref. |
|---|---|---|---|---|---|---|
| **SmH$_3$** $P6_3/mmc$ | 41.68(33) | 70(4) | 4 | M | silicone oil PTM | [2] |
| **SmH$_3$** $Fm$-3m | 38.69(33) | 80(4) | 4 | M | silicone oil PTM | [2] |
| **EuH$_2$** $Pnma$ | 42.6 | 39.9(5) | 4 | B-M | He PTM | [3] |
| **EuH$_2$** $P6_3/mmc$ | 39.8(1) | 44.9(9) | 4 | B-M | He PTM | [3] |
| **EuH$_{2+x}$** $I4/mmm$ | 38.8(2) | 69(2) | 4 | B-M | H$_2$ PTM; $I4/m$ phase shows similar compressibility | [3] |
| **Eu$_8$H$_{46}$** $Pm$-3n | 26.3(1) | 471(70) | 4 | B-M | DFT – close to exp. data; $V_0$ at 100 GPa | [4] |
| **EuH$_9$** $P6_3/mmc$ | 31.9(1) | 594(70) | 4 | B-M | DFT – close to exp. data; $V_0$ at 100 GPa | [4] |
| **EuH$_9$** $F$-43m | 31.3(1) | 699(27) | 4 | B-M | DFT – close to exp. data; $V_0$ at 100 GPa | [4] |
| **TbH$_3$** $P6_3/mmc$ | 39.69 | 81 | 4 | M | no PTM | [5] |
| **TbH$_3$** $Fm$-3m | 38.03 | 96 | 4 | M | no PTM | [5] |
| **DyH$_3$** $P6_3/mmc$ | 38.53 | 82 | 5.1 | M | silicone oil PTM | [6] |
| **DyH$_3$** $Fm$-3m | 35.54 | 119 | 1.9 | M | silicone oil PTM | [6] |
| **DyH$_3$** $Fm$-3m | 35.4(3) | 85(3) | 4 | B-M | H$_2$ PTM | [7] |
| **HoH$_3$** $P6_3/mmc$ | 37.69 | 87(3) | 4 | M | silicone oil PTM | [8] |
| **HoH$_3$** $Fm$-3m | 36.20 | 90(2) | 4 | M | silicone oil PTM | [8] |
| **ErH$_3$** $P6_3/mmc$ | 37.03 | 77(4) | 4 | M | silicone oil PTM | [2] |
| **ErH$_3$** $Fm$-3m | 35.70 | 81(3) | 4 | M | silicone oil PTM | [2] |
| **TmH$_3$** $P$-3c1 | 35.69 | 84.94 | 3.22 | B-M | DFT | [9] |
| **TmH$_3$** $Fm$-3m | 32.06 | 103.36 | 3.86 | B-M | DFT | [9] |
| **YbH$_2$** $Pnma$ | 35.63(7) | 40.2(22) | 4.75(45) | M | 4:1 MeOH:EtOH PTM | [10] |
| **YbH$_2$** $P6_3/mmc$ | 30.4(1) | 138(3) | 0 | M | 4:1 MeOH:EtOH PTM | [10] |

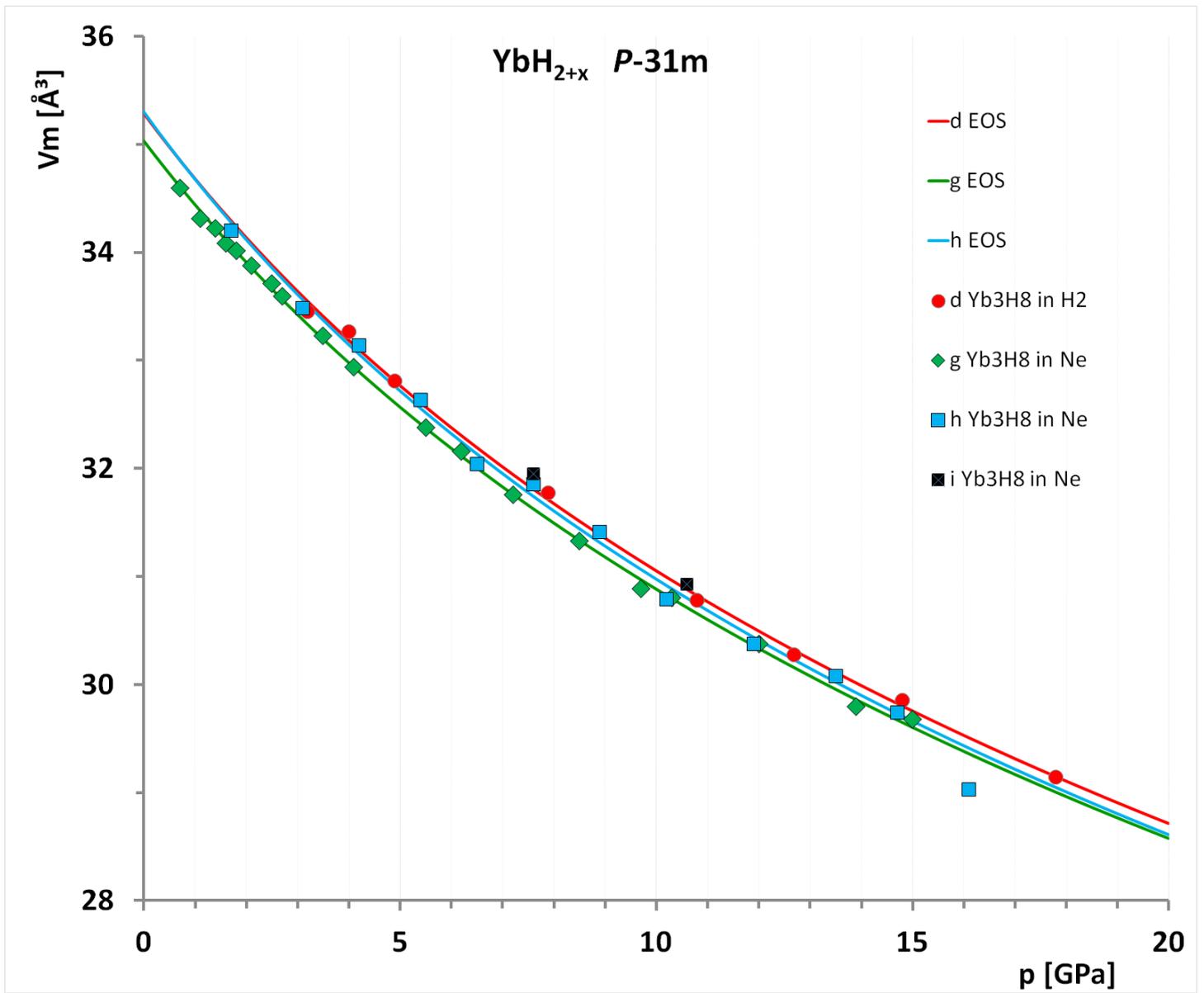

Figure S30. P–31m phase of $Yb_3H_8$ compressed in $H_2$ and Ne PTM. Comparison of the experimental data and EOS fitted.

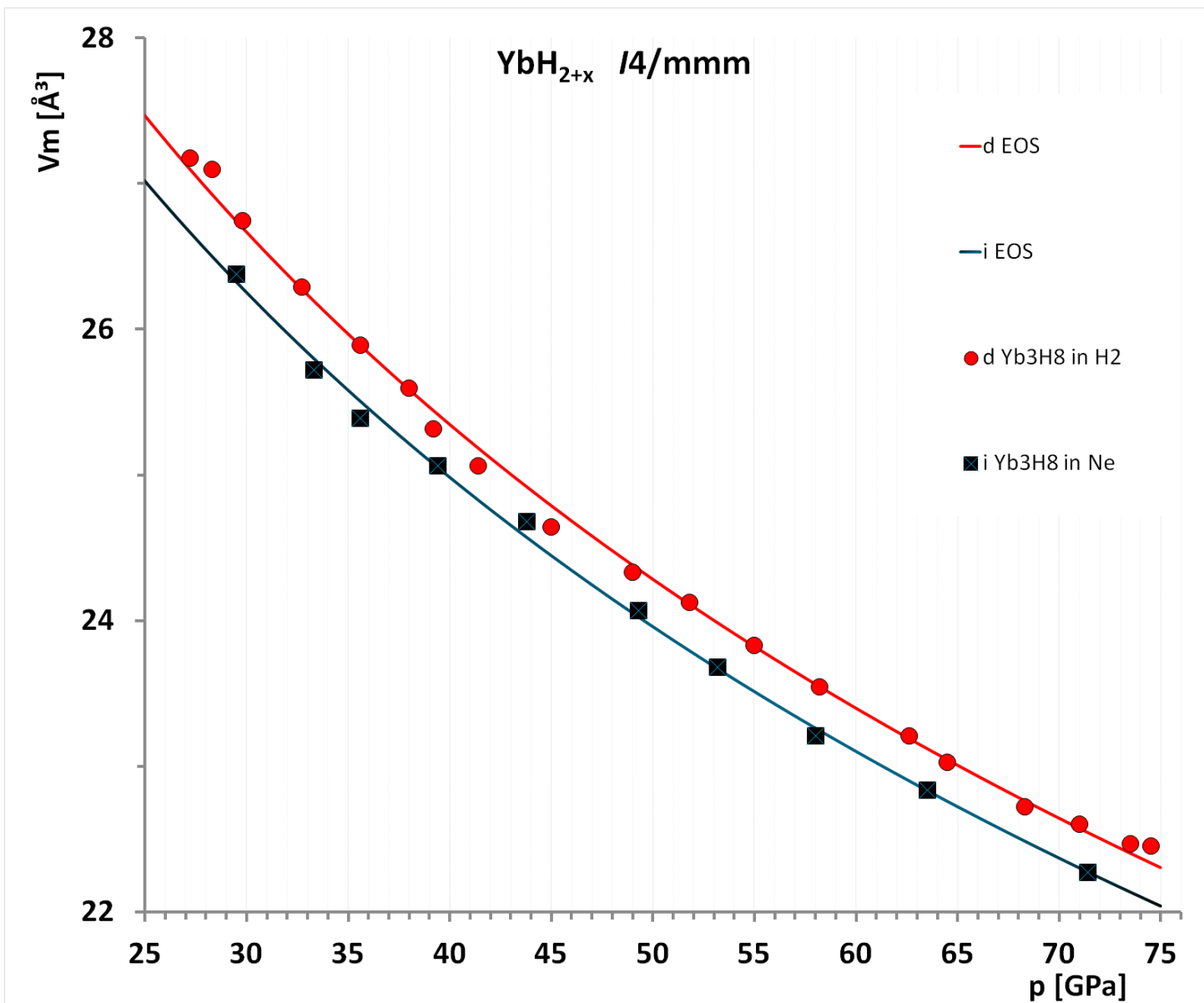

Figure S31. *I4/mmm* phase of $Yb_3H_8$ compressed in $H_2$ and Ne PTM. Comparison of the experimental data and EOS fitted.

*Table S3. Refined lattice parameters for Yb$_3$H$_8$ compressed in H$_2$ PTM. Run (d).*

| p [GPa] | YbH$_{2+x}$ phase | a [Å] | a (e.s.d.) | c [Å] | c (e.s.d) | V/Z [Å$^3$] | c/a | cwRp [%] |
|---|---|---|---|---|---|---|---|---|
| 3.2 | P-31m | 6.2625 | 9 | 8.8642 | 16 | 33.452 | 1.415 | 5.57 |
| 4.0 | | 6.2491 | 4 | 8.8526 | 12 | 33.265 | 1.417 | 4.62 |
| 4.9 | | 6.21623 | 13 | 8.8237 | 5 | 32.809 | 1.419 | 3.34 |
| 7.9 | | 6.1535 | 4 | 8.7196 | 10 | 31.771 | 1.417 | 3.27 |
| 10.8 | | 6.0898 | 5 | 8.6246 | 9 | 30.777 | 1.416 | 3.61 |
| 12.7 | | 6.0497 | 4 | 8.5950 | 16 | 30.269 | 1.421 | 3.75 |
| 14.8 | | 6.0441 | 5 | 8.4926 | 4 | 29.853 | 1.405 | 4.00 |
| 17.8 | | 5.9762 | 8 | 8.4788 | 18 | 29.139 | 1.419 | 3.86 |
| 14.8 | I4/m | 7.7780 | 3 | 4.8923 | 3 | 29.597 | 0.629 | 4.00 |
| 17.8 | | 7.6724 | 3 | 4.9259 | 3 | 28.997 | 0.642 | 3.86 |
| 20.6 | | 7.6386 | 2 | 4.8978 | 3 | 28.578 | 0.641 | 5.65 |
| 21.7 | | 7.6180 | 4 | 4.8849 | 5 | 28.349 | 0.641 | 7.52 |
| 24.0 | | 7.5837 | 4 | 4.8612 | 5 | 27.958 | 0.641 | 6.83 |
| 27.2 | I4/mmm | 3.3679 | 3 | 4.7916 | 7 | 27.175 | 1.423 | 6.81 |
| 28.3 | | 3.3613 | 4 | 4.7966 | 8 | 27.097 | 1.427 | 7.02 |
| 29.8 | | 3.3470 | 5 | 4.7748 | 7 | 26.745 | 1.427 | 6.50 |
| 32.7 | | 3.3270 | 9 | 4.7498 | 13 | 26.288 | 1.428 | 8.59 |
| 35.6 | | 3.3100 | 13 | 4.726 | 2 | 25.889 | 1.428 | 11.18 |
| 38.0 | | 3.2974 | 7 | 4.7082 | 9 | 25.596 | 1.428 | 7.26 |
| 39.2 | | 3.2837 | 8 | 4.6955 | 11 | 25.315 | 1.430 | 7.72 |
| 41.4 | | 3.2735 | 9 | 4.6778 | 13 | 25.063 | 1.429 | 7.76 |
| 45.0 | | 3.2545 | 11 | 4.6537 | 18 | 24.645 | 1.430 | 8.42 |
| 49.0 | | 3.2393 | 4 | 4.6375 | 9 | 24.331 | 1.432 | 8.38 |
| 51.8 | | 3.2296 | 13 | 4.6258 | 17 | 24.124 | 1.432 | 8.66 |
| 55.0 | | 3.2162 | 11 | 4.6075 | 14 | 23.830 | 1.433 | 7.34 |
| 58.2 | | 3.2027 | 13 | 4.5912 | 19 | 23.547 | 1.434 | 8.51 |
| 62.6 | | 3.1869 | 14 | 4.570 | 2 | 23.207 | 1.434 | 9.10 |
| 64.5 | | 3.1778 | 13 | 4.560 | 2 | 23.024 | 1.435 | 8.82 |
| 68.3 | | 3.1642 | 16 | 4.5392 | 18 | 22.724 | 1.435 | 6.86 |
| 71.0 | | 3.1575 | 4 | 4.5341 | 10 | 22.602 | 1.436 | 8.22 |
| 73.5 | | 3.1510 | 4 | 4.5261 | 10 | 22.469 | 1.436 | 7.67 |
| 74.5 | | 3.1506 | 10 | 4.5236 | 17 | 22.451 | 1.436 | 8.31 |

Table S4. Refined lattice parameters for Yb$_3$H$_8$ compressed in Ne PTM. Run (g).

| p [GPa] | YbH$_{2+x}$ phase | a [Å] | a (e.s.d.) | c [Å] | c (e.s.d) | V/Z [Å$^3$] | c/a | cwRp [%] |
|---|---|---|---|---|---|---|---|---|
| 0.7 | P-31m | 6.3317 | 6 | 8.967 | 3 | 34.592 | 1.416 | 5.31 |
| 1.1 | | 6.3189 | 8 | 8.9311 | 18 | 34.314 | 1.413 | 5.47 |
| 1.4 | | 6.3144 | 5 | 8.9195 | 12 | 34.221 | 1.413 | 4.78 |
| 1.6 | | 6.3062 | 9 | 8.9072 | 12 | 34.085 | 1.412 | 4.82 |
| 1.8 | | 6.3024 | 9 | 8.8999 | 12 | 34.016 | 1.412 | 4.97 |
| 2.1 | | 6.2944 | 9 | 8.8869 | 10 | 33.880 | 1.412 | 4.81 |
| 2.5 | | 6.2836 | 7 | 8.8737 | 11 | 33.714 | 1.412 | 4.43 |
| 2.7 | | 6.2704 | 4 | 8.8793 | 16 | 33.594 | 1.416 | 4.78 |
| 3.5 | | 6.2485 | 8 | 8.8444 | 17 | 33.228 | 1.415 | 5.31 |
| 4.1 | | 6.2297 | 5 | 8.8207 | 16 | 32.940 | 1.416 | 5.40 |
| 5.5 | | 6.1930 | 5 | 8.7739 | 15 | 32.380 | 1.417 | 5.89 |
| 6.2 | | 6.1862 | 7 | 8.7334 | 16 | 32.160 | 1.412 | 5.43 |
| 7.2 | | 6.1605 | 9 | 8.696 | 2 | 31.757 | 1.412 | 6.06 |
| 8.5 | | 6.1270 | 7 | 8.6720 | 18 | 31.326 | 1.415 | 6.23 |
| 9.7 | | 6.0967 | 6 | 8.6347 | 17 | 30.883 | 1.416 | 6.26 |
| 10.3 | | 6.0908 | 9 | 8.6283 | 18 | 30.801 | 1.417 | 6.44 |
| 12 | | 6.0617 | 7 | 8.5905 | 20 | 30.374 | 1.417 | 6.73 |
| 13.9 | | 6.0427 | 3 | 8.4800 | 14 | 29.795 | 1.403 | 6.30 |
| 15 | | 6.0333 | 9 | 8.4730 | 13 | 29.678 | 1.404 | 6.47 |
| 16.0 | I4/m | 7.7369 | 5 | 4.9373 | 7 | 29.554 | 0.638 | 4.61 |
| 17.6 | | 7.6878 | 5 | 4.9117 | 7 | 29.029 | 0.639 | 5.82 |
| 19.3 | | 7.6529 | 4 | 4.8981 | 5 | 28.687 | 0.640 | 4.58 |
| 21 | | 7.6211 | 3 | 4.8752 | 5 | 28.316 | 0.640 | 3.41 |
| 22.6 | | 7.5853 | 3 | 4.8494 | 6 | 27.902 | 0.639 | 3.79 |
| 24.8 | | 7.5378 | 7 | 4.8156 | 7 | 27.361 | 0.639 | 4.36 |
| 26 | | 7.5088 | 4 | 4.7985 | 5 | 27.055 | 0.639 | 4.27 |
| 27.2 | | 7.4926 | 4 | 4.7883 | 5 | 26.881 | 0.639 | 4.71 |
| 28.9 | I4/mmm | 3.3381 | 4 | 4.7733 | 7 | 26.594 | 1.430 | 5.34 |
| 29.8 | | 3.332 | 4 | 4.7637 | 6 | 26.444 | 1.430 | 5.12 |
| 30.6 | | 3.324 | 3 | 4.7535 | 5 | 26.261 | 1.430 | 5.73 |
| 32.3 | | 3.3154 | 2 | 4.7425 | 6 | 26.064 | 1.430 | 5.48 |
| 33.3 | | 3.3079 | 4 | 4.7316 | 6 | 25.887 | 1.430 | 5.15 |
| 34.6 | | 3.3036 | 7 | 4.7252 | 10 | 25.785 | 1.430 | 6.95 |
| 35.7 | | 3.296 | 8 | 4.7148 | 14 | 25.610 | 1.430 | 7.35 |
| 37.9 | | 3.2882 | 7 | 4.7042 | 13 | 25.432 | 1.431 | 7.09 |
| 39.3 | | 3.2794 | 7 | 4.6923 | 13 | 25.232 | 1.431 | 7.21 |
| 41.5 | | 3.2681 | 7 | 4.6783 | 16 | 24.983 | 1.432 | 7.92 |

Table S5. Refined lattice parameters for Yb$_3$H$_8$ compressed in Ne PTM. Run (i).

| p [GPa] | YbH$_{2+x}$ phase | a [Å] | a (e.s.d.) | c [Å] | c (e.s.d) | V/Z [Å$^3$] | c/a | cwRp [%] |
|---|---|---|---|---|---|---|---|---|
| 7.6 | P-31m | 6.15735 | 9 | 8.7581 | 3 | 31.951 | 1.422 | 4.32 |
| 10.6 | | 6.0946 | 2 | 8.6523 | 7 | 30.925 | 1.420 | 6.83 |
| 17.1 | I4/m | 7.7117 | 3 | 4.8952 | 5 | 29.112 | 0.635 | 5.56 |
| 19.3 | | 7.6438 | 7 | 4.8827 | 8 | 28.528 | 0.639 | 8.70 |
| 23.9 | | 7.5509 | 3 | 4.8080 | 4 | 27.413 | 0.637 | 6.50 |
| 29.5 | I4/mmm | 3.3469 | 2 | 4.7099 | 6 | 26.380 | 1.407 | 7.85 |
| 33.3 | | 3.3054 | 2 | 4.7086 | 5 | 25.722 | 1.425 | 8.73 |
| 35.6 | | 3.2930 | 5 | 4.6825 | 4 | 25.388 | 1.422 | 7.99 |
| 39.4 | | 3.2797 | 5 | 4.6601 | 8 | 25.063 | 1.421 | 11.59 |
| 43.8 | | 3.2619 | 6 | 4.6386 | 10 | 24.677 | 1.422 | 8.88 |
| 49.3 | | 3.2344 | 4 | 4.6011 | 6 | 24.067 | 1.423 | 10.51 |
| 53.2 | | 3.2185 | 2 | 4.5714 | 5 | 23.677 | 1.420 | 8.70 |
| 58 | | 3.1957 | 5 | 4.5453 | 7 | 23.209 | 1.422 | 8.22 |
| 63.5 | | 3.17532 | 19 | 4.5299 | 7 | 22.837 | 1.427 | 10.74 |
| 71.4 | | 3.1497 | 2 | 4.4900 | 7 | 22.272 | 1.426 | 12.51 |

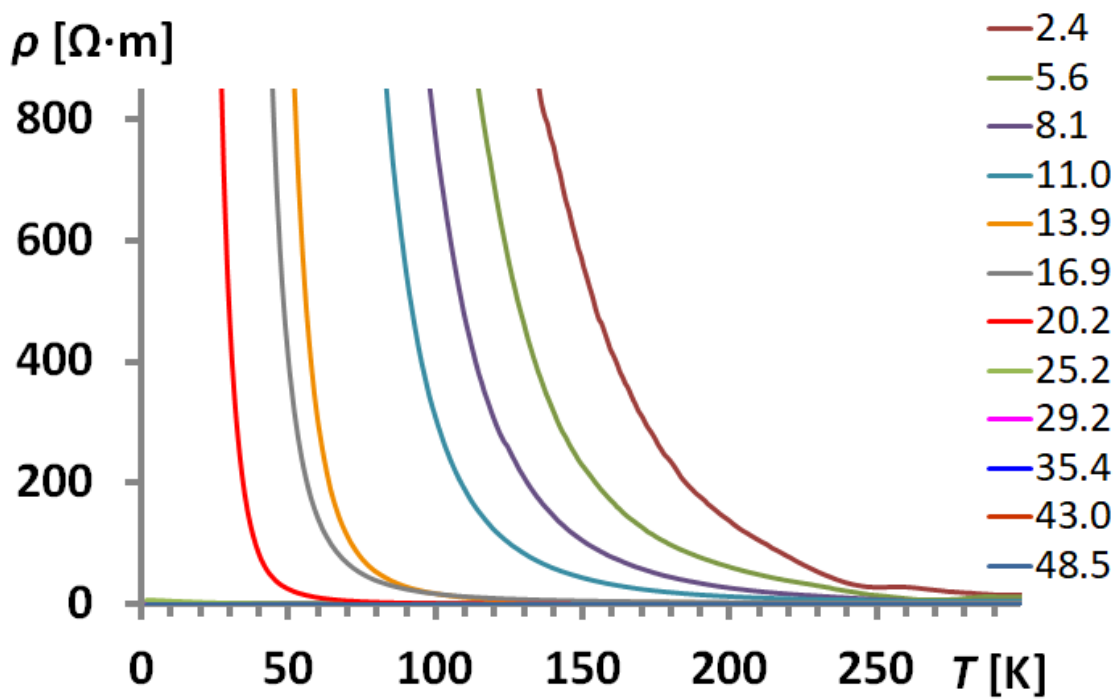
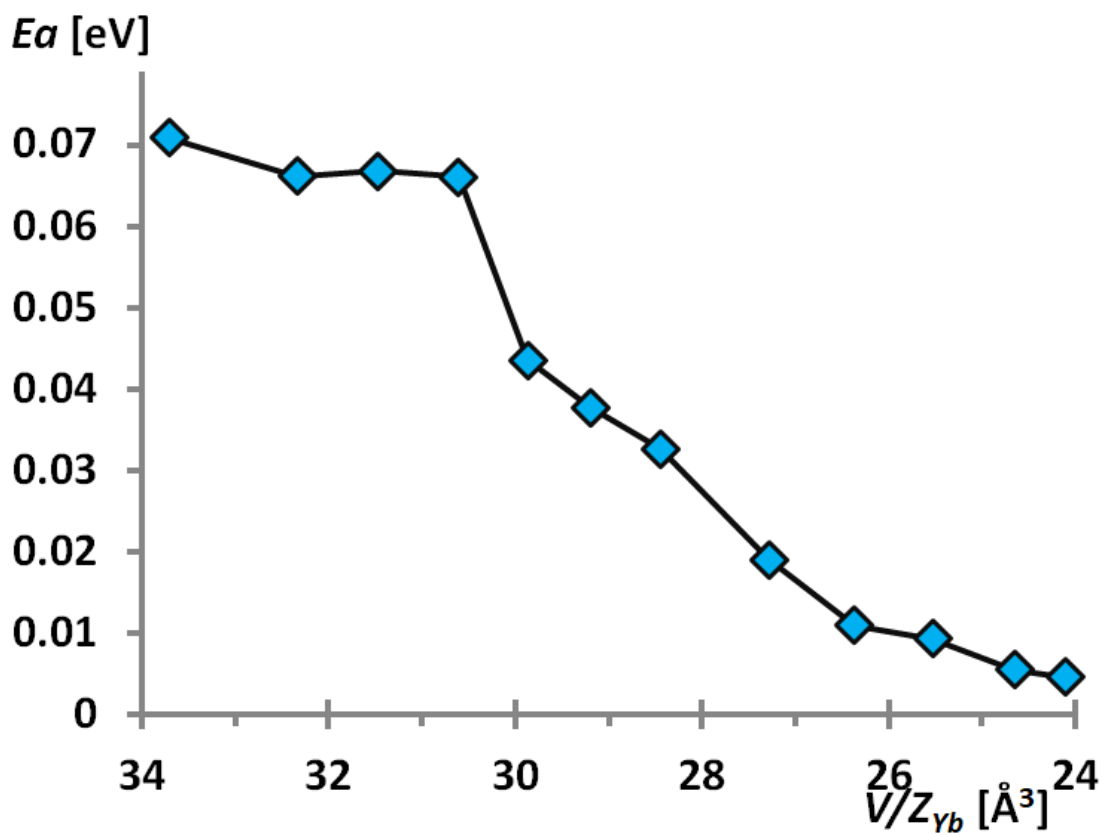

Figure S32. The resistivity vs. T [K] for Yb$_3$H$_8$ compressed in NaCl presented in a linear scale indicating semiconductor-like behavior (top). Evolution of the activation energy of the electronic transport in the function of volume per ytterbium atom (bottom).

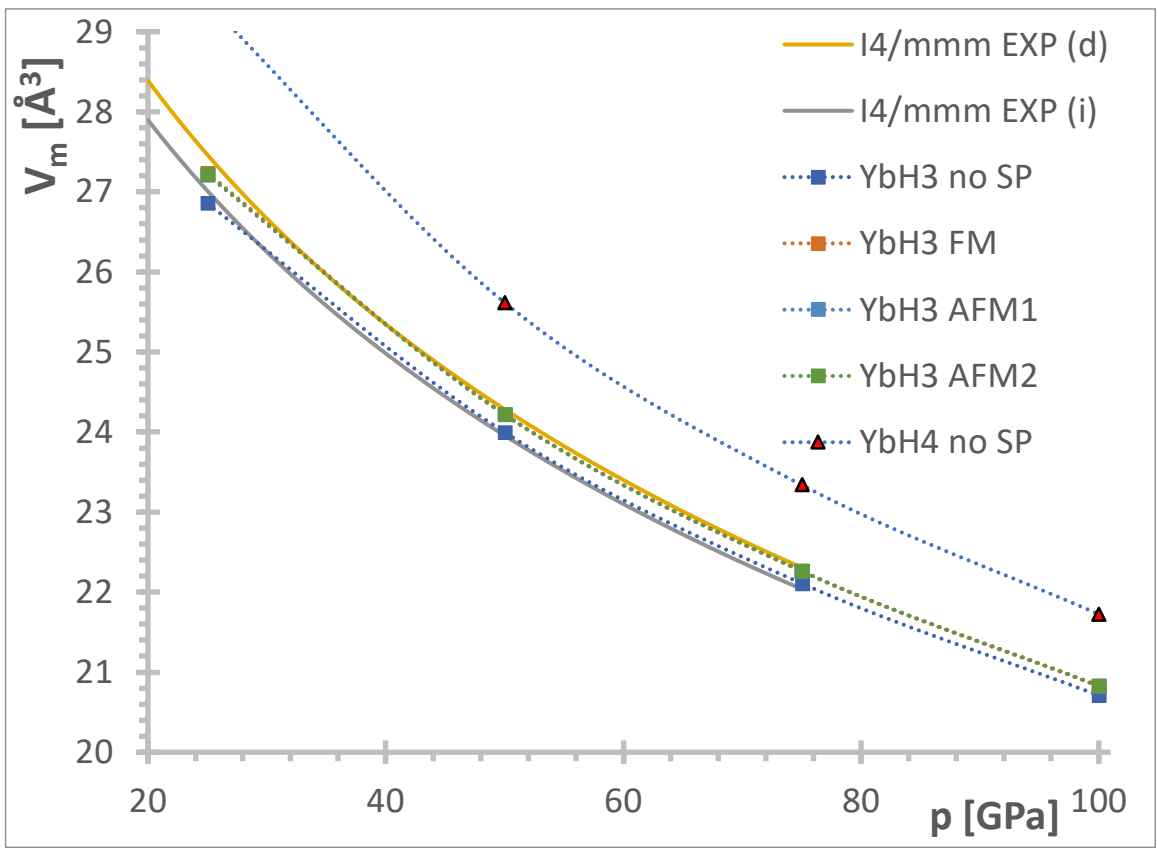
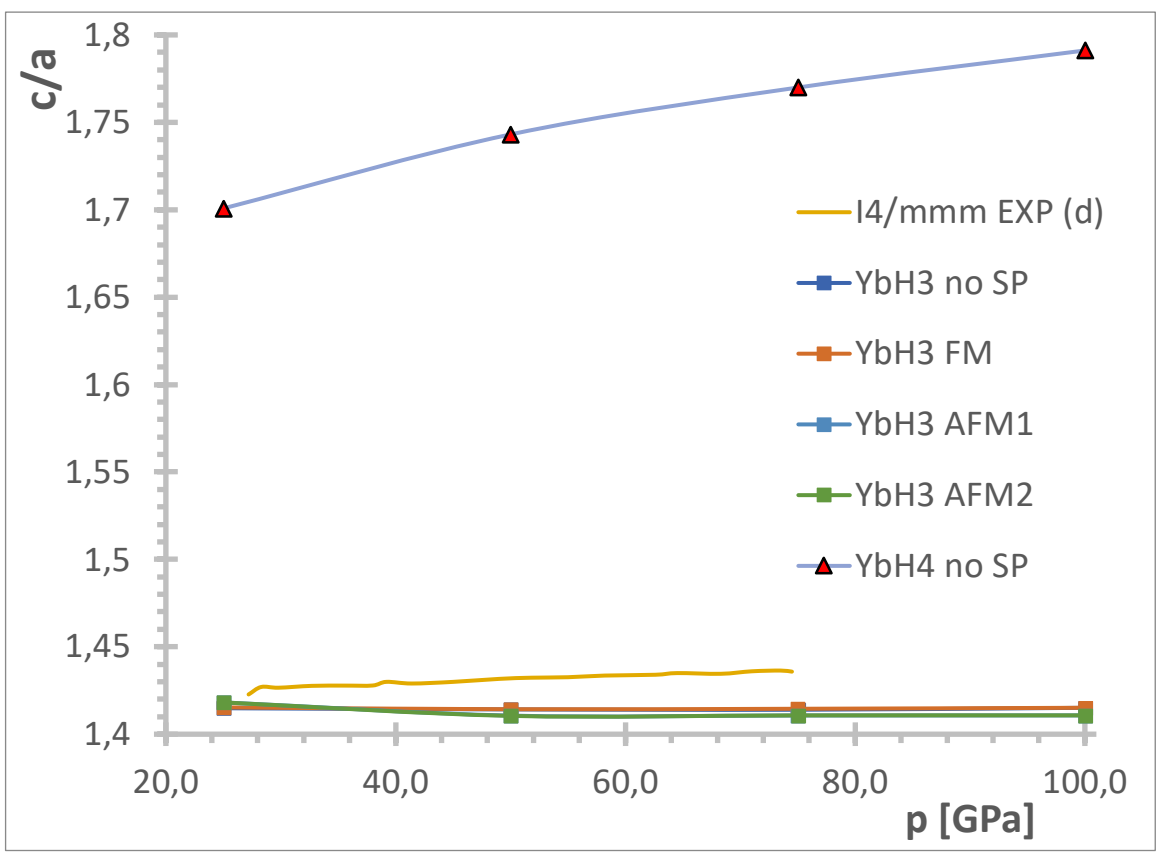

Figure S33. Top – The calculated molar volume of YbH$_3$ and YbH$_4$ (both I4/mmm) compared to the experimental results. Bottom – the calculated c/a ratio as compared to the experimental results.

# CIF files: basic parameters, experimental data

## Yb$_3$H$_8$ P-31m

Yb3H8 in Ne, 1.7 GPa (run (h)); LeBail fit (*cf. Fig. S16*)

```
#========================================================================
# CRYSTAL DATA
#----------------------------------------------------------------------
data_ phase_1

_chemical_formula_sum          'H2.67 Yb1'
_cell_length_a               6.30366(9)
_cell_length_b               6.30366(9)
_cell_length_c               8.9455(3)
_cell_angle_alpha              90
_cell_angle_beta               90
_cell_angle_gamma             120
_cell_volume                 307.835(12)
_space_group_name_H-M_alt      'P -3 1 m'
_space_group_IT_number          162

loop_
_space_group_symop_operation_xyz
  'x, y, z'
  '-x, -y, -z'
  '-y, x-y, z'
  'y, -x+y, -z'
  '-x+y, -x, z'
  'x-y, x, -z'
  '-y, -x, -z'
  'y, x, z'
  '-x+y, y, -z'
  'x-y, -y, z'
  'x, x-y, -z'
  '-x, -x+y, z'

loop_
  _atom_site_label
  _atom_site_occupancy
  _atom_site_fract_x
  _atom_site_fract_y
  _atom_site_fract_z
  _atom_site_adp_type
  _atom_site_U_iso_or_equiv
  _atom_site_type_symbol
  Yb1   1.0   0.000000   0.000000   0.000000    Uiso  0.025800 Yb
  Yb2   1.0   0.333333   0.666667   0.000000    Uiso  0.025800 Yb
  Yb3   1.0   0.3484(7)  0.000000   0.3292(13)  Uiso  0.025800 Yb
  H1    1.0   0.000000   0.000000   0.644000    Uiso  0.019000 H
  H2    1.0   0.333333   0.666667   0.219000    Uiso  0.019000 H
  H3    1.0   0.322000   0.000000   0.581200    Uiso  0.019000 H
  H4    1.0   0.356000   0.000000   0.075700    Uiso  0.019000 H
  H5    1.0   0.236400   0.000000   0.831900    Uiso  0.019000 H
```

**"YbH₃" *I*4/m**

Yb + H₂, sample heated *in situ*, *ca.* 12.5 GPa (run (c)); Rietveld fit (*cf. Fig. S8*)

```
#====================================================================
# CRYSTAL DATA
#--------------------------------------------------------------------
data_ phase_2

_chemical_formula_sum            'H3 Yb1'
_cell_length_a                   7.8268(3)
_cell_length_b                   7.8268(3)
_cell_length_c                   5.0150(3)
_cell_angle_alpha                90
_cell_angle_beta                 90
_cell_angle_gamma                90
_cell_volume                     307.21(2)
_space_group_name_H-M_alt        'I 4/m'
_space_group_IT_number           87

loop_
_space_group_symop_operation_xyz
 'x, y, z'
 '-x, -y, -z'
 '-x, -y, z'
 'x, y, -z'
 '-y, x, z'
 'y, -x, -z'
 'y, -x, z'
 '-y, x, -z'
 'x+1/2, y+1/2, z+1/2'
 '-x+1/2, -y+1/2, -z+1/2'
 '-x+1/2, -y+1/2, z+1/2'
 'x+1/2, y+1/2, -z+1/2'
 '-y+1/2, x+1/2, z+1/2'
 'y+1/2, -x+1/2, -z+1/2'
 'y+1/2, -x+1/2, z+1/2'
 '-y+1/2, x+1/2, -z+1/2'

loop_
  _atom_site_label
  _atom_site_occupancy
  _atom_site_fract_x
  _atom_site_fract_y
  _atom_site_fract_z
  _atom_site_adp_type
  _atom_site_U_iso_or_equiv
  _atom_site_type_symbol
  Yb1    1.0   0.000000    0.000000    0.000000   Uiso 0.010000 Yb
  Yb2    1.0   0.4109(7)   0.2013(13)  0.000000   Uiso 0.010000 Yb
  H1     1.0   0.000000    0.000000    0.500000   Uiso 0.010000 H
  H2     1.0   0.091400    0.307700    0.000000   Uiso 0.010000 H
  H3     1.0   0.202200    0.107800    0.239600   Uiso 0.010000 H
  H4     1.0   0.000000    0.500000    0.250000   Uiso 0.010000 H
```

**"YbH₃" *I*4/mmm**

Yb3H8 in Ne, 28.9 GPa (run (g)); LeBail fit (*cf. Fig. S19*). The H occupancies may vary.

```
#====================================================================
# CRYSTAL DATA
#--------------------------------------------------------------------
data_ phase_3

_chemical_formula_sum          'H3 Yb1'
_cell_length_a                 3.3381(4)
_cell_length_b                 3.3381(4)
_cell_length_c                 4.7733(7)
_cell_angle_alpha              90
_cell_angle_beta               90
_cell_angle_gamma              90
_cell_volume                   53.190(12)
_space_group_name_H-M_alt      'I 4/m m m'
_space_group_IT_number         139

loop_
_space_group_symop_operation_xyz
 'x, y, z'
 '-x, -y, -z'
 '-x, -y, z'
 'x, y, -z'
 '-y, x, z'
 'y, -x, -z'
 'y, -x, z'
 '-y, x, -z'
 '-x, y, -z'
 'x, -y, z'
 'x, -y, -z'
 '-x, y, z'
 'y, x, -z'
 '-y, -x, z'
 '-y, -x, -z'
 'y, x, z'
 'x+1/2, y+1/2, z+1/2'
 '-x+1/2, -y+1/2, -z+1/2'
 '-x+1/2, -y+1/2, z+1/2'
 'x+1/2, y+1/2, -z+1/2'
 '-y+1/2, x+1/2, z+1/2'
 'y+1/2, -x+1/2, -z+1/2'
 'y+1/2, -x+1/2, z+1/2'
 '-y+1/2, x+1/2, -z+1/2'
 '-x+1/2, y+1/2, -z+1/2'
 'x+1/2, -y+1/2, z+1/2'
 'x+1/2, -y+1/2, -z+1/2'
 '-x+1/2, y+1/2, z+1/2'
 'y+1/2, x+1/2, -z+1/2'
 '-y+1/2, -x+1/2, z+1/2'
 '-y+1/2, -x+1/2, -z+1/2'
 'y+1/2, x+1/2, z+1/2'

loop_
 _atom_site_label
 _atom_site_occupancy
```

```
_atom_site_fract_x
_atom_site_fract_y
_atom_site_fract_z
_atom_site_adp_type
_atom_site_U_iso_or_equiv
_atom_site_type_symbol
Yb1    1.0   0.000000    0.000000    0.000000   Uiso  0.020000 Yb
H1     1.0   0.000000    0.000000    0.500000   Uiso  0.020000 H
H2     1.0   0.000000    0.500000    0.250000   Uiso  0.020000 H
```

Yb3H8 in H$_2$, 73.5 GPa (run (d)); LeBail fit (*cf. Fig. S14*). The H occupancies may vary.

```
#======================================================================
# CRYSTAL DATA
#----------------------------------------------------------------------
data_ phase_3

_chemical_formula_sum           'H3 Yb1'
_cell_length_a                  3.1510(4)
_cell_length_b                  3.1510(4)
_cell_length_c                  4.5261(10)
_cell_angle_alpha               90
_cell_angle_beta                90
_cell_angle_gamma               90
_cell_volume                    44.938(12)
_space_group_name_H-M_alt       'I 4/m m m'
_space_group_IT_number          139

loop_
_space_group_symop_operation_xyz
 'x, y, z'
 '-x, -y, -z'
 '-x, -y, z'
 'x, y, -z'
 '-y, x, z'
 'y, -x, -z'
 'y, -x, z'
 '-y, x, -z'
 '-x, y, -z'
 'x, -y, z'
 'x, -y, -z'
 '-x, y, z'
 'y, x, -z'
 '-y, -x, z'
 '-y, -x, -z'
 'y, x, z'
 'x+1/2, y+1/2, z+1/2'
 '-x+1/2, -y+1/2, -z+1/2'
 '-x+1/2, -y+1/2, z+1/2'
 'x+1/2, y+1/2, -z+1/2'
 '-y+1/2, x+1/2, z+1/2'
 'y+1/2, -x+1/2, -z+1/2'
```

'y+1/2, -x+1/2, z+1/2'
'-y+1/2, x+1/2, -z+1/2'
'-x+1/2, y+1/2, -z+1/2'
'x+1/2, -y+1/2, z+1/2'
'x+1/2, -y+1/2, -z+1/2'
'-x+1/2, y+1/2, z+1/2'
'y+1/2, x+1/2, -z+1/2'
'-y+1/2, -x+1/2, z+1/2'
'-y+1/2, -x+1/2, -z+1/2'
'y+1/2, x+1/2, z+1/2'

loop_
 _atom_site_label
 _atom_site_occupancy
 _atom_site_fract_x
 _atom_site_fract_y
 _atom_site_fract_z
 _atom_site_adp_type
 _atom_site_U_iso_or_equiv
 _atom_site_type_symbol
Yb1   1.0   0.000000   0.000000   0.000000   Uiso   0.012700   Yb
H1    1.0   0.000000   0.000000   0.500000   Uiso   0.038000   H
H2    1.0   0.000000   0.500000   0.250000   Uiso   0.038000   H